\documentclass[%
 reprint,
 superscriptaddress,
nofootinbib,
 amsmath,amssymb,
 aps,
 prd,
]{revtex4-2}

\usepackage{graphicx}                   
\usepackage{dcolumn}                    
\usepackage{bm}                         
\usepackage{hyperref}                   
\usepackage{physics}                    
\usepackage[noabbrev]{cleveref}         
\usepackage{braket}                     
\usepackage{mathtools}                  
\usepackage{siunitx}                    
\usepackage{multirow,booktabs,xfrac}
\usepackage{tikz}           
\usepackage{subcaption}
\usepackage{svg}
\usepackage{array}

\usepackage{personal-macros}

\newcommand{\scA}{\Sigma} 
 
\svgpath{{./Figures/svg/}}
\svgsetup{inkscapepath=./Figures/svgpdf}
\def\pdfpath {./Figures/pdf}
\def\pngpath {./Figures/png}
\def\epspath {./Figures/eps}
\graphicspath{{./}{\pdfpath/}{\epspath/}{\pngpath/}}
\DeclareGraphicsExtensions{.pdf,.eps,.png}

\begin{document}

\preprint{}

\title{Multiple breaking patterns in the Brout-Englert-Higgs effect\\ beyond perturbation theory}

\author{Elizabeth Dobson}
\email{elizabeth.dobson@uni-graz.at}
\author{Axel Maas}
\email{axel.maas@uni-graz.at}
\author{Bernd Riederer}
\email{bernd.riederer@uni-graz.at}
\affiliation{Institute of Physics, NAWI Graz, University of Graz, Universitätsplatz 5, A-8010 Graz, Austria}

\date{\today}

\begin{abstract}
  In many BSM theories, especially GUTs, introducing a Brout-Englert-Higgs effect allows for multiple breaking patterns of the gauge symmetry. The possibility to select a particular pattern is usually decisive for the phenomenological viability of a theory. Beyond perturbation theory it is necessary to replace the Brout-Englert-Higgs effect by a manifestly gauge-invariant description. We study the simplest case with multiple breaking patterns, an \SU[3] Yang-Mills theory coupled to a single scalar `Higgs' field in the adjoint representation, on the lattice. We find that only one pattern remains at fixed parameters and gauge-fixing strategy, and that the associated quantum effective potential emerges from a non-trivial interplay of many aspects.
\end{abstract}

\maketitle

\section{Introduction}\label{sec:introduction}

Many candidate theories for physics beyond the standard model (BSM) consist of a gauge group and matter fields in representations of the gauge group such that multiple little groups are possible. Especially in grand-unified theories (GUTs) \cite{Langacker:1980js,O'Raifeartaigh:1986vq}, the usual phenomenology requires the choice of a particular little group, which coincides with the standard model gauge group\footnote{Tumbling theories \cite{Raby:1979my} will face the same issues eventually.}. This is usually done via a suitable gauge-fixing and a minimization of the corresponding (quantum) effective potential \cite{Bohm:2001yx,O'Raifeartaigh:1986vq}, which typically offers multiple degenerate solutions at tree level. Often, these degeneracies are lifted perturbatively beyond tree level, yielding a unique little group with the desired properties.

This approach, however, is not gauge invariant. Even in the presence of a Brout-Englert-Higgs (BEH) effect, Becchi-Rouet-Stora-Tyutin (BRST) invariance is insufficient \cite{Fujikawa:1982ss} to guarantee gauge independence of the results, due to the Gribov-Singer ambiguity \cite{Gribov:1977wm,Singer:1978dk}. Since physics must be independent of the choice of gauge fixing \cite{Frohlich:1980gj,Frohlich:1981yi,Banks:1979fi}, the physical, fully gauge-invariant spectrum needs to be built from gauge-invariant composite operators \cite{Frohlich:1980gj,Frohlich:1981yi,Banks:1979fi}. This will generally yield a different spectrum than the gauge-dependent one of the elementary degrees of freedom \cite{Maas:2015gma,Maas:2017xzh,Sondenheimer:2019idq}, but can be described by an augmented perturbation theory \cite{Maas:2017xzh,Sondenheimer:2019idq,Maas:2020kda,Dudal:2020uwb}, implementing the Fröhlich-Morchio-Strocchi (FMS) mechanism \cite{Frohlich:1980gj,Frohlich:1981yi}. This effect and its description using augmented perturbation theory has been supported in lattice calculations \cite{Maas:2016ngo,Maas:2018xxu,Afferrante:2020hqe,Greensite:2020lmh,Dobson:2021sgl} and a review of the situation can be found in \cite{Maas:2017wzi}. This naturally raises the question whether the selection of the little group is also affected, and whether the selection of the physical, gauge-invariant realization by means of a BEH effect is possible at all \cite{Maas:2017xzh}. If not, what determines the dynamics and the spectrum of the theory? These are important questions to address, given the phenomenological relevance of GUTs.

A definitive answer requires a fully non-perturbative approach, which can determine the physical and gauge-fixed quantities equally well. Lattice gauge theory is one such possibility. While there have been various investigations which could access these questions in the past \cite{Olynyk:1984pz,Lee:1985yi,Olynyk:1985tr,Olynyk:1985cd,Kajantie:1998yc}, they have either focused only on the physical results or used perturbative means to cover the gauge-fixed sector. The aim here is to cover both aspects on equal footing.

We therefore concentrate on the simplest model which exhibits the aforementioned properties: an \SU[3] gauge theory, coupled to a single scalar `Higgs' field in the adjoint representation. Lacking the strong sector, it is not a viable candidate for a full GUT, but it is the simplest non-Abelian theory with multiple little groups, while being at the same time computationally accessible.

We start by summarizing the usual perturbative construction and the obstructions for this theory in section \ref{s:patterns}. We then discuss the structures of the symmetries of this theory in detail in section \ref{s:symmetry}, and the implication for different gauge choices in section \ref{s:gauge}. The technical aspects of the lattice implementation are given in section \ref{s:lattice}. This section can be skipped if the reader is not interested in these details. We examine the phase structure and symmetries of the theory in section \ref{s:results}, and analyse the implications of our results. In this context we also see that the gauge-dependent degrees of freedom do indeed behave as expected in most parts of the phase diagram, and we do not have an artificial contradiction by erroneously calculating at an effectively strong coupling.

Our results show that only a single breaking pattern survives for a given set of parameters and gauge-fixing strategy. We also determine the full quantum effective potential to see that this does not just come about by lifting a degeneracy between different minima, but that the other minima are found to be eliminated altogether. However, we find that the choice of gauge fixing strongly affects which options are possible. In particular, unitary gauge and Landau--'t Hooft gauge show substantially different behaviours, and there is subtlety in precisely defining a suitable gauge. We thus conclude in section \ref{s:summary} with the somewhat pessimistic outlook that there is much less freedom in choosing the dynamics of a GUT than seems to be possible at the perturbative level.

We supplement this work with several technical appendices, including an outlook how to generalize the present constructions beyond \SU[3], which is still a somewhat special case. Some preliminary results can also be found in \cite{Dobson:2021sgl}.

\section{Multiple breaking patterns at tree-level}\label{s:patterns}

Our adjoint \SU[3] gauge-scalar theory can be described by the Lagrangian
\begin{align}\label{eqn:general_lagrangian}
  \cL        & = -\frac{1}{4}W_{\mu\nu}W^{\mu\nu}+2\tr[(D_\mu\Sigma)^\dag (D^\mu\Sigma)]- V(\Sigma)\nonumber \\
  W^{\mu\nu} & =\pd^\mu W^\nu-\pd^\nu W^\mu + ig [W^\mu,W^\nu]\,,
\end{align}
where $W^{\mu\nu}$ is the gauge field strength tensor and $V(\Sigma)$ is a gauge-invariant potential that allows for a BEH effect. The scalar field $\Sigma$ is in the adjoint representation of \SU[3], and is acted on by the covariant derivative $D_\mu$ as
\begin{equation}
  (D_\mu\scA)(x) = (\pd_\mu\scA)(x) + i g[W_\mu(x),\scA(x)]\,.
\end{equation}
Under a gauge transformation $G$, the scalar transforms as
\begin{equation}\label{eq:adj_gt}
  \scA(x)\to G(x)\scA(x)G^\dag(x)\,.
\end{equation}
For later convenience, it is also useful to define the fields in terms of the corresponding $\RR^8$ coefficient vectors as
\begin{equation}\label{eqn:vec}
  \Sigma=\Sigma^a T_a \qquad W_\mu = W_\mu^a T_a \qqtext{with} T_a = \frac{\lambda_a}{2}\,,
\end{equation}
where the $T^a$ are the generators and $\lambda_a$ the Gell-Mann matrices. The components of the scalar transform homogeneously, as $\Sigma_i(x)\mapsto G_{ij}(x)\Sigma_j(x)$.

Apart from the gauge symmetry, the kinetic part also has a global \ZZ[2] symmetry under the transformation $\scA(x)\mapsto -\scA(x)$. The potential is chosen to leave this global symmetry intact,
\begin{equation}\label{eqn:adj_potential}
  V = -\mu^2\tr\qty[\Sigma^2] + \lambda\qty(\tr\qty[\Sigma^2])^2\,.
\end{equation}
We start by following the path of the standard BEH mechanism using this expression. That is, the scalar field is split into a non-vanishing vacuum expectation value (vev) which points into a certain direction $\Sigma_0$ with absolute value $w>0$, and fluctuations $\sigma$ around this vev, which are assumed to be small\footnote{We note that we are implicitly using that only a subset of all possible breaking patterns can occur for a potential renormalizable by power counting \cite{O'Raifeartaigh:1986vq}. For the present \SU[3] case, however, this potential exhausts all possible cases. This is no longer true for $N>3$. Higher-order tree-level potentials, which may be considered also for GUTs as in the standard model case \cite{Gies:2014xha,Eichhorn:2015kea}, can in principle realize all possible subgroups, according to Morse theory \cite{O'Raifeartaigh:1986vq}. This may also happen at the level of the quantum effective potential.}:
\begin{equation}\label{eqn:adj_beh}
  \Sigma\qty(x) = \ev{\Sigma} + \sigma\qty(x) = w\Sigma_0 + \sigma\qty(x)\,.
\end{equation}
Note that by specifying $w\Sigma_0$ we completely fix the direction of the vev. This has far-reaching consequences to be discussed below.

\begin{figure}[t!]
  \centering
  \includesvg[width=.8\linewidth]{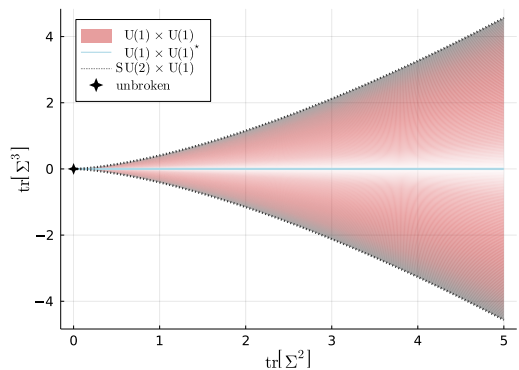}
  \caption{
    Different breaking patterns in terms of the matrix invariants $\tr\qty[\Sigma^2]$ and $\tr\qty[\Sigma^3]$.
    Note that there is no two-dimensional region in this plot with a symmetry pattern other than $\U[1]\times\U[1]$. The colour gradient is added for later convenience.
  }
  \label{fig:su3break}
\end{figure}

\begin{table*}[t!]
  \begin{tabular}{@{}l@{\hspace{4em}}c@{\hspace{4em}}c@{\hspace{4em}}c@{}}\toprule
                                                             & \multicolumn{3}{@{}c@{}}{Breaking pattern}                                                                                                           \\ \cmidrule(r){2-4}
                                                             & $\SU[2]\times\U[1]$                                & $\U[1]\times\U[1]^\star$                   & $\U[1]\times\U[1]$ (generic)                       \\ \midrule
    eigenvalues 
                                                             & $\{\lambda,\lambda, -2\lambda\}$                   & $\{\lambda, -\lambda, 0\}$                 & $\{\lambda_1, \lambda_2, -(\lambda_1+\lambda_2)\}$ \\
    \addlinespace[.5em]
    \multirow{2}{*}{vev-alignment $\qty(\Sigma_3,\Sigma_8)$} & $\pm\qty(0,1)$                                     & $\pm\qty(1,0)$                             & \multirow{2}{*}{other}                             \\
                                                             & $\pm\qty(\sqrt{3}/2,\pm1/2)$                       & $\pm\qty(1/2,\pm\sqrt{3}/2)$               &                                                    \\
    \addlinespace[.5em]
    breaking angle $\theta_0$                                & $\tfrac{\qty(2n+1)\pi}{6} \text{ for }n=0,\dots,5$ & $\tfrac{2n\pi}{6} \text{ for }n=0,\dots,5$ & other                                              \\
    \addlinespace[.5em]
    gauge boson degeneracy                                   & $\qty(4,4)$                                        & $\qty(2,4,2)$                              & $\qty(2,2,2,2)$                                    \\
    \addlinespace[.5em]
    scalars $(0,m_H)$                                        & $\qty(3,1)$                                        & $\qty(1,1)$                                & $\qty(1,1)$                                        \\ \bottomrule
  \end{tabular}
  \caption{Symmetry breaking patterns as a function of the direction of the chosen vev, i.e. $\theta_0$. In the general case there are three distinct non-zero eigenvalues and the vev is invariant under $\U[1]\times\U[1]$. For six special values of $\theta_0$ the eigenvalues degenerate, leading to a $\SU[2]\times\U[1]$ symmetry. For a further six values, the determinant is zero, resulting in a special case of the $\U[1]\times\U[1]$ pattern, denoted by $\U[1]\times\U[1]^\star$, which is invariant under the transformation $\theta_0\mapsto -\theta_0$. Note that the overall scale of $\Sigma$ is given by $w$ and that the ordering of the eigenvalues and the respective massless gauge boson modes differs in the different $\theta_0$ sectors. However, the number of massless and massive modes is fixed as stated here.
  }
  \label{tab:adj_patterns}
\end{table*}

The different breaking patterns originate from the existence of more than one unitarily in-equivalent\footnote{That is, there is no gauge transformation (\ref{eq:adj_gt}), which can map them into each other.} direction for $w\Sigma_0$.
Since $\Sigma_0$ is diagonalizable and traceless, the most general vev can be parameterized by an angle $\theta_0$ in the two-dimensional Cartan of \SU[3], with
\begin{equation}\label{eqn:adj_breaking_angle}
  \Sigma_0 = \cos\qty(\theta_0) T_3 + \sin\qty(\theta_0)T_8\,.
\end{equation}
Both $w$, and the eigenvalues of $\Sigma_0$, can in principle be obtained from the scalar field without explicitly gauge-fixing by using the matrix invariants\footnote{Note that in the present theory the relation $\tr\qty[\Sigma^3]=3\det\qty[\Sigma]$ holds and can alternatively be used as has been done in \cite{Dobson:2021sgl}. However, with having a generalization to \SU[N] groups in mind it makes more sense to only use quantities which are renormalizable by power-counting for any $N$.} $\tr\qty[\Sigma^3]$ and $\tr\qty[\Sigma^2]$ \cite{Olynyk:1984pz}.
For the specific definition of the vev-direction in \cref{eqn:adj_breaking_angle}, the angle $\theta_0$ and $w$ can be obtained from
\begin{align}
  w               & = \sqrt{2\tr\qty[\Sigma^2]}  \label{eqn:w_gauge_invariant}                                       \\
  \sin(3\theta_0) & = \sqrt{6}\frac{\tr\qty[\Sigma^3]}{\tr\qty[\Sigma^2]^{3/2}}\,. \label{eqn:theta_gauge_invariant}
\end{align}
However, interpreting the angle $\theta_0$ in terms of a vev-direction already requires implicit gauge-fixing, to obtain the relation (\ref{eqn:theta_gauge_invariant}). Therefore, one should keep in mind that whenever angles or breaking patterns are involved, the gauge has been fixed, either implicitly or explicitly.

Depending on $\Sigma_0$, the corresponding eigenvalues can have different degeneracies, giving either a breaking pattern of $\SU[2]\times\U[1]$ or $\U[1]\times\U[1]$. Which one is realized depends on the choice of the vev-direction \cite{Maas:2017xzh}. \Cref{tab:adj_patterns} lists the possible patterns. In the plane of matrix invariants, as shown in  \cref{fig:su3break}, it is visible that the two breaking patterns $\SU[2]\times\U[1]$ and $\U[1]\times\U[1]^{\star}$ from \cref{tab:adj_patterns} play a special role.

\begin{figure*}[th!]
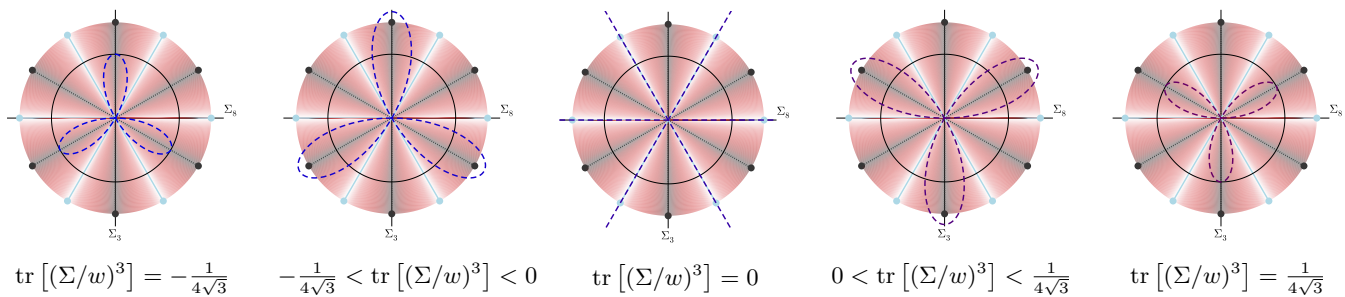

  \centering
  \begin{subfigure}{0.18\textwidth}
    \includesvg[width=\hsize]{{circle_1}}
    \caption*{$\tr\qty[(\Sigma/w)^3]=-\tfrac{1}{4\sqrt{3}}$}
  \end{subfigure}
  \hfill
  \begin{subfigure}{0.18\textwidth}
    \includesvg[width=\hsize]{{circle_2}}
    \caption*{\mbox{$-\tfrac{1}{4\sqrt{3}}<\tr\qty[(\Sigma/w)^3]<0$}}
  \end{subfigure}
  \hfill
  \begin{subfigure}{0.18\textwidth}
    \includesvg[width=\hsize]{{circle_3}}
    \caption*{$\tr\qty[(\Sigma/w)^3]=0$}
  \end{subfigure}
  \hfill
  \begin{subfigure}{0.18\textwidth}
    \includesvg[width=\hsize]{{circle_4}}
    \caption*{$0<\tr\qty[(\Sigma/w)^3]<\tfrac{1}{4\sqrt{3}}$}
  \end{subfigure}
  \hfill
  \begin{subfigure}{0.18\textwidth}
    \includesvg[width=\hsize]{{circle_5}}
    \caption*{$\tr\qty[(\Sigma/w)^3]=\tfrac{1}{4\sqrt{3}}$}
  \end{subfigure}
  \caption{The additional discrete symmetries of the breaking patterns visualized as intersection points of the unit circle with $\tr\qty[(\Sigma/w)^3]$. The coloured sectors correspond to the $2\pi/3$ periodicity of the system in $\theta_0$. For the $\SU[2]\times\U[1]$ pattern (leftmost and rightmost panels) there are three intersection points, yielding a \ZZ[3] symmetry. Likewise, the $\U[1]\times\U[1]^{\star}$ (middle panel) special case has six intersection points, and thus a $\ZZ[6]\sim S_3$ symmetry. The other cases correspond to $\U[1]\times\U[1]$ with six intersection points, but only a \ZZ[3] symmetry.}
  \label{fig:su3intersects}
\end{figure*}

Because \cref{eqn:theta_gauge_invariant} yields a $2\pi/3$ periodicity in $\theta_0$, there is a further freedom in choosing $\theta_0$. The full range of $2\pi$ decomposes into 6 sectors of size $\pi/3$, two pairs always related by the global $\ZZ[2]$ symmetry. This corresponds to the $3!=6$ possible permutations of the eigenvalues of $\Sigma_0$ and yields additional unbroken discrete groups, as shown in figure \ref{fig:su3intersects}. Thus, so far the breaking patterns are actually $\SU[2]\times\U[1]\times \ZZ[3]$, $\U[1]\times\U[1]\times \ZZ[3]$, and  $\U[1]\times\U[1]\times \ZZ[6]$, where the latter is the one referred to as $\U[1]\times\U[1]^{\star}$. The existence of these discrete groups will be relevant for the possible choices of gauges below. Since the interval $[-\pi/6,\pi/6]$ is the most convenient for later we choose it as the fundamental domain in the following.

Inserting (\ref{eqn:adj_breaking_angle}) into the potential (\ref{eqn:adj_potential}) shows that the minimum of the potential is independent of $\theta_0$, and thus all values of $\theta_0$ correspond to a possible choice at tree-level, as long as $w=\pm\mu/\sqrt{\lambda}$. Note that both values of $w$ are mapped into each other under the global $\ZZ[2]$ symmetry. At the same time, there is no gauge transformation, which has the same effect, as (\ref{eq:adj_gt}) leaves the sign of $\Sigma$ unchanged. Additionally, it should be mentioned that only in the \SU[3]-theory the angle-independence of the potential is sufficient to show that all angles do minimize the potential\footnote{If the potential were to have an additional $\ZZ[2]$-breaking term $\sim\tr[\Sigma^3]$, then this would no longer be the case.}. This is due to the enhanced \O[8]-symmetry of the potential and cannot be assumed for the general \SU[N] case. A more detailed discussion on this is given in \cref{app:sun}.

At tree-level, the two different breaking patterns lead to drastically different spectra for the elementary particles, which can be sorted into the representations of the unbroken subgroups. The massive gauge bosons appear either as oppositely charged pairs of the \U[1] subgroups or as doublets of an unbroken \SU[2]. As a consequence, the massive gauge bosons always come in pairs, with masses
\begin{subequations}\label{eqn:boson_masses}
  \begin{align}
    m_1 & = gw\cos\qty(\theta_0)                                                      \\
    m_2 & = \frac{gw\left(\cos\qty(\theta_0)+\sqrt{3}\sin\qty(\theta_0)\right)}{2}    \\
    m_3 & = \frac{gw\left(\cos\qty(\theta_0)-\sqrt{3}\sin\qty(\theta_0)\right)}{2}\,.
  \end{align}
\end{subequations}
Thus, for special values of $\theta_0$ some doublets can degenerate, or a doublet can have vanishing mass. There are always two massless gauge bosons corresponding to the unbroken subgroups, i.e. one for each unbroken \U[1]. In the $\SU[2]\times\U[1]$ case there is a triplet for the unbroken \SU[2]. Thus, there are 2 and 4 massless modes for the $\U[1]\times\U[1]$ and $\SU[2]\times\U[1]$ patterns, respectively. In addition, there is always a scalar with independent mass $m_H$, as well as one or three remaining massless scalars for the $\U[1]\times\U[1]$ and $\SU[2]\times\U[1]$ patterns, respectively. Of course, in the unbroken case all gauge bosons (and scalars) are degenerate.

The gauge-invariant composite spectrum from augmented perturbation theory is different from the elementary one, and also differs between the breaking patterns with and without \SU[2] subgroup. See \cite{Maas:2017xzh} for details.

\section{Gauge symmetry, vacuum expectation value, and global symmetry}\label{s:symmetry}

As the condition in \cref{eqn:adj_beh} is gauge-dependent \cite{Lee:1974zg,Frohlich:1980gj,Frohlich:1981yi,Maas:2012ct}, its implementation requires gauge-fixing, a process which leads to some subtleties. In the present case there are two particular questions to answer. First: is there a BEH effect at all, i.e. is $w\neq 0$? Second: if there is a BEH effect, what is the value of $\theta_0$?

At tree-level, the presence of a BEH effect is determined by a condition on $w$ alone, with no constraint on $\theta_0$. However, beyond tree-level, the questions need to be phrased more carefully. The issue is that $\Sigma_0$ is not gauge-invariant. Furthermore, it is known from the standard-model case of an \SU[2] gauge theory with a fundamental scalar doublet that the question of $|w|>0$ is potentially dependent on the gauge in which it is calculated \cite{Lee:1974zg,Caudy:2007sf,Maas:2012ct}. This may also apply to $\theta_0$.

All of this can be traced back to the fact that the BEH effect is really just fixing the gauge in presence of a particular type of potential, such that
\begin{equation}\label{eqn:beh}
  \left\langle \frac{1}{V}\int \dd[4]{x}\Sigma\qty(x)\right\rangle=w\Sigma_0
\end{equation}
can hold\footnote{This will only work in a suitably chosen renormalization scheme.}. Thus, the right question to be asked is whether there exists a gauge, for a fixed set of tree-level parameters $(g,\mu,\Lambda)$, such that \cref{eqn:beh} holds for a given set of values $(w,\theta_0)$. Currently, we do not have the means to check all possible gauges, so we will concentrate here on whether this is true for unitary gauge and (minimal\footnote{The Gribov-Singer ambiguity seems to be not quantitatively relevant in presence of a BEH effect (i.e.\ $|w|>0$) in Landau gauge, despite its qualitative relevance to the asymptotic state space by invalidating the perturbative BRST construction \cite{Fujikawa:1982ss}. Hence, the specification of how Gribov copies are treated is likely not relevant. However, this is so far only circumstantial evidence \cite{Maas:2010nc,Lenz:2000zt}.} \cite{Maas:2011se}) Landau--'t Hooft gauge.

There is already one immediate issue in \cref{eqn:beh}: it is not invariant under a global $\ZZ[2]$ transformation. This is at first sight nothing dramatic, as there are other cases where the gauge condition in a BEH effect breaks global symmetries \cite{Maas:2017wzi}. However, there always remains an unbroken global diagonal symmetry of the same type. This diagonal symmetry originates from the fact that a global transformation can be undone by a gauge transformation, such that the gauge condition remains invariant. In the present case, changes of the gauge condition by a \ZZ[2]-transformation cannot be undone by a gauge transformation in general, since (\ref{eq:adj_gt}) cannot change the sign of $\Sigma$. The only exception is when $\Sigma_0$ has the special $\U[1]\times \U[1]^\star$ form in table \ref{tab:adj_patterns}, for which a $\ZZ[2]$ transformation is equivalent to reordering the eigenvalues. Otherwise, for every space-time averaged value of $\Sigma$, the path integration also contains one with opposite sign, and the two necessarily average to zero.

Thus, strictly speaking, the gauge condition (\ref{eqn:beh}) can in general not be satisfied. Hence, except where it is possible to fix to the $\U[1]\times \U[1]^\star$ pattern, no BEH effect would be possible at all. Of course, as soon as the global $\ZZ[2]$ symmetry is spontaneously or explicitly broken, this is again possible, and we will discuss this situation below.

If we step away from the desire to emulate perturbation theory by implementing (\ref{eqn:beh}) for a moment, it is helpful to reconsider what a BEH effect truly is. In fact, a BEH effect is long-range order in the direction of the scalar field. This can be detected using the gauge-fixed scalar field, e.g.\ by a suitable renormalized expression like \cite{Caudy:2007sf,Maas:2017wzi}
\begin{equation}\label{eqn:adj_gauge_order_lat}
  \left\langle \left(\frac{1}{V}\int_V \dd[4]{x} \Sigma(x)\right)^2\right\rangle\,,
\end{equation}
which will not vanish in the infinite-volume limit $V\to\infty$ in the presence of long-range ordering. Therefore, long-range order and the physical content of the BEH effect can be detected while foregoing the implementation of (\ref{eqn:beh}). The physical consequences, like massive gauge bosons, then remain, but are generated non-perturbatively \cite{Maas:2012ct,Maas:2013aia}.

\begin{figure}[t!]
  \centering
  \includesvg[width=\linewidth]{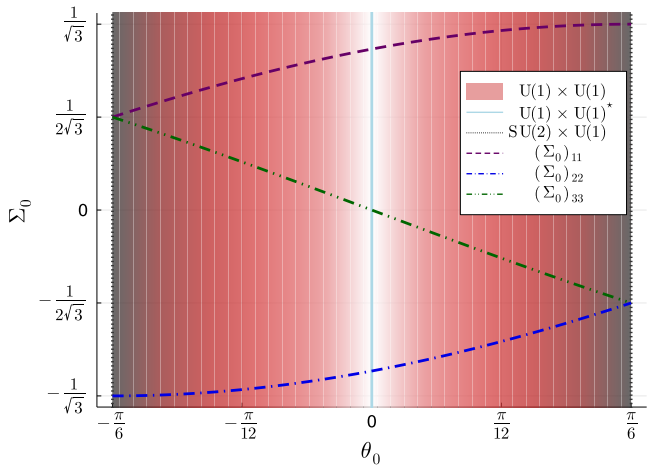}
  \caption{Normalized components of the vev as a function of $\theta_0$ in the fundamental domain and the corresponding breaking patterns for specific values.}
  \label{fig:comp}
\end{figure}

Another option to resolve the issue is possible due to the periodicity of the breaking angle.  In fact there exists for every field configuration a gauge transformation such that $\theta_0$ can be rotated into the interval between $-\pi/6$ and $\pi/6$, by performing a permutation of the eigenvalues. Thus, by constraining $\Sigma_0$ to lie in this interval, we still break the $\ZZ[2]$ symmetry, but there now exists an unbroken diagonal subgroup. That subgroup contains a $\ZZ[2]$ transformation followed by a suitably chosen element of the $S_3$ permutation subgroup of \SU[3], which transfers the vev back into this $\theta_0$ interval. Therefore, the action of this diagonal subgroup changes the value of $\theta_0$ to $-\theta_0$\footnote{Note, that the sign-flip of the angle only holds for the specific definition of the fundamental region above. In general this transformation mirrors the angle around the centre of the chosen domain.}. However, there is no preference for $\theta_0$ to be greater or smaller than zero with regard to the breaking pattern itself. Thus, the distribution will be symmetric, manifesting the $\ZZ[2]$ symmetry of the theory. Nevertheless, this does not imply that $\Sigma_0$ itself is symmetric, as can be seen in \cref{fig:comp}. It is thus well possible to have different vevs, even when averaging over a symmetric distribution of $\theta_0$. Additionally, this subgroup still conserves the fact that the $\U[1]\times\U[1]^\star$ pattern is intrinsically $\ZZ[2]$-invariant.

This allows now a useful definition of a BEH effect in this theory:

\emph{A BEH effect occurs in a fixed gauge if and only if, after performing a gauge transformation for each configuration such as to fulfil (\ref{eqn:beh}) with $\theta_0$ between $-\pi/6$ and $\pi/6$, the quantity (\ref{eqn:adj_gauge_order_lat}) (or equivalently $|w|$) does not vanish in the infinite volume limit.}

The resulting value of $\Sigma_0$ then determines the breaking pattern possible for the given set of parameters --- i.e.\ the direction of $\Sigma_0$, and thus the breaking pattern, cannot be predetermined. It is that value of $\Sigma_0$ around which gauge-fixed expansion techniques need to expand.

Only in the case of breaking pattern coexistence there is the possibility to choose a pattern. To identify coexistence, it is necessary to determine the quantum effective action and determine its structure. Since the vev (\ref{eqn:beh}) has by construction two independent parameters, the quantum effective action needs likewise two sources. To obtain a power-countable renormalizable structure, the quantum effective action can be constructed in the following way. Define after gauge-fixing the two components
\begin{align}
  \Sigma_3 & = 2\tr\qty[\Sigma T_3] \\
  \Sigma_8 & = 2\tr\qty[\Sigma T_8]
\end{align}
and introduce two constant sources $j_3$ and $j_8$. The source-dependent partition function then defines the free energy $W(j_3,j_8)$
\begin{equation}
  e^{W\qty(j_3,j_8)}=\int{\cal D}A_\mu\,{\cal D}\Sigma\, e^{iS+\int \dd[d]{x}\left(j_3\Sigma_3+j_8\Sigma_8\right)}.
\end{equation}
Performing a Legendre transformation with respect to $j_3$ and $j_8$ yields the quantum effective action as a function of the classical (space-time independent) fields $\sigma_3$ and $\sigma_8$,
\begin{align}
  \Gamma\qty(\sigma_3,\sigma_8) & =\sigma_3 j_3(\sigma_3)+\sigma_8 j_8(\sigma_8)-W\qty(j_3(\sigma_3),j_8(\sigma_8)) \\
  \sigma_i                      & =\frac{\partial W}{\partial j_i}\,,
\end{align}
which of course requires resolving the classical fields $\sigma_i$ as a function of the sources. In the classical limit this will reproduce the classical action. The quantum effective action is necessarily a convex function, and can therefore not have multiple minima. Multiple breaking patterns will indicate themselves by flat sections, which is essentially what can be expected from a Maxwell construction \cite{Rivers:1987hi}. It should be noted that the free energy and the quantum effective action is perturbatively an analytical function of the classical fields \cite{Bohm:2001yx}. But the quantum effective action does not need to be an analytic function beyond perturbation theory: for example, non-analyticities occur in Yang-Mills theory in minimal Landau gauge \cite{Maas:2013sca}.

It now remains to uncover what is possible. As it turns out, this will indeed depend on the gauge choice.

\section{Gauge choices}\label{s:gauge}

We will consider two gauge choices, Landau--'t Hooft gauge and unitary gauge, the two extremes of the $R_\xi$ condition for $\xi=0$ and $\xi\to\infty$, respectively.

\subsection{Unitary gauge}

Consider for the moment unitary gauge. It is defined by fixing the direction of the vev locally, which in the present case translates to diagonalizing $\Sigma(x)$. Since diagonalization is invariant under permutations of the eigenvalues it is possible to impose strong ordering, in the sense that the diagonal elements are sorted by size\footnote{Different kinds of ordering schemes simply puts the local breaking angles into different sectors of size $\pi/3$.}. As a consequence the vacuum expectation value
\begin{align}\label{eqn:vev}
  \left\langle \frac{1}{V}\int_V \dd[4]{x} \Sigma(x)\right\rangle
\end{align}
is necessarily non-vanishing if $\Sigma(x)$ is non-zero in any sizeable fraction of the volume $V$. The forceful alignment of space-time-adjacent fields enforces maximal long-range order on the scalar field. As a consequence, if the scalar field has a non-vanishing amplitude at all, it will have a vacuum expectation value. As will be seen, this is then the case throughout the phase diagram. Thus, unitary gauge enforces a BEH effect for all values of the parameters. In principle this is also expected to happen in other theories with a BEH effect.

That appears counter-intuitive at first, as it is known that there are strongly-interacting regions within the phase diagram, essentially being a QCD-like theory. The reason for this behaviour is, of course, a trade-off when gauge-fixing. By enforcing order on the scalar field, any disorder is transferred to the gauge fields, which therefore will strongly fluctuate. Thus, while a vacuum condensate is enforced in this way, the strong gauge-field fluctuations, as well as possible local amplitude fluctuations of the Higgs field, will invalidate any perturbative calculations. Especially, despite the presence of a vev, there may be no mass generation for the gauge bosons.

This can be remedied, but only at the expense of deviating from the perturbative gauge condition (\ref{eqn:beh}).  Instead of enforcing a strong ordering of the eigenvalues, we note that the breaking patterns do not rely on the ordering of the eigenvalues as listed in table \ref{tab:adj_patterns}. However, long-range ordering implies that the ordering of the eigenvalues need to be correlated over long distances. Thus, we fix to unitary gauge, but only admit gauge transformations from the coset \SU[3]$/S_3$, where $S_3$ is the permutation group of order 3. In practice, diagonalization algorithms do not ensure such an ordering. To enforce it, we determine before diagonalization locally an angle $\theta\qty(x)$ in analogy to (\ref{eqn:theta_gauge_invariant}). We then reorder the eigenvalues after diagonalization such that the sign of the local angle $\theta_0\qty(x)$ coincides with the original up to a rotation of the domain. This leaves us in the end with local $\Sigma_0\qty(x)$-matrices, as in \cref{eqn:adj_breaking_angle}, with angles in the region $\theta_0\qty(x)\in[0,\pi/6]$ for $\theta\qty(x) \ge 0$ and $\theta_0\qty(x)\in(\pi/2,7\pi/6]$ for $\theta\qty(x) < 0$. This means that on a local level the global $\ZZ[2]$ is still preserved. When calculating the expectation value over several configurations, although restricting the ordering in some sense, the preservation of the (diagonal) \ZZ[2] symmetry implies again that for every configuration there exists another configuration with exactly opposite ordering. Thus, in this case (\ref{eqn:vev}) vanishes. This allows to unambiguously identify the presence of long-range ordering in this gauge using (\ref{eqn:adj_gauge_order_lat}). The finally obtained expectation value of the angle $\theta_0$ is again restricted to the fundamental domain, due to the properties of the $\arcsin$-function, although the local angles do not lie within this domain.

What should be noted is that both prescriptions effectively remove the discrete $\ZZ[3]$ groups (see figure \ref{fig:su3intersects}). However, the first prescription also explicitly breaks the global $\ZZ[2]$, while the second does not. As a consequence of the explicit breaking of the global $\ZZ[2]$, the vev is enforced everywhere. Thus, such a prescription would only be allowed, if either boundary conditions or the embedding of the theory would require such an explicit breaking, while the second one is always applicable.

However, all expansion schemes \cite{Bohm:2001yx,Maas:2017xzh} rely on a unique value for the vacuum expectation value. Thus, to eventually obtain configurations, which are fixed to the same gauge as in the continuum, requires, after establishing that for a given set of lattice parameters long-range ordering persists, the eigenvalues to agree with (\ref{eqn:adj_beh}). Thus, this is a two-step process. Of course, when a BEH effect is possible, this procedure will yield the same result as enforcing long-range order. Differences will only arise for the case of non-BEH-like dynamics, especially for QCD-like dynamics.

\subsection{Landau--'t Hooft gauge}

An alternative and independent approach is to use (minimal) Landau--'t Hooft gauge. In that case, the gauge fields are first locally transformed to (minimal) Landau gauge, which leaves the global gauge freedom unaffected. The corresponding gauge transformation is then applied to the local scalar fields. The space-time averaged expectation value of the scalar field is then used to obtain a global gauge transformation as in the unitary gauge, i.e.\ diagonalization  with ordered eigenvalues. This global transformation is then applied to the local gauge and scalar fields.

In this case it is not necessary to artificially preserve the diagonal \ZZ[2]-symmetry locally while diagonalizing, since Landau gauge enforces maximum smoothness on the gauge fields locally \cite{Cucchieri:1995pn}. This avoids to a large extent that local correlations are transferred to the gauge fields and thus keeps local correlations between the scalar fields intact. As a consequence, it is not possible to obtain a non-vanishing vacuum expectation value everywhere, but only where long-range ordering exists.

\section{Lattice implementation}\label{s:lattice}

\subsection{Action and configurations}

To investigate the phase diagram, we perform lattice simulations. To this end, we use the Wilson action for the theory \cite{Montvay:1994cy}
\begin{subequations}\label{eqn:adj_lattice_lagrangian}
  \begin{align}
    S= & \sum_{x}\left(  \beta\sum_{\mu<\nu}\qty[1-\frac{1}{3}\Re\qty[\tr\qty[U_{\mu\nu}\qty(x)]]]  \right. \label{eqn:adj_lattice_lagrangian_gauge}                       \\
       & + \gamma\qty[2\tr\qty[\Sigma\qty(x)^2] - 1]^2+  2\tr\qty[\Sigma\qty(x)^2]\label{eqn:adj_lattice_lagrangian_scalar}                                                \\
       & \left.- 4\kappa\sum_{\mu=1}^{4} \tr\qty[\Sigma\qty(x)U_{\mu}\qty(x)\Sigma\qty(x+\hat{\mu})U_{\mu}\qty(x)^{\dagger}]\right) \label{eqn:adj_lattice_lagrangian_int}
  \end{align}
\end{subequations}
where the tree-level couplings are related to the continuum tree-level couplings by
$\beta = 6/g^2$, $a^2\mu^2=(1-2\gamma)/\kappa - 8$ and $\lambda=2\gamma/\kappa^2$. The plaquette is given by $U_{\mu\nu}\qty(x) = U_{\mu}\qty(x)U_{\nu}\qty(x+\hat{\mu})U_{\mu}^{\dagger}\qty(x+\hat{\nu})U_{\nu}^{\dagger}\qty(x)$ and the links are related to the gauge fields as $U_{\mu}\qty(x)=\exp{i a W_{\mu}^a T^a}$.

We note that thereby the gauge degrees of freedom are group-valued, which adds another global \ZZ[3] (center-) symmetry in the gauge sector of the lattice theory. In the continuum limit, only terms in the action remain, which are trivially invariant under that symmetry, and it can thus not play a dynamical role. Of course, our present theory is potentially trivial, and may therefore not have an (interacting) continuum limit. In this case, however, the terms sensitive to the symmetry become sub-leading for the long-range, low energy physics, which determine the phase diagram \cite{Hasenfratz:1986za}. Thus, while we monitored this symmetry using Polyakov loops, we did not use it as a dynamical information in the following.

The generated configurations where obtained on symmetric lattices $N^4$ with sizes $N=4$, 6, 8, 10, 12, 14, 16, 20 and 24. The simulations used a heatbath algorithm for the link updates \cite{Cabibbo:1982zn} with an additional Metropolis step to accommodate for the interaction term. The scalar updates were performed as a generalized pseudo-heatbath method like the one proposed in \cite{Knechtli:1999tw}. Here we decided to solve the resulting cubic equation in this update without any approximation to obtain a higher acceptance rate in areas where $\kappa$ becomes large. Further details on the methods used can be found in \cref{app:algorithms}.

A configuration is obtained after one full sweep, which consists of 5 pure link sweeps followed by 1 scalar sweep and overrelaxation sweeps for both the links and the scalar fields. To assure decorrelation sufficient configurations for thermalization and in between measurements have been dropped, yielding an autocorrelation time of 1 for local quantities like the plaquette, where possible. Additionally, individual runs have been performed and combined. We note that in many cases the global structure of the phase diagram, especially with respect to the $\ZZ[2]$ symmetry, could be determined reasonably well\footnote{Note that we cannot exclude currently that at volumes beyond our computational resources this changes again, as has been observed in the \SU[2] theory with a fundamental Higgs \cite{Bonati:2009pf}.} already with volumes up to $14^4$. Larger volumes have been used to further test this insight as well as to study places where this was not the case.

A serious issue has been critical slowing down, especially in the BEH-like regime at small values of the scalar self-coupling $\gamma$. Similar effects were previously seen in the \SU[2] case \cite{Afferrante:2020hqe}. There, a massless physical vector mode was observed, which is potentially the origin of this problem. Likewise, in the current theory such physical massless vector boson modes are also expected, as well as additional massless scalar modes towards the infinite-volume limit \cite{Maas:2017xzh}. They especially are expected in the $\SU[2]\times\U[1]$ case which, as will be seen, is indeed the one seeming to emerge towards $\gamma\to 0$. Thus, the emergence of strong autocorrelations is consistent. At the moment, only sufficiently long Monte Carlo trajectories seem to be suitable to address these autocorrelations. Details of this problem are relegated to \cref{a:crit}. The results in section\footnote{All numerical analysis results have been obtained using the Julia programming language \cite{Julia:2017} with heavy usage of the packages \cite{JuliaPlots:2022,JuliaUnROOT:2022}. Analytic and arbitrary precision results have been obtained using Mathematica \cite{Mathematica}. For data storing and analysis of the propagators the ROOT framework \cite{Brun:1997pa} has been used.} \ref{s:results} will be given in a way including the effects of critical slowing down as discussed in the appendix.

\subsection{Observables and gauge fixing}

To determine the status of the global $\ZZ[2]$ symmetry we used \cite{Maas:2017wzi,Langfeld:2004vu}
\begin{equation}\label{eqn:z2order}
  \mathcal{O}_{\ZZ[2]}=\left\langle\qty(\frac{1}{V}\sum_x \tr\qty[\Sigma(x)^3])^2\right\rangle\,.
\end{equation}
This quantity approaches zero for $V\to\infty$ like a power-law if the global $\ZZ[2]$ symmetry is unbroken. If it approaches a constant, the symmetry is meta-stable, and will break when applying an arbitrarily small explicit breaking. However, without such a breaking, the symmetry remains intact, and any quantity not invariant under it will vanish \cite{Maas:2017wzi}. We will use this information below. As noted above, the gauge condition breaks the $\ZZ[2]$ symmetry and the gauge symmetry to a common subgroup. However, as the order parameter (\ref{eqn:z2order}) is gauge-invariant, it is insensitive to it, and will always give the status of the global symmetry alone.

Additionally, it should be noted that a locally normalized version of the quantity in \cref{eqn:z2order} would make it a measure for `\ZZ[2]-magnetization', like in the Ising model, and thus be even better suited to study the behaviour. However, while this modification changes the values of the order parameter quantitatively the absolute value tells us something about the severity of the \ZZ[2]-breaking. Therefore, we obtained both versions in the actual simulations and in fact they show the same behaviour.

Gauge-fixing to unitary gauge can be done straightforwardly, as it only requires the diagonalization of a $3\times 3$ matrix, which can still be done analytically following \cite{Kopp:2006wp}. Suitably (not) sorting the eigenvalues can be done as previously described. The only obstruction would be if the scalar field locally vanishes, leading to gauge defects \cite{Greensite:2006ns,Ripka:2003vv}. On a finite lattice in a simulation, the field will never be exactly zero with any appreciable probability, so this does not need to be explicitly checked.

Determining the Higgs vacuum expectation value is done using the observables
\begin{align}
  w\Sigma_0     & = \left\langle\frac{1}{V}\sum_x\Sigma\qty(x)\right\rangle \label{eqn:lvev}             \\
  \mathcal{O}_w & = \sqrt{2\expval{\frac{1}{V^2}\tr\qty[\qty(\sum_x\Sigma\qty(x))^2]}}\label{eqn:lvev2},
\end{align}
where $\mathcal{O}_w$, as discretization of (\ref{eqn:adj_gauge_order_lat}), is used to detect long-range ordering. Note that the order parameter has been re-scaled to match the usual definition of the vev in \cref{eqn:w_gauge_invariant}. The corresponding values for $w$ and $\theta_0$ can then be obtained by inserting $\Sigma_0$ into \cref{eqn:w_gauge_invariant} and \cref{eqn:theta_gauge_invariant}, respectively. For the angle this yields values in the fundamental domain. For $\mathcal{O}_w$ it is necessary to determine its value for $V\to\infty$. Only if it is non-zero a vacuum expectation value can exist.

For (minimal) Landau--'t Hooft gauge first the gauge fields are transformed to minimal Landau gauge using stochastic overrelaxation following \cite{Suman:1993mg,Cucchieri:1995pn}. Afterwards, the scalar field is correspondingly gauge transformed. To ensure the global gauge condition (\ref{eqn:beh}) the space-time average $\sum_x \Sigma(x)$ is determined per configuration, and then diagonalized in the same way as for unitary gauge. The required global gauge transformation is then applied locally to the links and the scalar fields \cite{Maas:2016ngo}. This yields the gauge-fixed configurations. The vacuum expectation value is determined as for unitary gauge using (\ref{eqn:lvev}). The deconstruction into length and angle is then performed as before. Likewise, a vev is only present if the infinite-volume limit is non-zero.

Concerning the breaking-pattern a deeper analysis of the results is required, and will be done in section \ref{s:results}.

\subsection{Quantum effective potential}

As noted before, the possibility for multiple breaking patterns is strongly tied to the quantum effective potential. Especially interesting is whether any structures exist indicative of a choice. At finite lattice spacing even multiple minima, which merge to a flat structure in the continuum limit, may be a signal. Thus, determining the quantum effective potential is an important step in understanding the problem at hand.

However, in practice there exists a problem. Because the quantum effective potential is defined in a fixed gauge as a function of the gauge-fixed classical fields, its correct determination requires the inclusion of the Faddeev-Popov determinant of the corresponding gauge choice in the Markov chain. This yields the practical problem that the determinant is in general gauges, and especially in Landau gauge, not positive definite, thus introducing a sign problem \cite{Mehta:2009zv,vonSmekal:2008es,vonSmekal:2007ns}. There is not yet an efficient solution known for this problem, and so an exact determination of the quantum effective potential is not possible.

To surpass this problem, we determine the quantum effective action by reweighting \cite{Maas:2013sca}, i.e. by determining the free energy as
\begin{equation}
  e^{W(j_3,j_8)}=\left\langle e^{\sum_x \left(j_3 \Sigma_3(x)+j_8\Sigma_8(x)\right)}\right\rangle_\text{gauge fixed}\,,
\end{equation}
using constant sources $j_3$ and $j_8$. In this way the back reaction of the source term on the weight of the configurations is neglected. This will work better the smaller the sources are. However, it can be expected to fail at the latest when the source term becomes of the same order as the action. Of course, this can happen for (drastically) different values of $j_3$ and $j_8$. Thus, for the constant sources used here, only some ellipsoidal patch of $j_3$-$j_8$ space will be possible. We will consider here this maximum patch size. The subsequent calculation of the quantum effective potential can then be done straightforwardly by a numerical Legendre transformation \cite{Maas:2013sca}. However, the exponential nature will in general require higher precision than native number formats, and thus it has been calculated using arbitrary precision arithmetic using Mathematica.

\subsection{Running coupling}

Our primary interest is in those cases, where a perturbative construction (augmented by the FMS mechanism) is possible. Also, we wanted to avoid situations where strong non-perturbative effects could alter the result. To check the nature of the interaction, we determined the running gauge coupling in the mini-MOM scheme \cite{vonSmekal:2009ae} individually for the various colours \cite{Maas:2018xxu}. This scheme allows to determine the gauge coupling just from the two-point correlation functions in the gauge sector and the ghost sector for Landau gauge,
\begin{align}
  \alpha^a(p) & =\frac{N}{2\pi\beta}Z^a(p)G^a(p)^2\label{rcoupling} \\
  Z^a(p)      & =\sum_\mu \frac{p^2}{3}D_{\mu\mu}^{aa}\nonumber     \\
  G^a(p)      & =p^2 D^{aa}_G\nonumber\,,
\end{align}
where there is no summation over $a$ and $D_{\mu\nu}$ and $D_G$ are the gauge boson propagator and the ghost propagator, respectively. The corresponding calculations have been done using the methods described in \cite{Maas:2010qw}. As a by-product, this therefore also allows determining, at least in principle, the mass spectrum of the gauge bosons, and to test it against the perturbative tree-level predictions \cite{Maas:2016ngo,Maas:2018xxu,Afferrante:2020hqe}, and identify possible degeneracies.

Of course, there are also the couplings to the scalars. However, if these are non-perturbatively strong, this should make itself felt in (substantial) deviations of the gauge sector from tree-level behaviour.

\section{Results}\label{s:results}

\subsection{Global symmetry}

\begin{figure}[h!]
  \centering
  \includesvg[width=\linewidth]{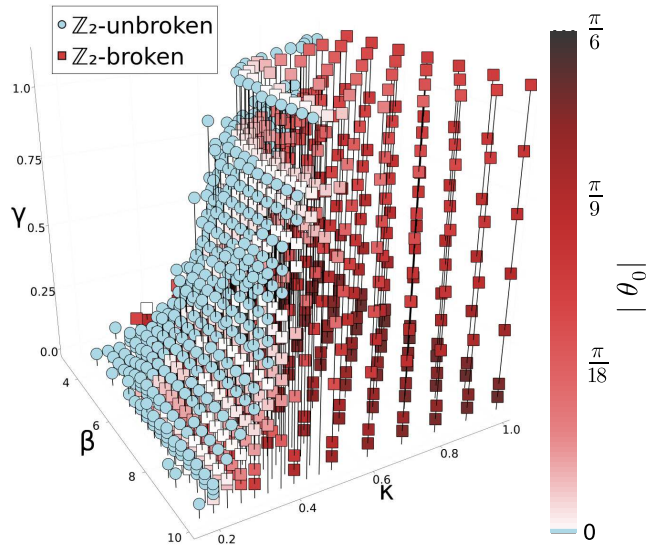}
  \caption{The status of the $\ZZ[2]$ symmetry in the phase diagram. Circles are unbroken, and squares are affected by spontaneous symmetry breaking. The colour scheme describes the expectation value of the absolute breaking angle and is the same as in all previous figures. Blue corresponds to $\U[1]\times\U[1]^\star$ or the region without BEH effect, the darkest red to $\SU[2]\times\U[1]$, and other to $\U[1]\times\U[1]$. See \cref{sec:results_implicit} for more details.}
  \label{fig:z2}
\end{figure}

\begin{figure}[t!]
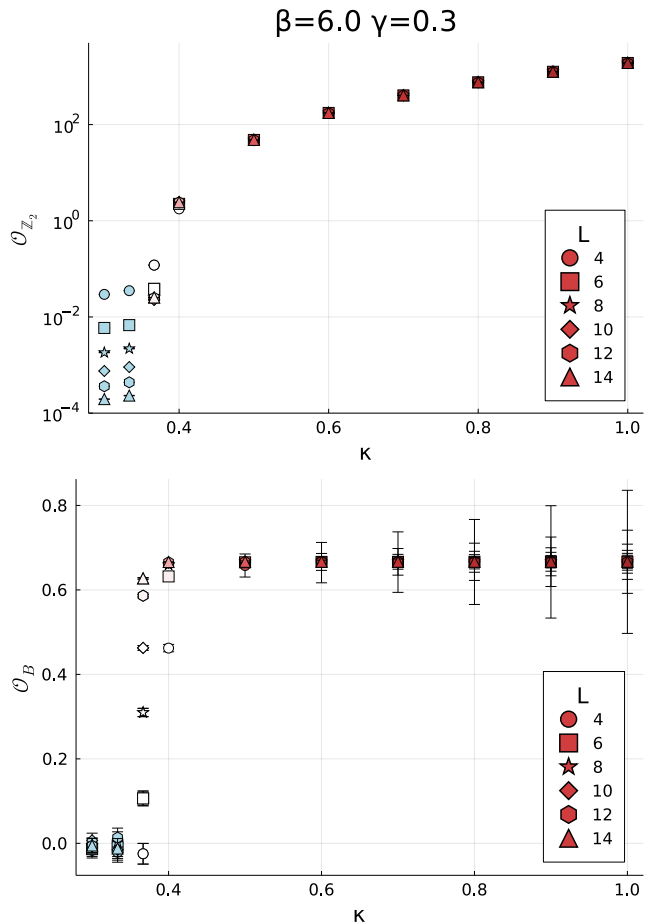

  \centering
  \includesvg[width=0.99\linewidth]{{Z2-parameter_6.0-x-0.3}}
  \hfill
  \includesvg[width=0.99\linewidth]{{Z2-parameter_binder_6.0-x-0.3}}
  \caption{The order parameter (\ref{eqn:z2order}) (top panel, logarithmic) and its Binder cumulant (bottom panel) at fixed $\beta=6$ and $\gamma=0.3$ as a function of $\kappa$. The colour scheme of the symbols is the same as in figure \ref{fig:z2}.}
  \label{fig:z2order}
\end{figure}

Most unambiguous is the situation with the global symmetry. We find a clear separation of the phase diagram into a region of intact $\ZZ[2]$ symmetry and a region where spontaneous symmetry breaking of the global $\ZZ[2]$ symmetry is possible\footnote{Note that the Osterwalder-Seiler-Fradkin-Shenker argumentation \cite{Osterwalder:1977pc,Fradkin:1978dv} does not apply to this theory, as it requires a full breaking of the gauge group by the BEH effect. Thus, a separation in different phases is possible. This has already been observed in other cases \cite{Wellegehausen:2011sc}.}. This is shown in figure \ref{fig:z2}. In \cref{fig:z2order} we also see clearly the different volume scaling behaviours of the order parameter above and below the phase transition around $\kappa\approx0.4$ for these parameters. While we did not attempt to investigate the phase boundary in detail, breaking a symmetry, accompanied by a local order parameter, requires necessarily a non-analyticity, and thus a phase transition. Thus, we find that the phase diagram separates clearly into two distinct phases.

The order of the phase transition would require a more detailed analysis along lines of constant physics. Cursory investigations of the order parameter and its Binder cumulant
\begin{equation}\label{eqn:binder}
  \mathcal{O}_B = 1 - \frac{\expval{\qty(\sum_x \tr\qty[\Sigma\qty(x)^3])^4}}{3\expval{\qty(\sum_x \tr\qty[\Sigma\qty(x)^3])^2}^2}\,,
\end{equation}
shown for an example in figure \ref{fig:z2order}, suggest that at least part of the phase boundary is of second order, and thus defines possible continuum limits of the theory. We also did not observe any double peak structures in the distribution of the order parameter, which would be expected for a first-order transition. However, the critical region appears to be relatively small, and thus substantial effort would be required for a full quantitative control. This will not be necessary for the remainder of this work.

A further decomposition in more phases than just the two, as was investigated for the \SU[2] adjoint case \cite{Baier:1986ni}, cannot be excluded with our results. However, we did not observe any further discontinuities in the \ZZ[2] order parameter except on the surface shown in figure \ref{fig:z2}.

\subsection{Breaking angle from implicit gauge-fixing}\label{sec:results_implicit}

As discussed in the introduction it is also possible to calculate the breaking angle $\theta_0$ directly from gauge-invariant quantities, i.e. the matrix invariants, as shown in \cref{eqn:theta_gauge_invariant}. From this we can define another quantity by
\begin{equation}\label{eqn:O_theta}
  \mathcal{O}_{\theta_0} = \ev{\abs{\frac{1}{3}\arcsin(\sqrt{6}\frac{\sum_x\tr\qty[\Sigma\qty(x)^3]}{\qty(\sum_x\tr\qty[\Sigma\qty(x)^2])^{3/2}})}}\,,
\end{equation}
which quantifies the possible breaking angle depending on the obtained configurations for a specific parameter set. The results for this are included as a colour scheme in \cref{fig:z2} and uses the same colours as in \crefrange{fig:su3break}{fig:comp}.

Before discussing the results for this quantity, a word of caution on this parameter. It needs to be emphasized that this quantity does not imply that the corresponding pattern is automatically realized, but rather should  be understood as ``a possible value of the breaking angle, if there exists a non-vanishing vev for this parameter set''. The reason why the breaking angle does not necessarily have to have this value is again twofold. First, it is possible that additional gauge-transformations applied to the scalar-field, like a Landau gauge, modify the space-time averaged scalar field and thus lead to a different value of the actual $\theta_0$. Second, the distribution of the parameter (\ref{eqn:O_theta}) needs to be monitored to make sure that it is indeed Gaussian and does not contain multiple peaks. For the present case this has been checked and indeed for all investigated parameter sets the distributions do not show multi-peak structures.

What can immediately be seen from \cref{eqn:O_theta} is that the angle directly relates to the \ZZ[2]-symmetry breaking term. If the \ZZ[2]-symmetry is unbroken the breaking angle must be zero and thus corresponds to the $\U[1]\times\U[1]^\star$ pattern. This is reflected in \cref{fig:z2}. In the \ZZ[2]-broken region the angle can be any non-zero value in the fundamental domain. However, the angle correlates with the absolute value of $\tr\qty[\Sigma^3]$, which in turn means that only if \ZZ[2]-breaking is strong enough the $\SU[2]\times\U[1]$ pattern can be realized. This happens only for $\gamma\to 0$ and $\kappa$ sufficiently large. Conversely, close to the \ZZ[2]-phase transition the breaking angle tends towards the $\U[1]\times\U[1]^\star$ pattern. Overall, the breaking angle obtained from implicitly gauge-fixed quantities suggests that specific breaking patterns may only be realized for certain parameter sets if the vev does not vanish in the infinite volume limit.

\subsection{Unitary gauge results}

\begin{figure}[h!]
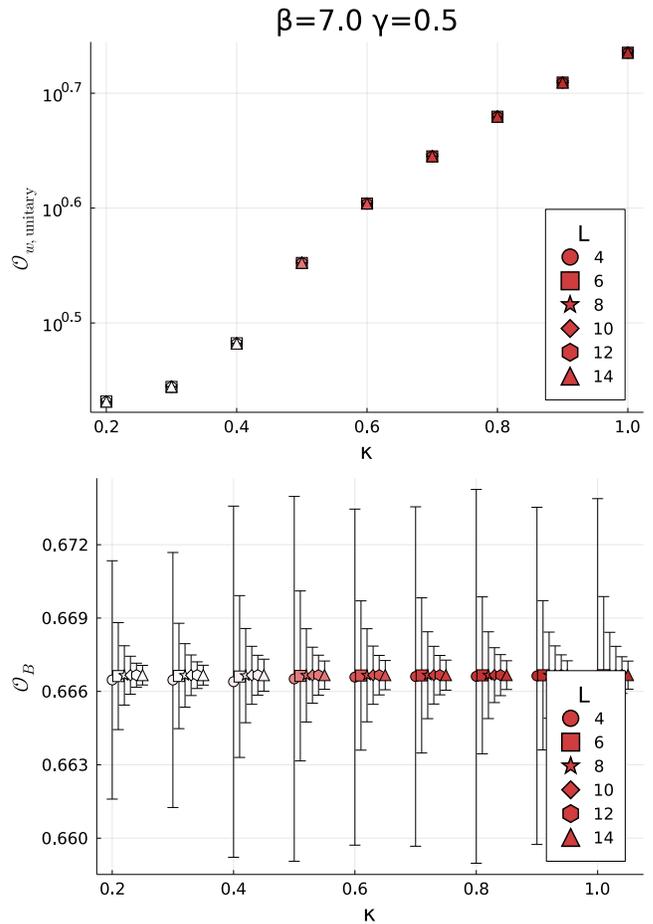

  \centering
  \includesvg[width=0.99\linewidth]{{vev-parameter_7.0-x-0.5}}
  \hfill
  \includesvg[width=0.99\linewidth]{{vev-parameter_binder_7.0-x-0.5}}
  \caption{The order parameter (\ref{eqn:lvev2}) in over-aligned unitary gauge (top panel, logarithmic) and its Binder cumulant (bottom panel) at fixed $\beta=7$ and $\gamma=0.5$ as a function of $\kappa$. The colour scheme of the symbols is the same as in figure \ref{fig:z2}. The angles are obtained individually for the gauge-fixed configurations. The data points for the binder cumulant have been shifted along the $\kappa$-axis for better visibility. }
  \label{fig:vevorder}
\end{figure}

\begin{figure*}[t!]
  \centering
  \begin{subfigure}{0.23\textwidth}
    \includegraphics[width=\linewidth,trim={1.5cm 0 4cm 1.2cm},clip]{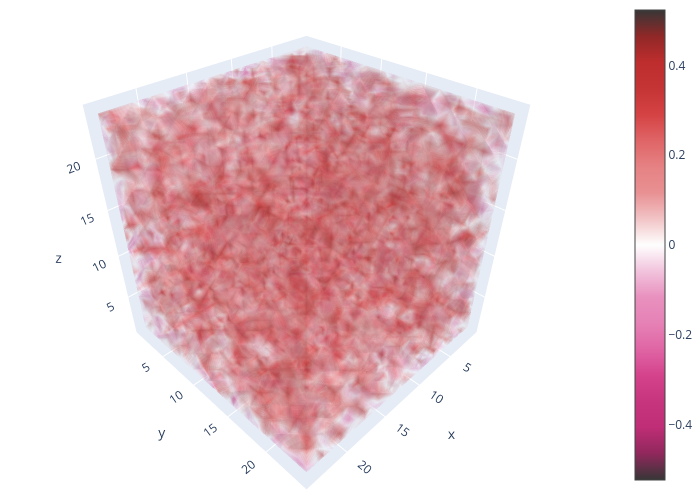}
    \caption{un-aligned\linebreak}
    \label{fig:conf_broken}
  \end{subfigure}
  \begin{subfigure}{0.23\textwidth}
    \includegraphics[width=\linewidth,trim={1.5cm 0 4cm 1.2cm},clip]{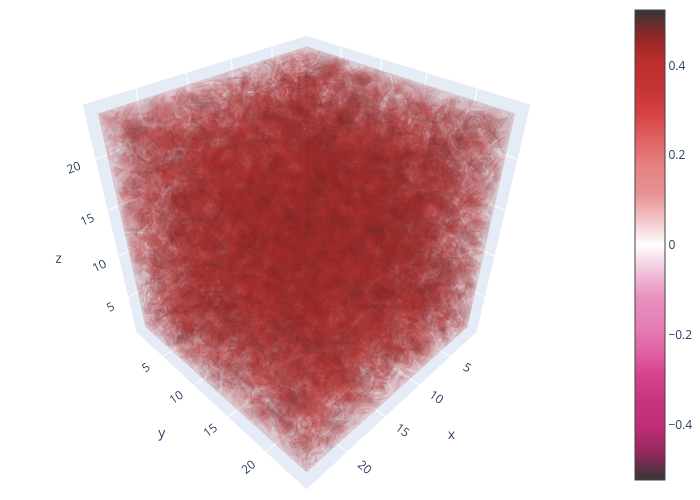}
    \caption{weakly aligned (\ZZ[2]-positive)}
    \label{fig:conf_bulk}
  \end{subfigure}
  \begin{subfigure}{0.23\textwidth}
    \includegraphics[width=\linewidth,trim={1.5cm 0 4cm 1.2cm},clip]{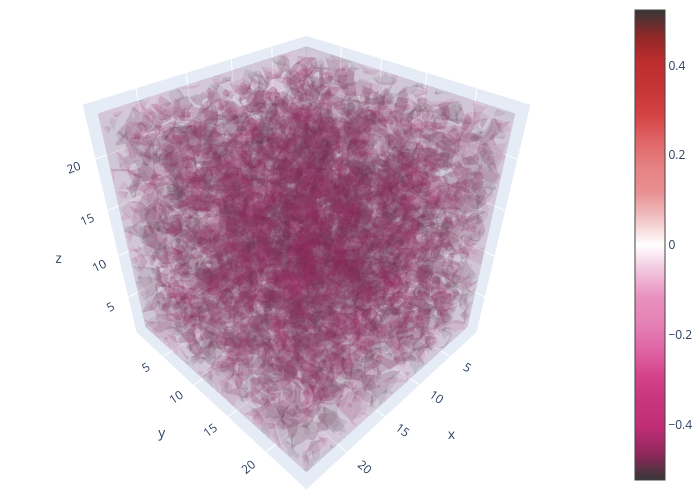}
    \caption{strongly aligned (\ZZ[2]-negative)}
    \label{fig:conf_deep_broken}
  \end{subfigure}
  \begin{subfigure}{0.23\textwidth}
    \includegraphics[width=\linewidth,trim={1.5cm 0 4cm 1.2cm},clip]{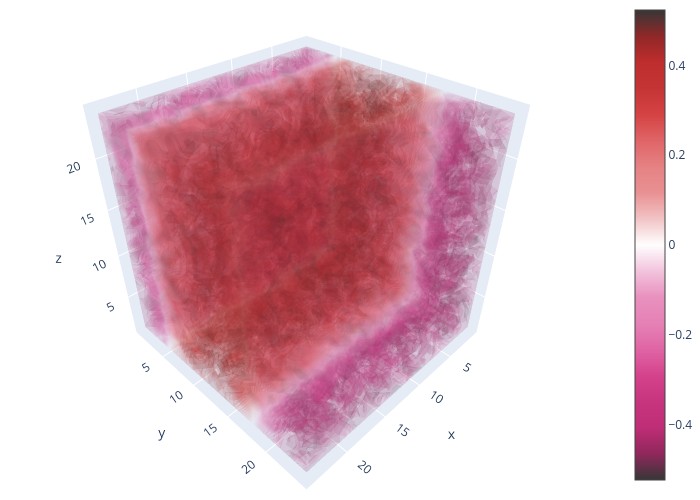}
    \caption{separately aligned\linebreak}
    \label{fig:conf_two_pats}
  \end{subfigure}
  \begin{subfigure}{0.05\textwidth}
    \includegraphics[width=\linewidth,trim={20cm 0 0 1.2cm},clip]{{conf_24-9.750000-0.950000-0.112500_t1}.png}
  \end{subfigure}
  \caption{The different kinds of spatial $\theta_0$ distributions for one time-slice on individual configurations. In the \ZZ[2]-unbroken phase and close to the phase transition all configurations behave like (a). Within the \ZZ[2]-broken phase the configurations show different strengths of alignment (b-c). Also, some configurations equilibrate to a combination of patterns (d). Note that we have separated the usual colour scale here into local angles with positive and negative signs.}
  \label{fig:conf}
\end{figure*}

\begin{figure*}[t!]
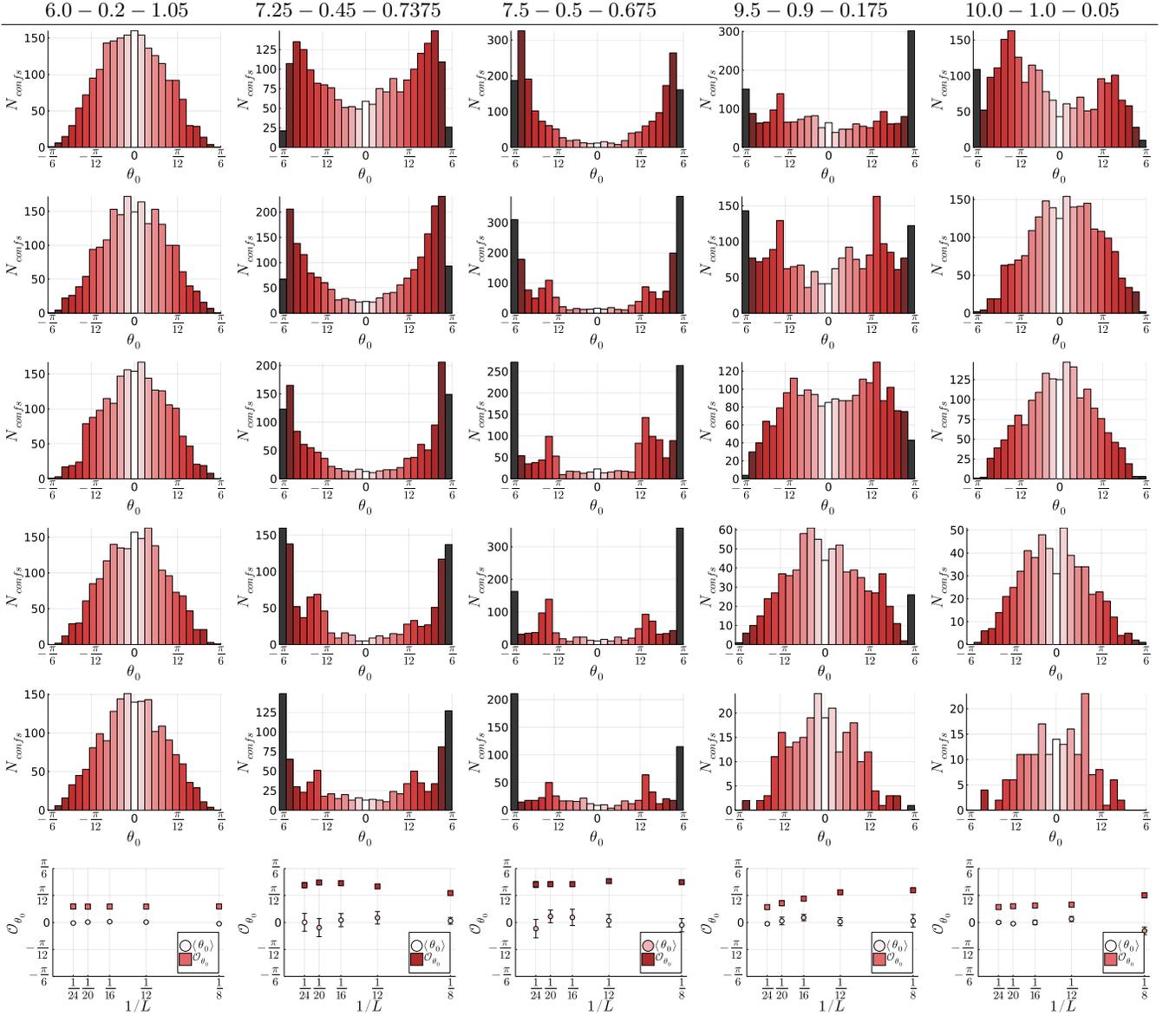

  \centering
  \begin{tabular}{ccccc}
    $6.0-0.2-1.05$                                                                &
    $7.25-0.45-0.7375$                                                            &
    $7.5-0.5-0.675$                                                               &
    $9.5-0.9-0.175$                                                               &
    $10.0-1.0-0.05$                                                                 \\\hline
    \includesvg[width=0.19\textwidth]{{theta_hist_8-6.000000-0.200000-1.050000}}  &
    \includesvg[width=0.19\textwidth]{{theta_hist_8-7.250000-0.450000-0.737500}}  &
    \includesvg[width=0.19\textwidth]{{theta_hist_8-7.500000-0.500000-0.675000}}  &
    \includesvg[width=0.19\textwidth]{{theta_hist_8-9.500000-0.900000-0.175000}}  &
    \includesvg[width=0.19\textwidth]{{theta_hist_8-10.000000-1.000000-0.050000}}   \\
    \includesvg[width=0.19\textwidth]{{theta_hist_12-6.000000-0.200000-1.050000}} &
    \includesvg[width=0.19\textwidth]{{theta_hist_12-7.250000-0.450000-0.737500}} &
    \includesvg[width=0.19\textwidth]{{theta_hist_12-7.500000-0.500000-0.675000}} &
    \includesvg[width=0.19\textwidth]{{theta_hist_12-9.500000-0.900000-0.175000}} &
    \includesvg[width=0.19\textwidth]{{theta_hist_12-10.000000-1.000000-0.050000}}  \\
    \includesvg[width=0.19\textwidth]{{theta_hist_16-6.000000-0.200000-1.050000}} &
    \includesvg[width=0.19\textwidth]{{theta_hist_16-7.250000-0.450000-0.737500}} &
    \includesvg[width=0.19\textwidth]{{theta_hist_16-7.500000-0.500000-0.675000}} &
    \includesvg[width=0.19\textwidth]{{theta_hist_16-9.500000-0.900000-0.175000}} &
    \includesvg[width=0.19\textwidth]{{theta_hist_16-10.000000-1.000000-0.050000}}  \\
    \includesvg[width=0.19\textwidth]{{theta_hist_20-6.000000-0.200000-1.050000}} &
    \includesvg[width=0.19\textwidth]{{theta_hist_20-7.250000-0.450000-0.737500}} &
    \includesvg[width=0.19\textwidth]{{theta_hist_20-7.500000-0.500000-0.675000}} &
    \includesvg[width=0.19\textwidth]{{theta_hist_20-9.500000-0.900000-0.175000}} &
    \includesvg[width=0.19\textwidth]{{theta_hist_20-10.000000-1.000000-0.050000}}  \\
    \includesvg[width=0.19\textwidth]{{theta_hist_24-6.000000-0.200000-1.050000}} &
    \includesvg[width=0.19\textwidth]{{theta_hist_24-7.250000-0.450000-0.737500}} &
    \includesvg[width=0.19\textwidth]{{theta_hist_24-7.500000-0.500000-0.675000}} &
    \includesvg[width=0.19\textwidth]{{theta_hist_24-9.500000-0.900000-0.175000}} &
    \includesvg[width=0.19\textwidth]{{theta_hist_24-10.000000-1.000000-0.050000}}  \\
    \includesvg[width=0.19\textwidth]{{theta_vol_6.000000-0.200000-1.050000}}     &
    \includesvg[width=0.19\textwidth]{{theta_vol_7.250000-0.450000-0.737500}}     &
    \includesvg[width=0.19\textwidth]{{theta_vol_7.500000-0.500000-0.675000}}     &
    \includesvg[width=0.19\textwidth]{{theta_vol_9.500000-0.900000-0.175000}}     &
    \includesvg[width=0.19\textwidth]{{theta_vol_10.000000-1.000000-0.050000}}
  \end{tabular}
  \caption{The $\theta_0$ distribution on individual configurations for increasing volumes (top to bottom) from deep in the $\ZZ[2]$ unbroken region across the phase boundary down to $\gamma\approx 0$ (left to right). The last row shows the volume behaviour of the total expectation value. The header of the column gives the respective simulation parameters $\beta-\kappa-\gamma$.}
  \label{fig:theta}
\end{figure*}

\begin{figure*}[t!]
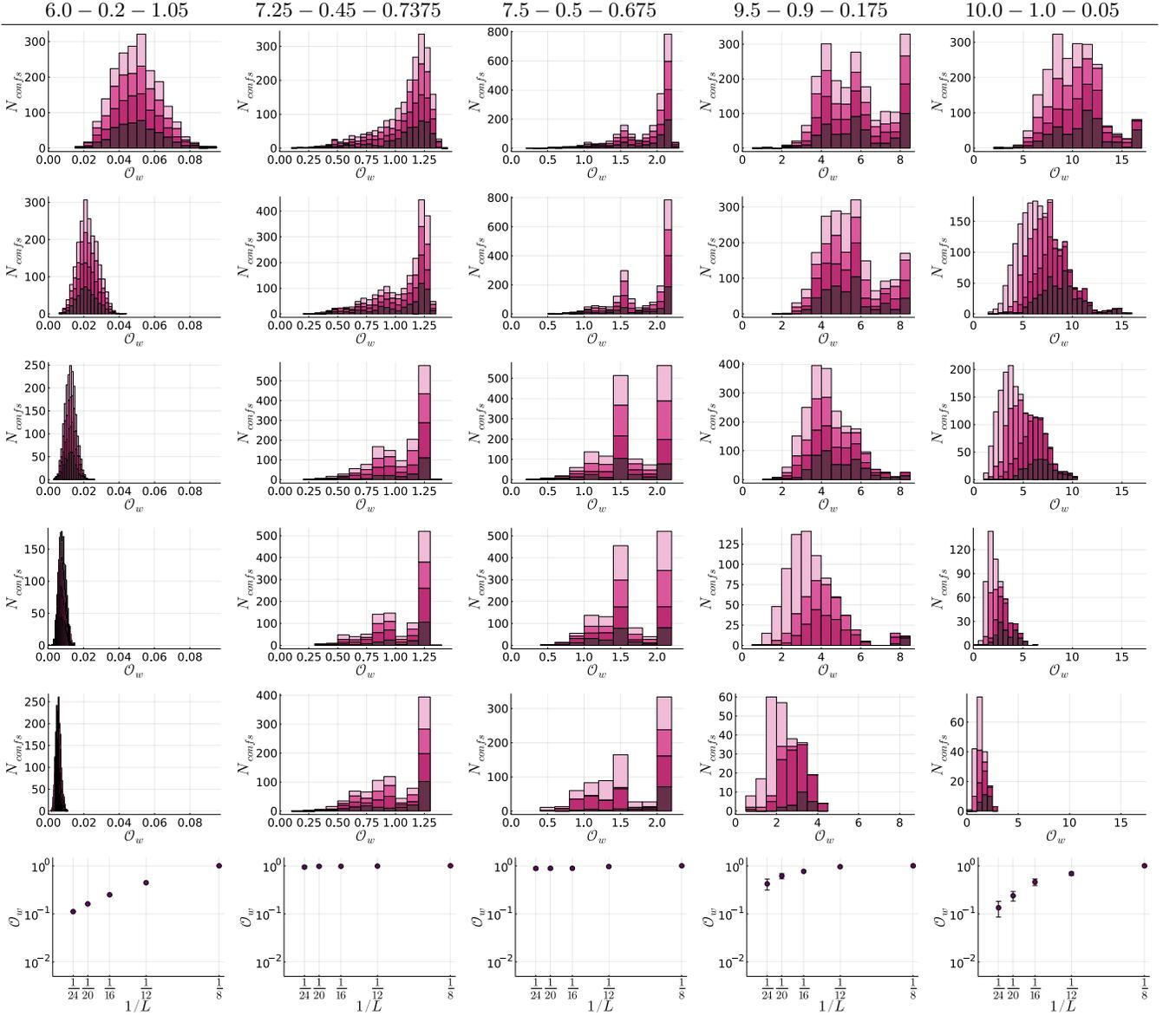

  \centering
  \begin{tabular}{ccccc}
    $6.0-0.2-1.05$                                                              &
    $7.25-0.45-0.7375$                                                          &
    $7.5-0.5-0.675$                                                             &
    $9.5-0.9-0.175$                                                             &
    $10.0-1.0-0.05$                                                               \\\hline
    \includesvg[width=0.19\textwidth]{{vev_hist_8-6.000000-0.200000-1.050000}}  &
    \includesvg[width=0.19\textwidth]{{vev_hist_8-7.250000-0.450000-0.737500}}  &
    \includesvg[width=0.19\textwidth]{{vev_hist_8-7.500000-0.500000-0.675000}}  &
    \includesvg[width=0.19\textwidth]{{vev_hist_8-9.500000-0.900000-0.175000}}  &
    \includesvg[width=0.19\textwidth]{{vev_hist_8-10.000000-1.000000-0.050000}}   \\
    \includesvg[width=0.19\textwidth]{{vev_hist_12-6.000000-0.200000-1.050000}} &
    \includesvg[width=0.19\textwidth]{{vev_hist_12-7.250000-0.450000-0.737500}} &
    \includesvg[width=0.19\textwidth]{{vev_hist_12-7.500000-0.500000-0.675000}} &
    \includesvg[width=0.19\textwidth]{{vev_hist_12-9.500000-0.900000-0.175000}} &
    \includesvg[width=0.19\textwidth]{{vev_hist_12-10.000000-1.000000-0.050000}}  \\
    \includesvg[width=0.19\textwidth]{{vev_hist_16-6.000000-0.200000-1.050000}} &
    \includesvg[width=0.19\textwidth]{{vev_hist_16-7.250000-0.450000-0.737500}} &
    \includesvg[width=0.19\textwidth]{{vev_hist_16-7.500000-0.500000-0.675000}} &
    \includesvg[width=0.19\textwidth]{{vev_hist_16-9.500000-0.900000-0.175000}} &
    \includesvg[width=0.19\textwidth]{{vev_hist_16-10.000000-1.000000-0.050000}}  \\
    \includesvg[width=0.19\textwidth]{{vev_hist_20-6.000000-0.200000-1.050000}} &
    \includesvg[width=0.19\textwidth]{{vev_hist_20-7.250000-0.450000-0.737500}} &
    \includesvg[width=0.19\textwidth]{{vev_hist_20-7.500000-0.500000-0.675000}} &
    \includesvg[width=0.19\textwidth]{{vev_hist_20-9.500000-0.900000-0.175000}} &
    \includesvg[width=0.19\textwidth]{{vev_hist_20-10.000000-1.000000-0.050000}}  \\
    \includesvg[width=0.19\textwidth]{{vev_hist_24-6.000000-0.200000-1.050000}} &
    \includesvg[width=0.19\textwidth]{{vev_hist_24-7.250000-0.450000-0.737500}} &
    \includesvg[width=0.19\textwidth]{{vev_hist_24-7.500000-0.500000-0.675000}} &
    \includesvg[width=0.19\textwidth]{{vev_hist_24-9.500000-0.900000-0.175000}} &
    \includesvg[width=0.19\textwidth]{{vev_hist_24-10.000000-1.000000-0.050000}}  \\
    \includesvg[width=0.19\textwidth]{{vev_vol_6.000000-0.200000-1.050000}}     &
    \includesvg[width=0.19\textwidth]{{vev_vol_7.250000-0.450000-0.737500}}     &
    \includesvg[width=0.19\textwidth]{{vev_vol_7.500000-0.500000-0.675000}}     &
    \includesvg[width=0.19\textwidth]{{vev_vol_9.500000-0.900000-0.175000}}     &
    \includesvg[width=0.19\textwidth]{{vev_vol_10.000000-1.000000-0.050000}}
  \end{tabular}
  \caption{The $\mathcal{O}_w$ distribution on individual configurations for increasing volumes (top to bottom) from deep in the $\ZZ[2]$ unbroken region across the phase boundary down to $\gamma\approx 0$ (left to right). The last row shows the volume behaviour of the total expectation value $\mathcal{O}_w$ normalized to the largest value (i.e.\ the smallest volume). The header of the column gives the respective simulation parameters $\beta-\kappa-\gamma$. Darker bins have been obtained later in the MC-history; see \cref{a:crit}.}
  \label{fig:vev}
\end{figure*}

\begin{figure*}[t!]
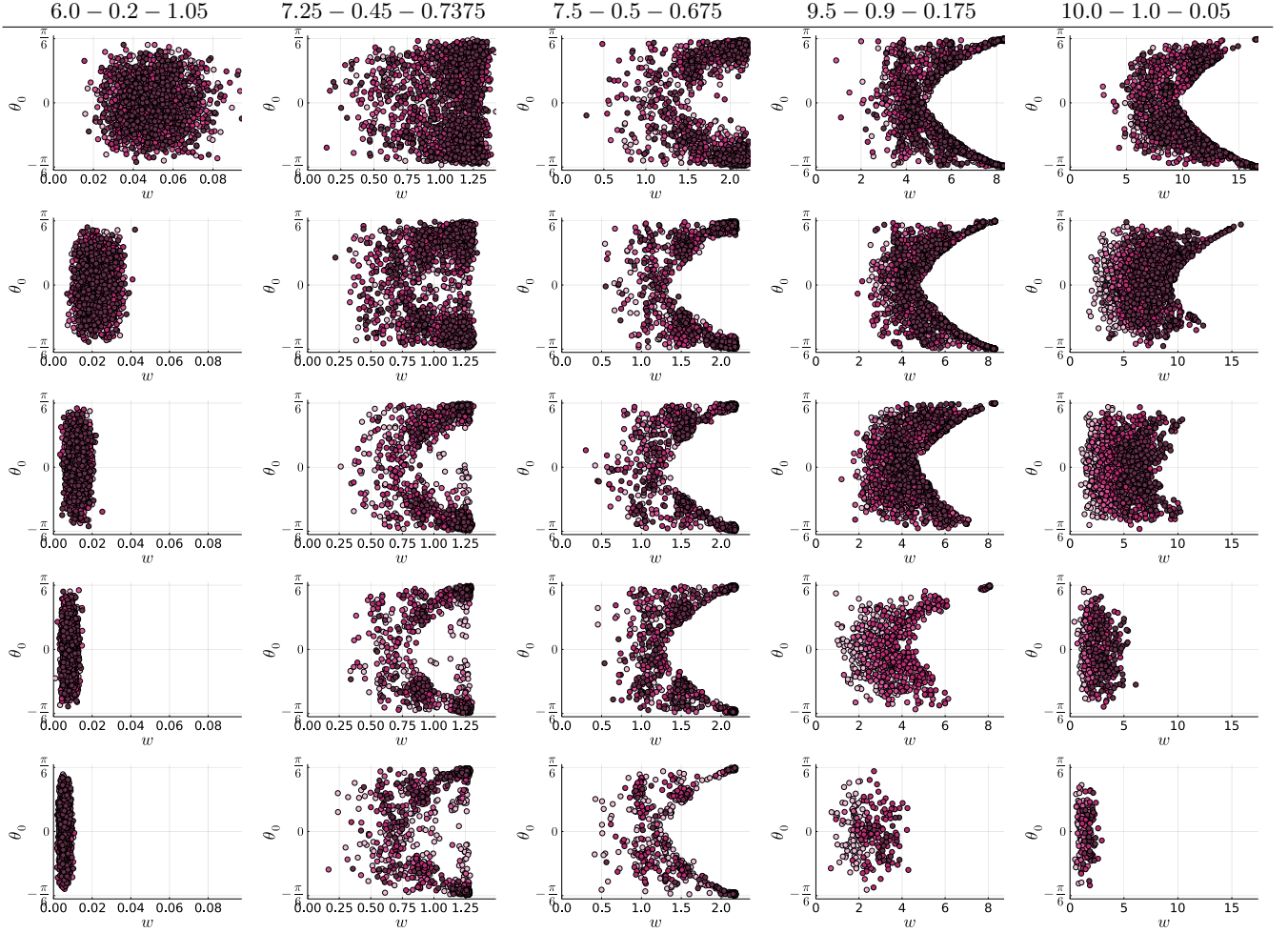

  \centering
  \begin{tabular}{ccccc}
    $6.0-0.2-1.05$                                                               &
    $7.25-0.45-0.7375$                                                           &
    $7.5-0.5-0.675$                                                              &
    $9.5-0.9-0.175$                                                              &
    $10.0-1.0-0.05$                                                                \\\hline
    \includesvg[width=0.19\textwidth]{{theta_vev_8-6.000000-0.200000-1.050000}}  &
    \includesvg[width=0.19\textwidth]{{theta_vev_8-7.250000-0.450000-0.737500}}  &
    \includesvg[width=0.19\textwidth]{{theta_vev_8-7.500000-0.500000-0.675000}}  &
    \includesvg[width=0.19\textwidth]{{theta_vev_8-9.500000-0.900000-0.175000}}  &
    \includesvg[width=0.19\textwidth]{{theta_vev_8-10.000000-1.000000-0.050000}}   \\
    \includesvg[width=0.19\textwidth]{{theta_vev_12-6.000000-0.200000-1.050000}} &
    \includesvg[width=0.19\textwidth]{{theta_vev_12-7.250000-0.450000-0.737500}} &
    \includesvg[width=0.19\textwidth]{{theta_vev_12-7.500000-0.500000-0.675000}} &
    \includesvg[width=0.19\textwidth]{{theta_vev_12-9.500000-0.900000-0.175000}} &
    \includesvg[width=0.19\textwidth]{{theta_vev_12-10.000000-1.000000-0.050000}}  \\
    \includesvg[width=0.19\textwidth]{{theta_vev_16-6.000000-0.200000-1.050000}} &
    \includesvg[width=0.19\textwidth]{{theta_vev_16-7.250000-0.450000-0.737500}} &
    \includesvg[width=0.19\textwidth]{{theta_vev_16-7.500000-0.500000-0.675000}} &
    \includesvg[width=0.19\textwidth]{{theta_vev_16-9.500000-0.900000-0.175000}} &
    \includesvg[width=0.19\textwidth]{{theta_vev_16-10.000000-1.000000-0.050000}}  \\
    \includesvg[width=0.19\textwidth]{{theta_vev_20-6.000000-0.200000-1.050000}} &
    \includesvg[width=0.19\textwidth]{{theta_vev_20-7.250000-0.450000-0.737500}} &
    \includesvg[width=0.19\textwidth]{{theta_vev_20-7.500000-0.500000-0.675000}} &
    \includesvg[width=0.19\textwidth]{{theta_vev_20-9.500000-0.900000-0.175000}} &
    \includesvg[width=0.19\textwidth]{{theta_vev_20-10.000000-1.000000-0.050000}}  \\
    \includesvg[width=0.19\textwidth]{{theta_vev_24-6.000000-0.200000-1.050000}} &
    \includesvg[width=0.19\textwidth]{{theta_vev_24-7.250000-0.450000-0.737500}} &
    \includesvg[width=0.19\textwidth]{{theta_vev_24-7.500000-0.500000-0.675000}} &
    \includesvg[width=0.19\textwidth]{{theta_vev_24-9.500000-0.900000-0.175000}} &
    \includesvg[width=0.19\textwidth]{{theta_vev_24-10.000000-1.000000-0.050000}}
  \end{tabular}
  \caption{The $\Sigma_0$ distribution on individual configurations for increasing volumes (top to bottom) from deep in the $\ZZ[2]$ unbroken region across the phase boundary down to $\gamma\approx 0$ (left to right). The header of the column gives the respective simulation parameters $\beta-\kappa-\gamma$. Darker points have been obtained later in the MC-history; see \cref{a:crit}.}
  \label{fig:theta_vs_vev}
\end{figure*}

\begin{figure*}[t!]
  \centering
  \begin{tabular}{ccccc}
    $6.0-0.2-1.05$                                                                                  &
    $7.25-0.45-0.7375$                                                                              &
    $7.5-0.5-0.675$                                                                                 &
    $9.5-0.9-0.175$                                                                                 &
    $10.0-1.0-0.05$                                                                                   \\\hline
    \includegraphics[width=0.19\textwidth]{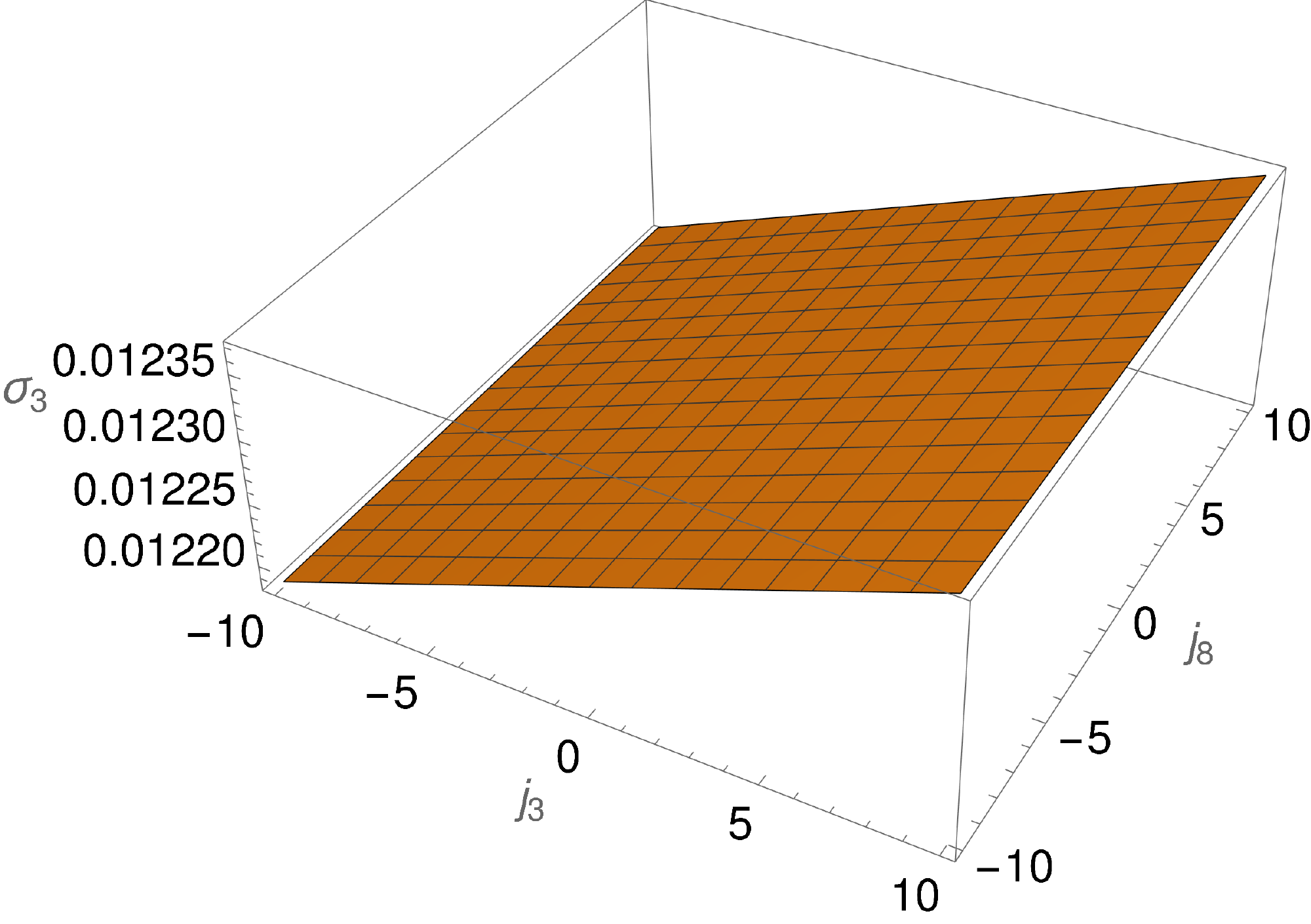}              &
    \includegraphics[width=0.19\textwidth]{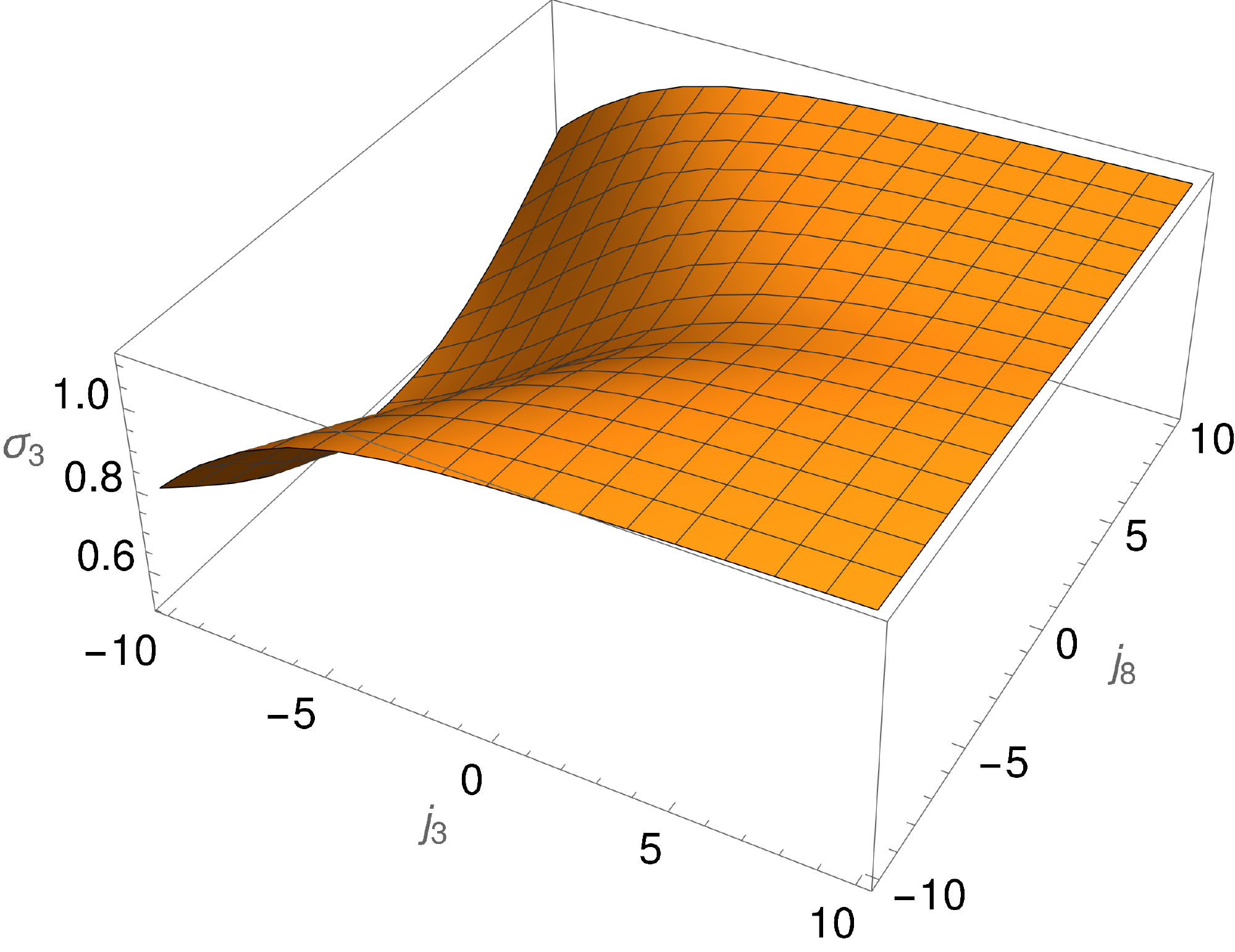}              &
    \includegraphics[width=0.19\textwidth]{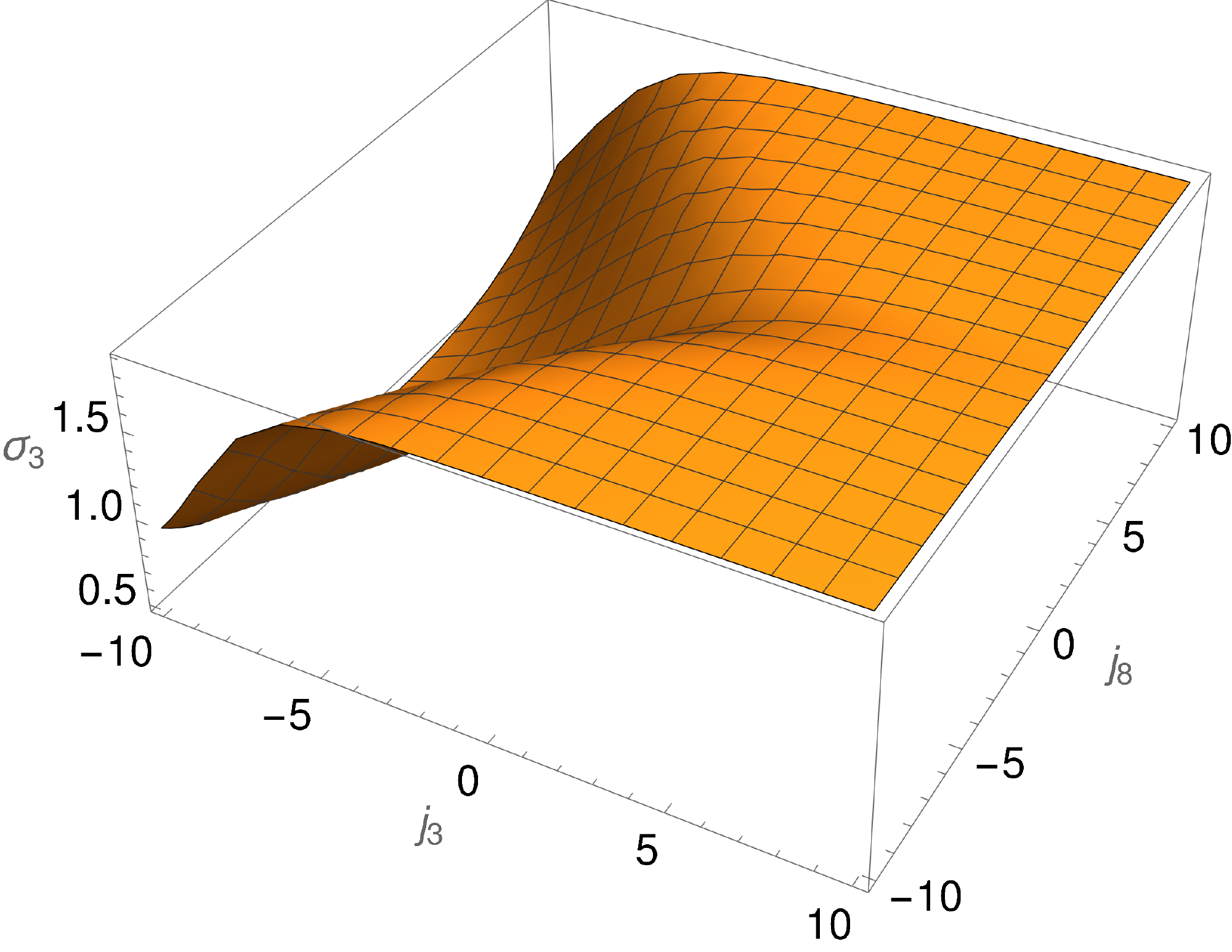}              &
    \includegraphics[width=0.19\textwidth]{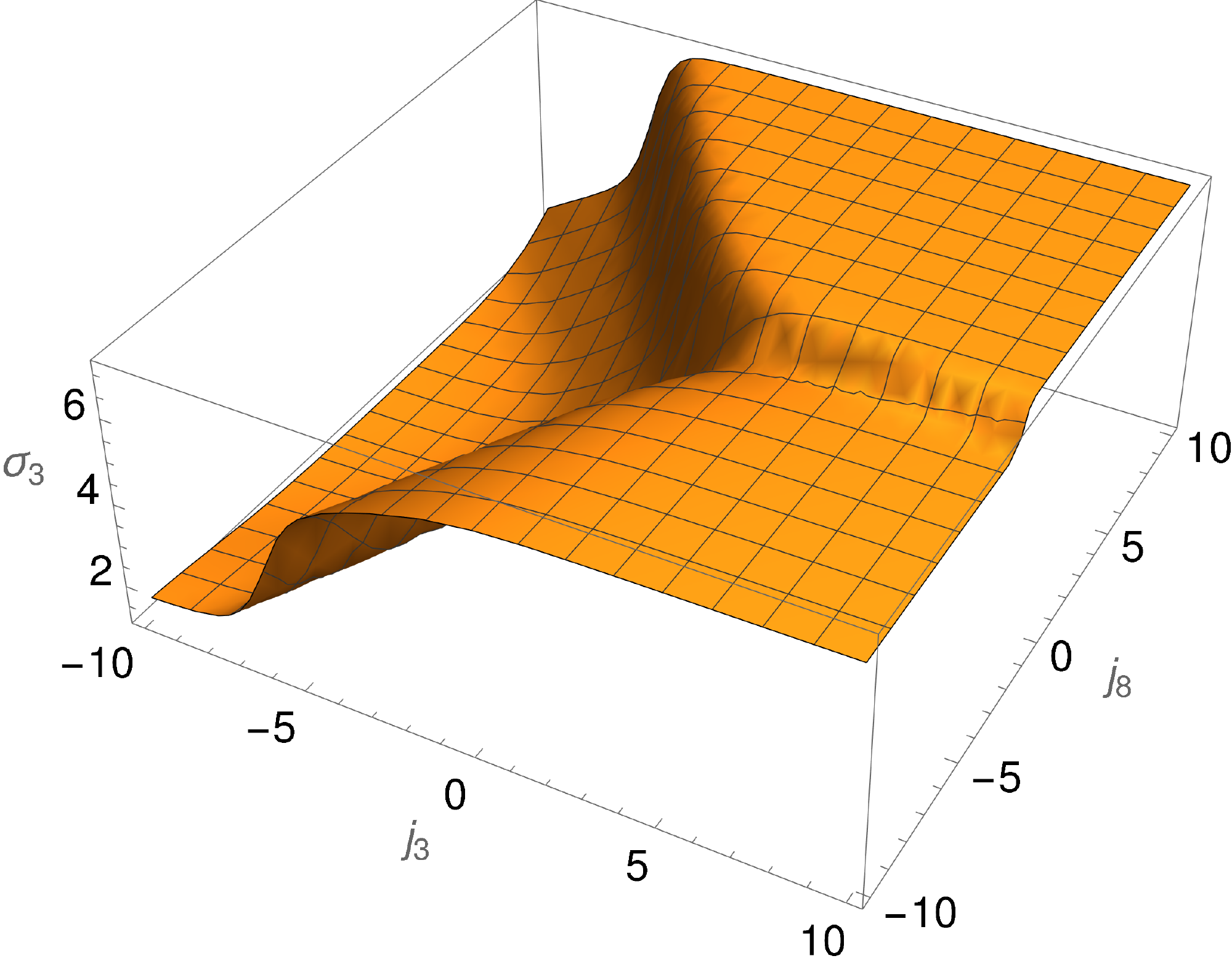}              &
    \includegraphics[width=0.19\textwidth]{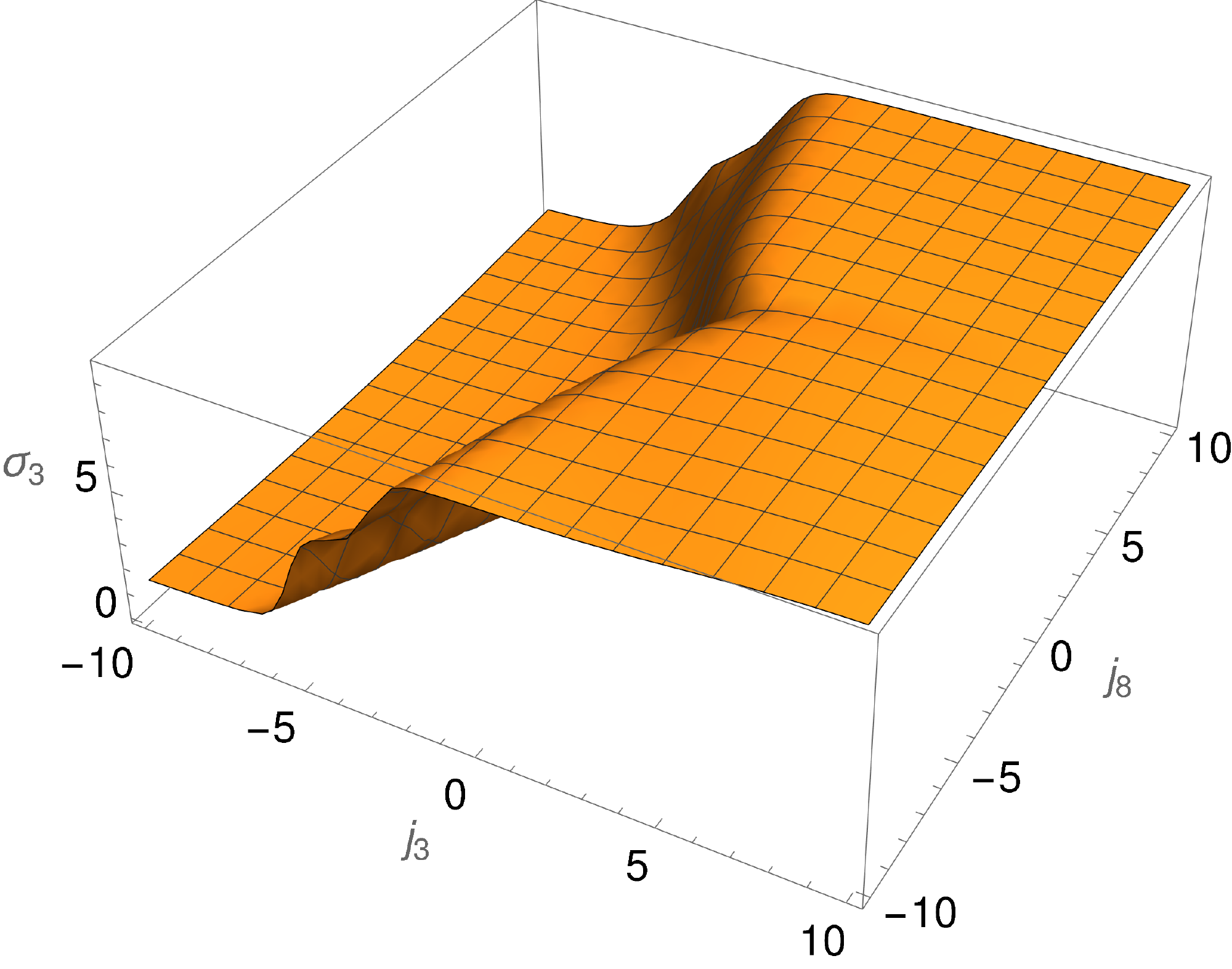}               \\
    \includegraphics[width=0.19\textwidth]{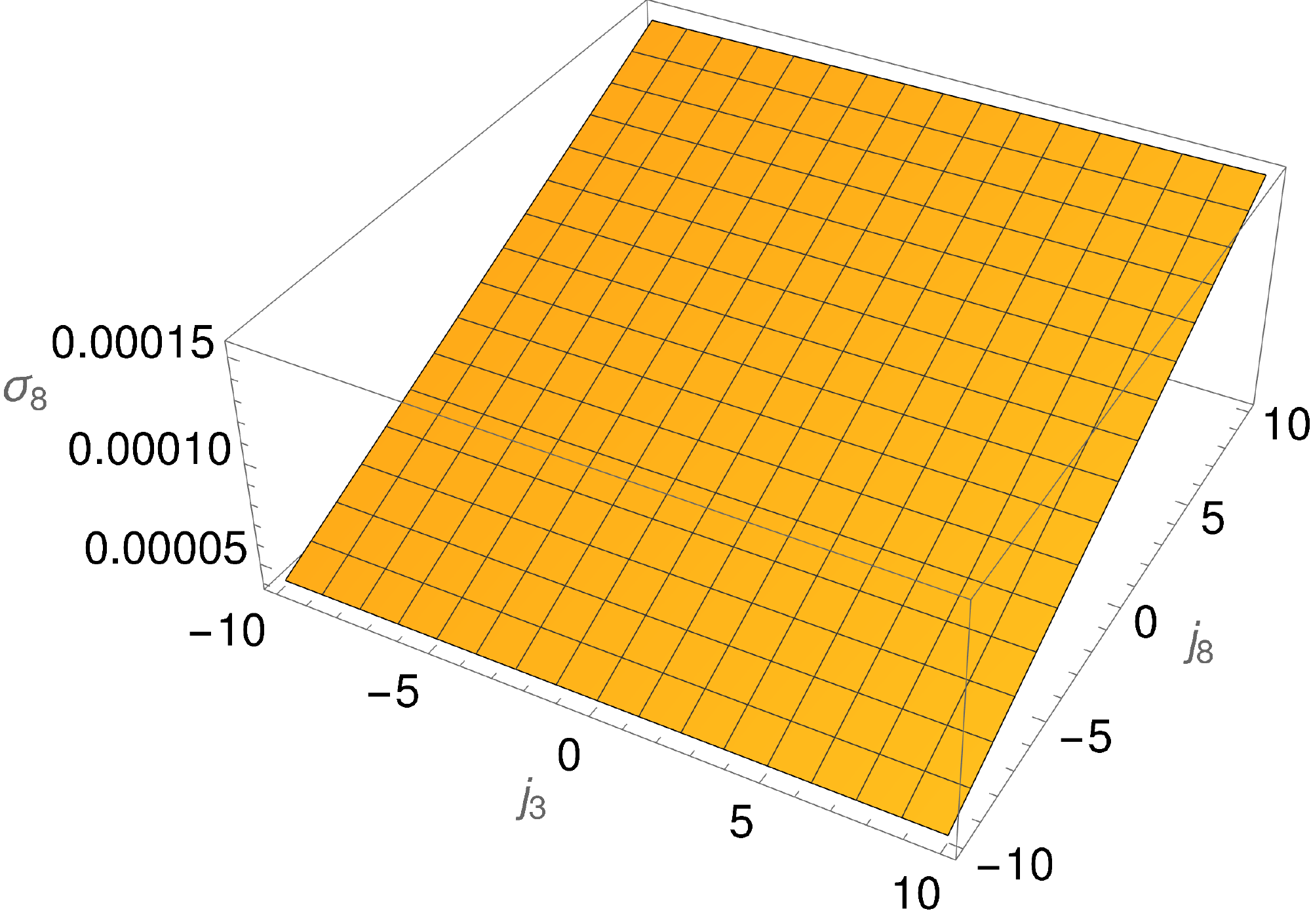}              &
    \includegraphics[width=0.19\textwidth]{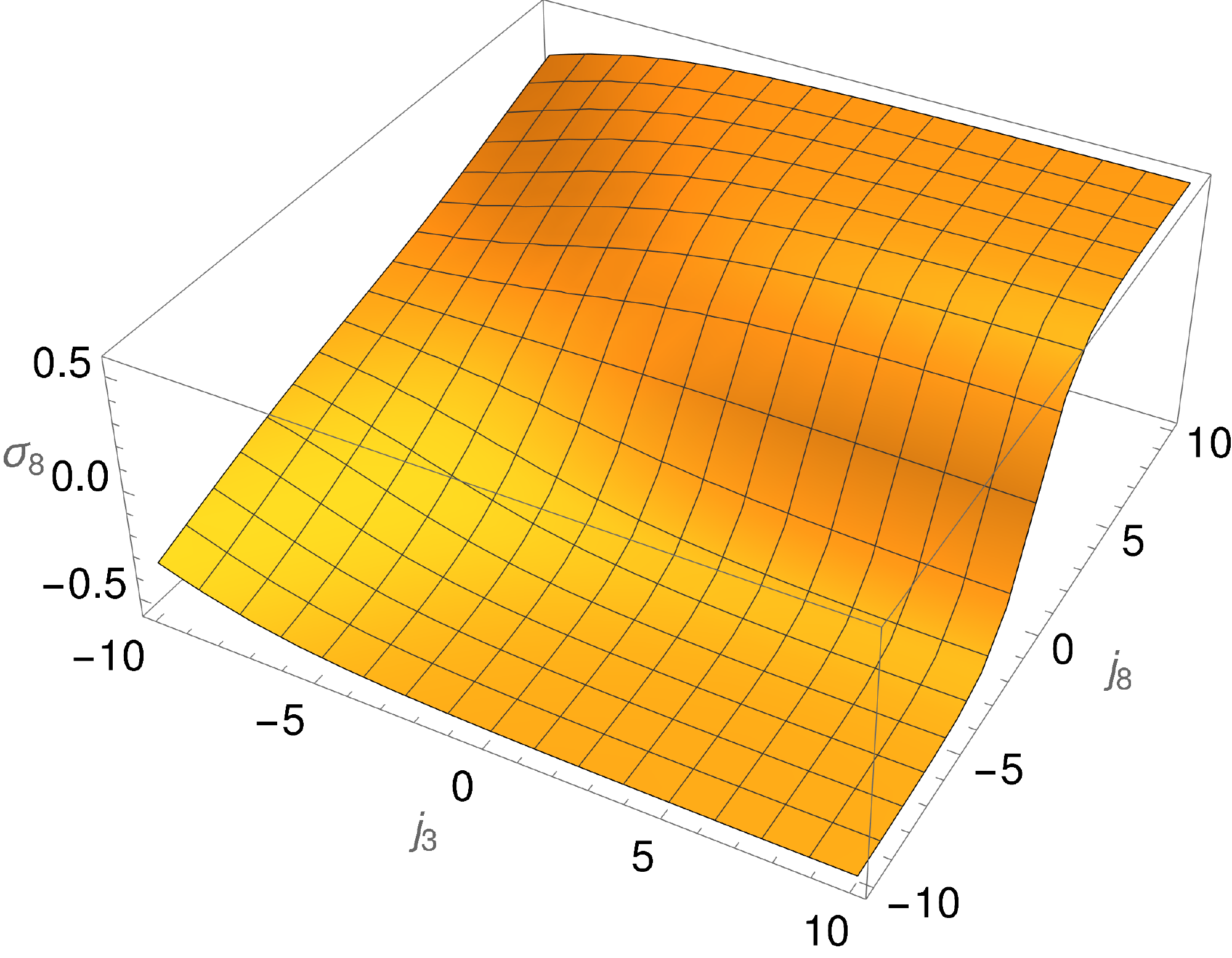}              &
    \includegraphics[width=0.19\textwidth]{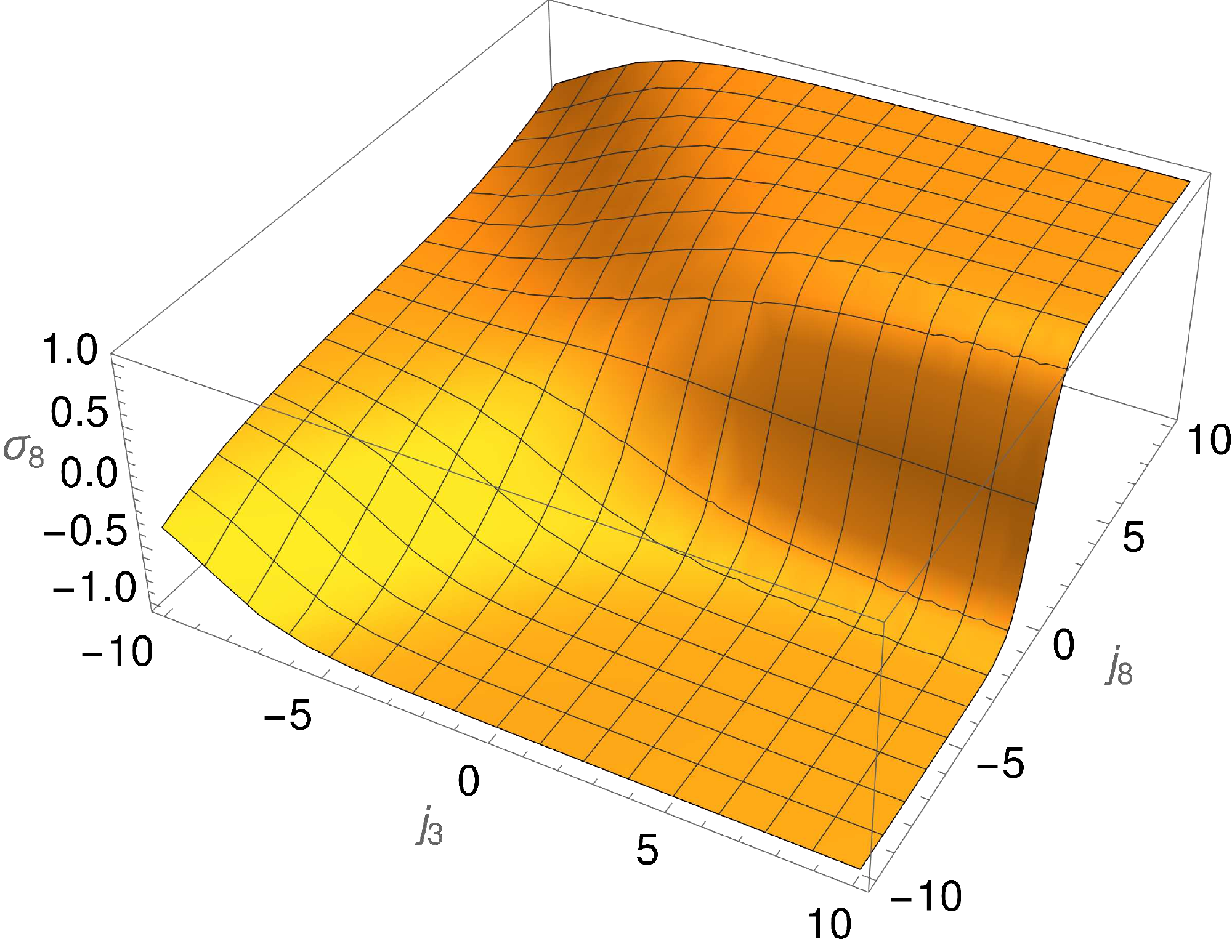}              &
    \includegraphics[width=0.19\textwidth]{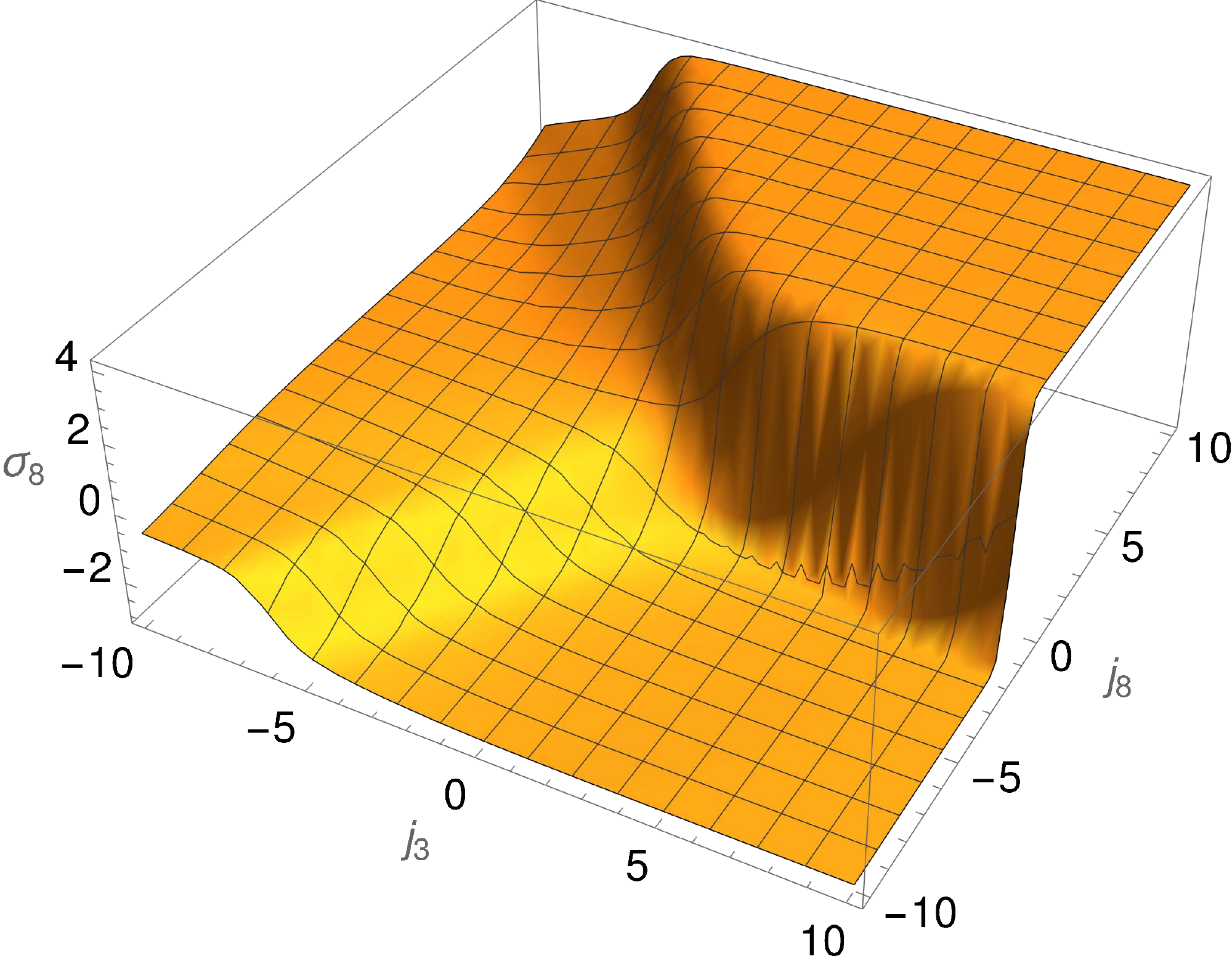}              &
    \includegraphics[width=0.19\textwidth]{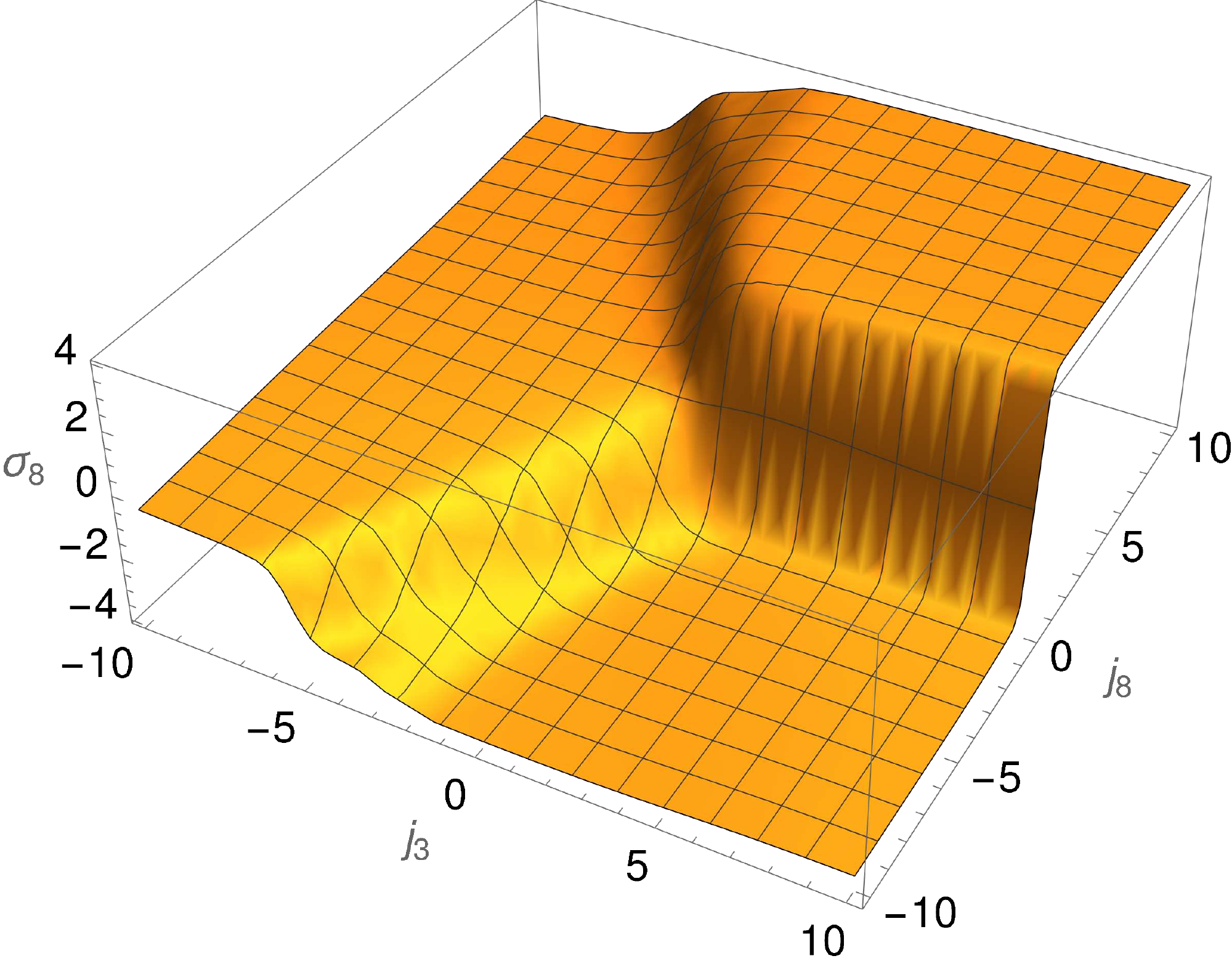}               \\
    \includegraphics[width=0.19\textwidth]{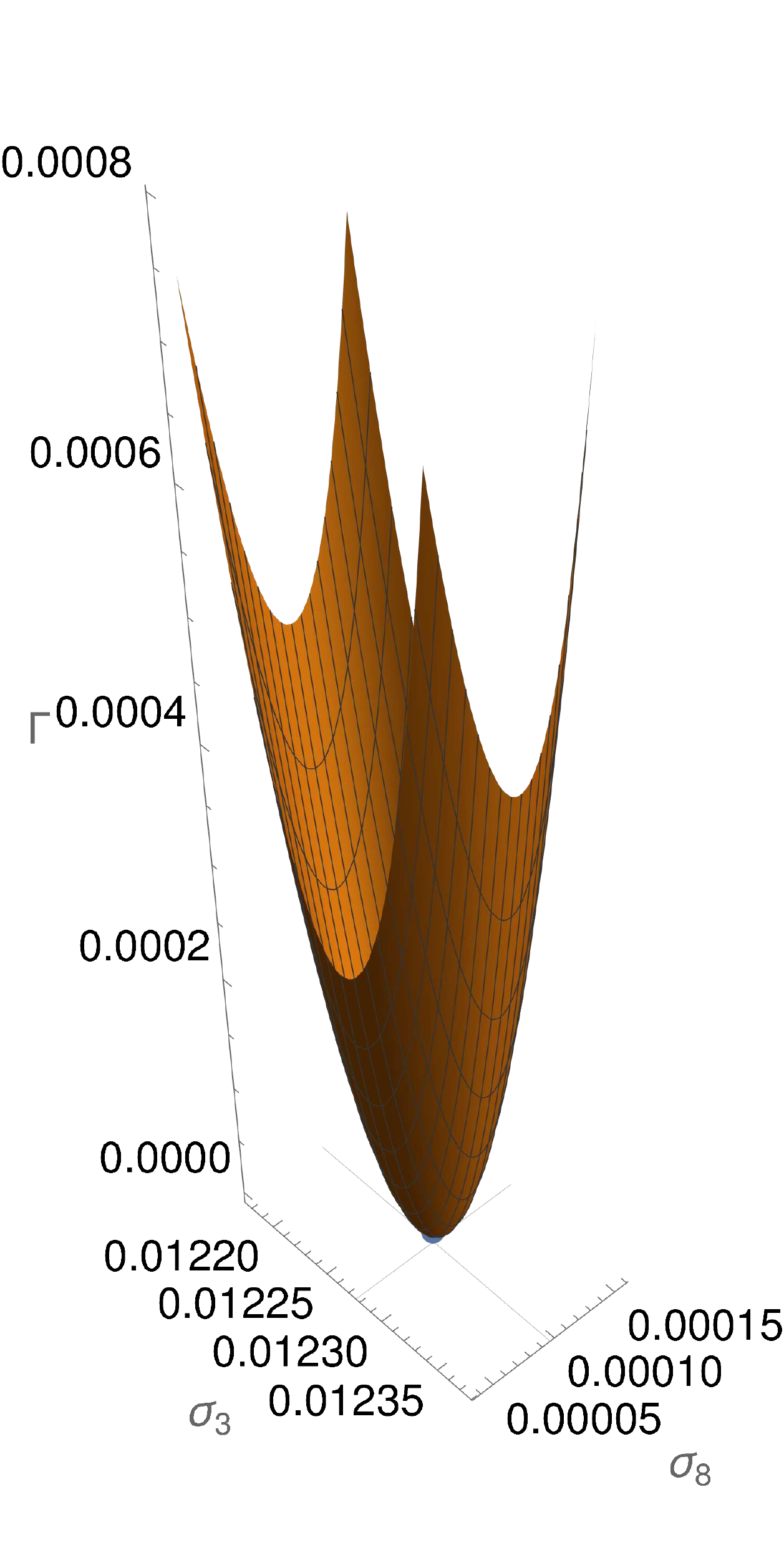} &
    \includegraphics[width=0.19\textwidth]{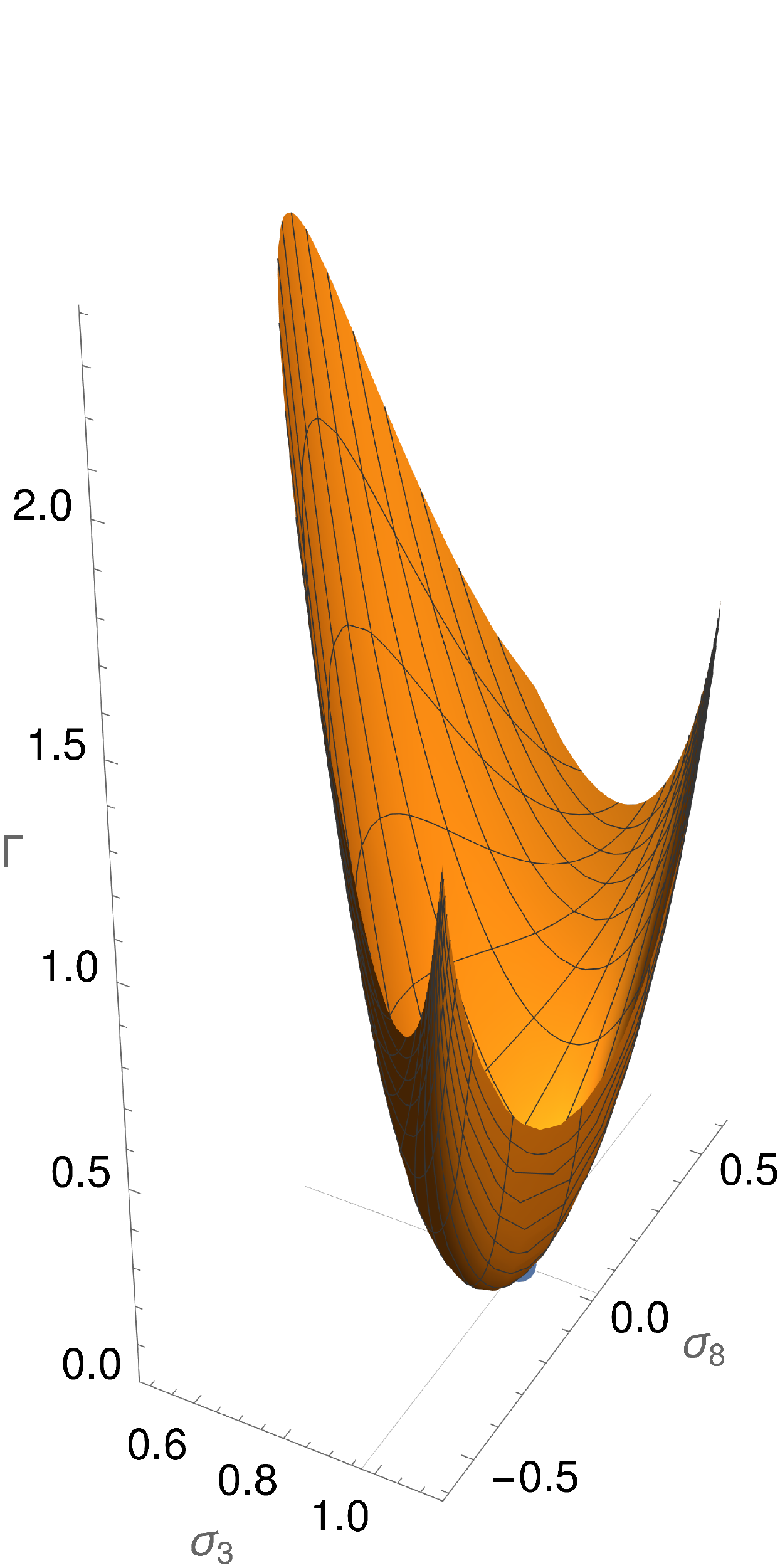} &
    \includegraphics[width=0.19\textwidth]{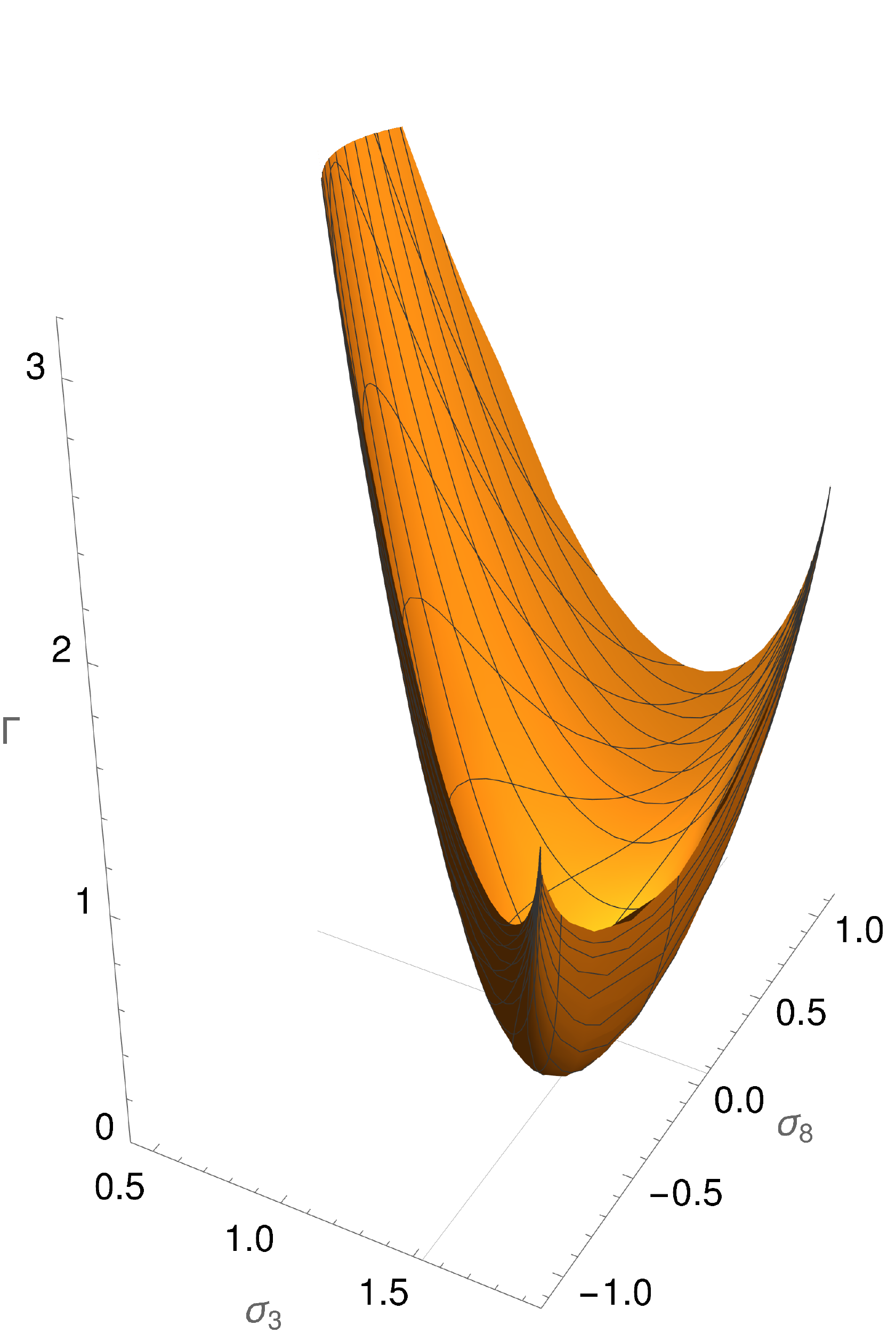} &
    \includegraphics[width=0.19\textwidth]{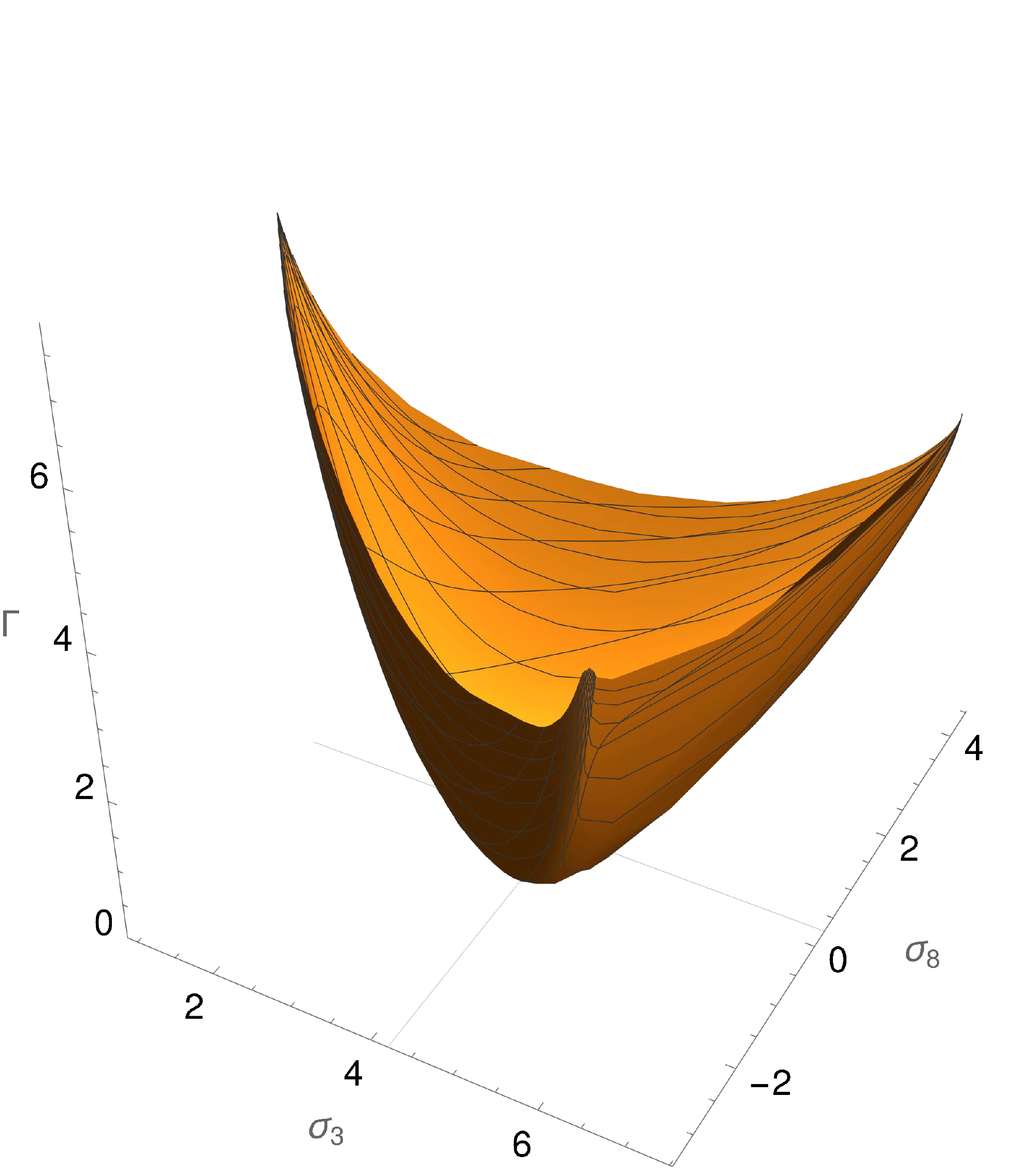} &
    \includegraphics[width=0.19\textwidth]{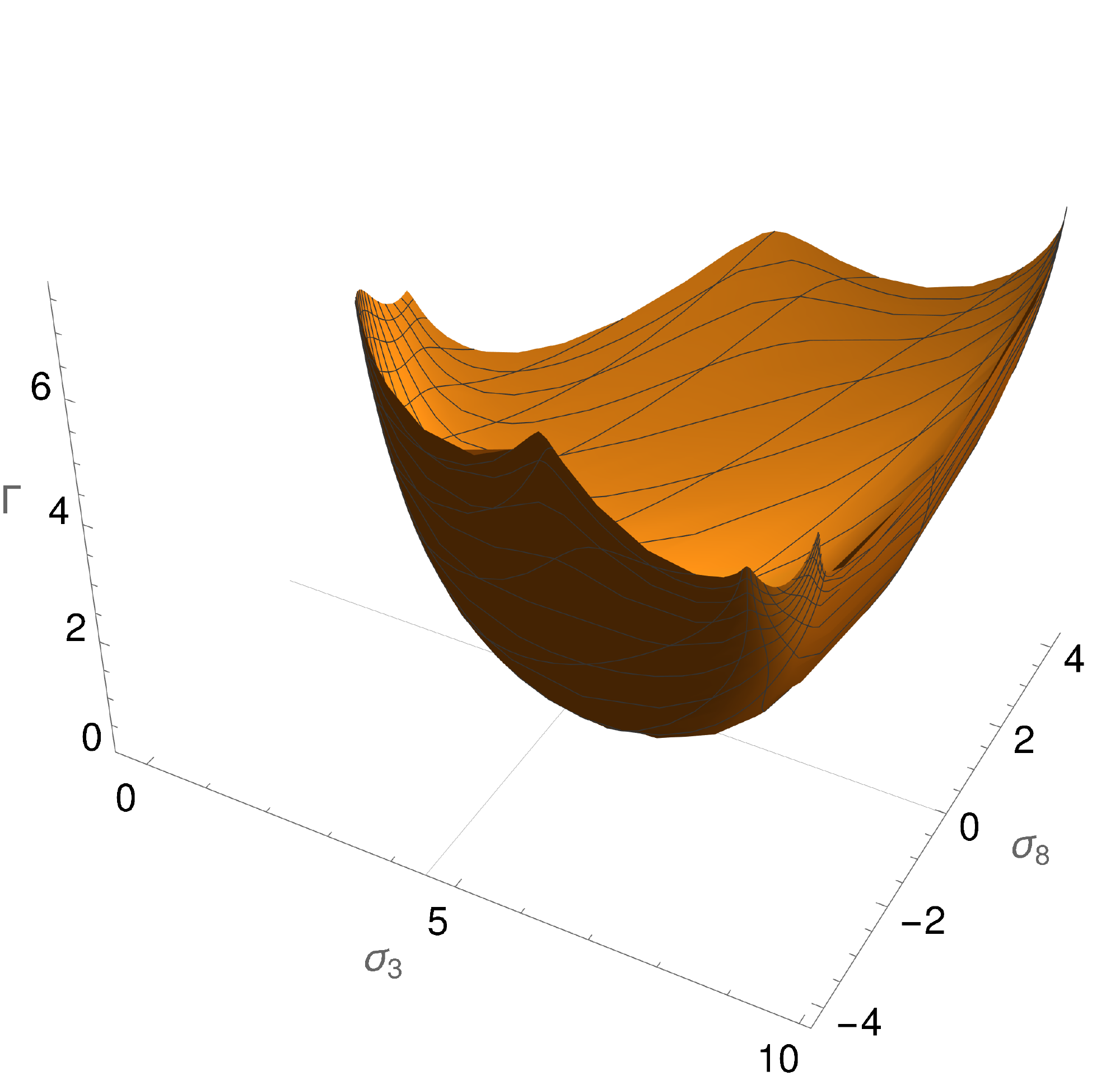}
  \end{tabular}
  \caption{The classical fields $\sigma_3$ (top) and $\sigma_8$ (middle) as a function of the sources, and the quantum effective potential $\Gamma$ (bottom; note the different axes ranges) as a function of the classical fields in various parts of the phase diagram for $V=16^4$. The header of the column gives the respective simulation parameters $\beta-\kappa-\gamma$. Note that smaller volumes have been chosen to reduce equilibration problems; see \cref{a:crit}.}
  \label{fig:qep}
\end{figure*}

As anticipated, imposing unitary gauge with a-priori strict ordering does indeed yield a vacuum expectation value everywhere in the phase diagram. It is therefore possible to force a vev in certain gauges. As we will see, this is not true in all gauges. Thus, the existence of a vev is indeed gauge-dependent and can be independent of the physical phase structure, as has already been observed in the \SU[2] fundamental case \cite{Greensite:2004ke,Caudy:2007sf,Maas:2012ct}.

There is, however, another aspect, which is remarkable. It is usually assumed that if physical quantities undergo a phase transition, then gauge-dependent quantities will also exhibit non-analyticities, even if the reverse is not true \cite{Maas:2017wzi}. However, in the present case, this is actually not the case. The vev and its Binder cumulant, shown in figure \ref{fig:vevorder}, exhibit no sign of a transition, in stark contrast to the \ZZ[2] order parameter shown in figure \ref{fig:z2order}. Thus, in general it cannot be expected that physical non-analyticities induce necessarily non-analyticities in gauge-fixed quantities.

When performing the not over-aligned unitary gauge-fixing, which preserves the \ZZ[2]-symmetry locally, we do find a different picture. We then find indeed that a vev only arises in some part of the phase diagram: It only occurs in the region where the global $\ZZ[2]$ symmetry is broken, at least within our sampling of the phase diagram. The vev shows thus a non-analyticity. In fact, the vev, and its Binder cumulant, show the same behaviour as the \ZZ[2] order parameter at coinciding values of the coupling, within our resolution. Thus, in contrast to the over-aligned gauge, here the gauge-dependent quantities follow the physical phase structure.

Additionally, this gauge-fixing procedure also allows us to plot the local angle distribution on the lattice. In \cref{fig:conf} we show exemplary configurations for a single time-slice in different parts of the phase diagram, which exhibit different strengths of alignment and corresponding average values. When going closer to $\gamma\approx 0$ the alignment becomes stronger and the average value approaches $\pm\pi/6$, i.e. the plot becomes darker. This is shown in \cref{fig:conf_deep_broken}. A similar formation of structures has also been observed in other systems with symmetry breaking \cite{Endrodi:2021kur}.

Finally, it should also be mentioned that in both cases the breaking angles obtained from the space-time-averaged gauge-fixed fields are indeed very similar to the ones obtained from \cref{eqn:O_theta}. This is a non-trivial result due to the interchange of the space-time average and the matrix powers. So indeed the long-range correlations in this case are suitably well described already without any explicit gauge-fixing.

The other results are then similar as the 't Hooft-Landau gauge ones. Given that the non-sorted/post-sorted unitary gauge cannot really be replicated in (augmented) perturbation theory, we therefore will consider this gauge no further.

\subsection{Landau--'t Hooft gauge}

In this gauge, we observe a vev whenever global $\ZZ[2]$ is broken. Beyond that, we find a more intricate situation.

When just calculating the angle from the vev, we again find a distribution very similar to the one shown in figure \ref{fig:z2}. What we observe is that it appears that throughout the $\ZZ[2]$ broken phase the angle $\theta_0$ is such that the breaking pattern is exclusively $\U[1]\times\U[1]$. Only close to the phase boundary to the unbroken phase it approaches $\U[1]\times\U[1]^\star$, and for $\gamma\to 0$ it approaches $\SU[2]\times\U[1]$. Whether it reaches the special values cannot be determined for certain numerically. This result paints a very different picture than at tree-level.

To understand it better, it is useful to resolve the situation on a per-configuration basis. We show sample distributions of the angle $\theta_0$ as a function of volume and for different points in the phase diagram in \cref{fig:theta}. We also show a more detailed resolution of the vev and $\theta_0$ distributions in figures \ref{fig:vev} and \ref{fig:theta_vs_vev}, for the unbroken phase, close to the phase boundary on both sides of the transition, inside the bulk of the broken phase, and close to $\gamma=0$, respectively. To interpret the plots, it is important to remember that the average vev is determined by (\ref{eqn:lvev}), and thus the per-configuration angles and vev do not necessarily average independently to it, due to their non-linear relation (\ref{eqn:vev}). Furthermore, critical slowing-down effects are very pronounced for $\gamma\to 0$, making the results on larger volumes increasingly unreliable, see appendix \ref{a:crit}. We therefore also indicate the information for the configuration-wise result from where in the Monte-Carlo trajectory they are obtained, and deduce (necessarily) parts of our findings on the observed trend. Estimating the amount of computing time for full thermalization on our largest volumes puts a different approach clearly beyond the current capabilities of our infrastructure.

The result is that we observe different distributions depending on the couplings. These distributions have large finite-volume effects, which complicate the analysis. We also observe that the $\theta_0$ distribution shows the expected $\ZZ[2]$-symmetric distribution around $\theta_0=0$. Given the meta-stability in the $\ZZ[2]$-broken phase this implies a drastic change of the results once the symmetry becomes explicitly or spontaneously broken.

In the $\ZZ[2]$-unbroken phase (the first columns of \crefrange{fig:theta}{fig:theta_vs_vev}), no preferential structure is observed, and both $\theta_0$ and $w$ are Gaussian distributed. As the volume increases, $w$ decreases towards zero, while the distribution of $\theta_0$ remains Gaussian. This is indeed the effect expected for the absence of a BEH effect. Thus, the $\ZZ[2]$-unbroken phase in this gauge can be interpreted as one in which no BEH effect prevails in the infinite-volume limit.

Outside the $\gamma\approx 0$ regime and inside the $\ZZ[2]$-broken phase, the picture is relatively similar (see the second and third columns of \crefrange{fig:theta}{fig:theta_vs_vev}). The $\theta_0$ distribution is strongly non-Gaussian, increasing with volume towards the $\theta_0=\pm\pi/6$ boundaries, and large absolute values of the angles correlate with large values of $w$. The change to this radically different distribution from the $\ZZ[2]$-unbroken phase is very abrupt. However, as the $\ZZ[2]$ alignment becomes weaker and weaker towards the phase boundary, as evidenced by the order parameter values, the pattern needs to align more and more towards being compatible with no such alignment, but maintaining the vev. This is only possible in the $\U[1]\times\U[1]^\star$ pattern, thus emerging towards the phase boundary.

This can also be understood in the following way. Because the $\ZZ[2]$ order parameter (\ref{eqn:z2order}) depends only on the sign of the determinant, any information about long-range order in the scalar field is lost, as this depends on the relative ordering of the eigenvalues. However, the $\U[1]\times\U[1]^\star$ case implies a vanishing determinant, due to the zero eigenvalue. Thus, in case of a long-range ordering of this type, \ZZ[2] symmetry would be necessarily intact. In the $\U[1]\times\U[1]$ case the determinant depends non-linearly on the relative sign of the two eigenvalues. If one of the eigenvalues is relatively large and the other one is asymmetrically distributed around zero, the determinant can fluctuate wildly in configurations, but long-range order is still possible for the eigenvalues. Hence, in total there is no way to decide a-priori from the status of the \ZZ[2] symmetry whether a BEH effect is present, at most a subset of the possible patterns can be deduced, as discussed in the previous section. Thus, it is highly non-trivial that a correspondence is found. In absence of the other effects, this yields that the $\U[1]\times\U[1]$ pattern prevails.

The situation changes when moving towards small $\gamma$ (shown in the fourth and last columns of \crefrange{fig:theta}{fig:theta_vs_vev}). The first thing to be noted is that for these setups the vev $w$ shows again a volume dependence. This is however not an indication of an unbroken phase in this case but rather an artefact of the equilibration problems we encounter with decreasing $\gamma$. For a more detailed discussion on this effect see \cref{a:crit}. Taking this into account we also see that the vev and $\theta_0$ then become less correlated once more compared to the previous cases. The absence of a correlation would be expected if there are more independent possibilities to orient the average value without an action penalty. This would be the case in the $\SU[2]\times\U[1]$ case. That is again consistent with the averaged result. In this case the determinant in the $\ZZ[2]$ order parameter (\ref{eqn:z2order}) is also long-range ordered. Thus, this pattern can only be realized if \ZZ[2] is broken. Intuitively, it is not surprising to find this pattern only at very small $\gamma$, where fluctuations of the scalar field are less restricted. There, the BEH effect is not strong enough to generate masses for all gauge bosons, and thus more remain massless. This is consistent with NLO estimates in three dimensions, where at small $\gamma$ also only the $\SU[2]\times\U[1]$ pattern \cite{Kajantie:1998yc} has been found.

However, in all cases the behaviour on individual configurations is thus very different from the averaged behaviour evidenced by figure \ref{fig:z2}. Though it should be mentioned that, as in the unitary gauge case, we observe that the breaking patterns obtained from the implicit gauge-fixing procedure do coincide with the ones obtained in this particular gauge. Especially the strong appearance of the $\SU[2]\times\U[1]$ angle $\theta_0=\pm\pi/6$ in individual configurations averaging to a $\U[1]\times\U[1]$ pattern eventually strongly suggests that the alignment per configuration is not particularly strong. However, this effect averages out, except when the potential becomes shallow enough for $\gamma\approx 0$. This view is also corroborated by the intricate space-time pattern in the individual configurations seen in \cref{fig:conf_two_pats} for unitary gauge. We clearly see there the formation of Weiss domains in a wave pattern. Similar observations have been made in other systems \cite{Endrodi:2021kur}. Although the referred configurations are obtained in unitary gauge the same considerations should carry over to Landau--'t Hooft gauge since these two gauges are only the opposing extremes of $R_\xi$.

\begin{figure*}[t!]
  \centering
  \begin{tabular}{cccc}
    \includegraphics[width=0.441\linewidth]{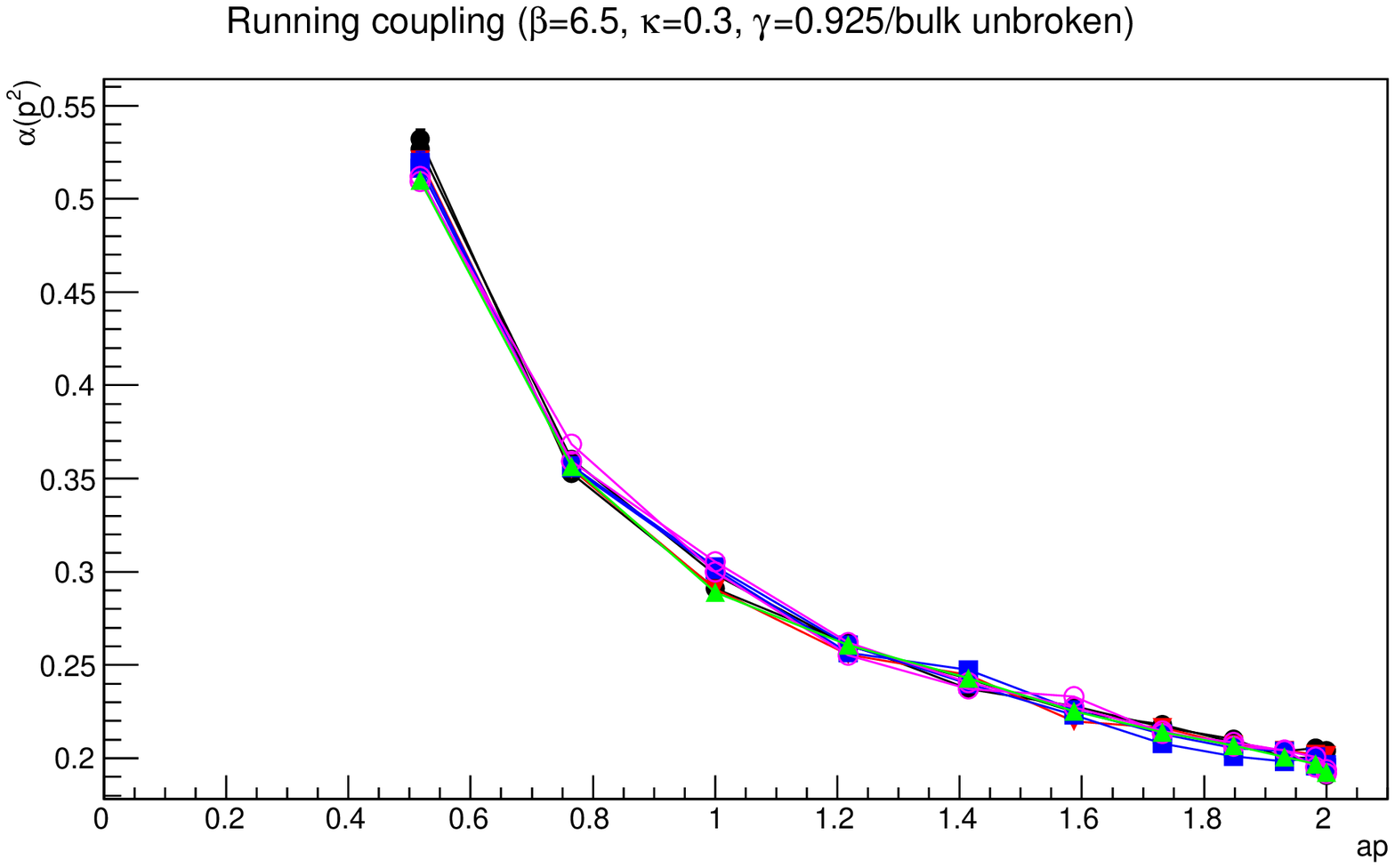}  &
    \includegraphics[width=0.441\linewidth]{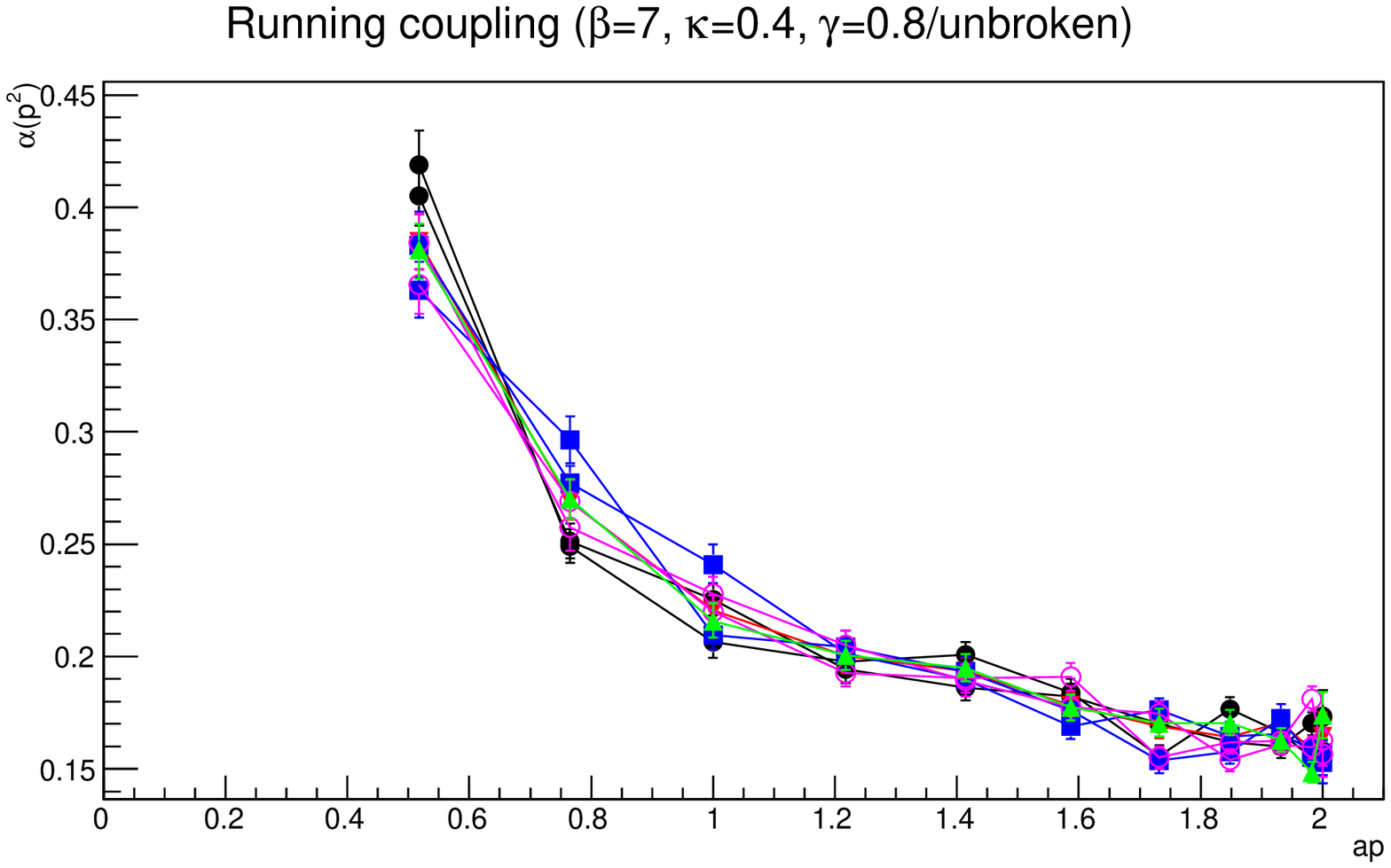}    & \\
    \includegraphics[width=0.441\linewidth]{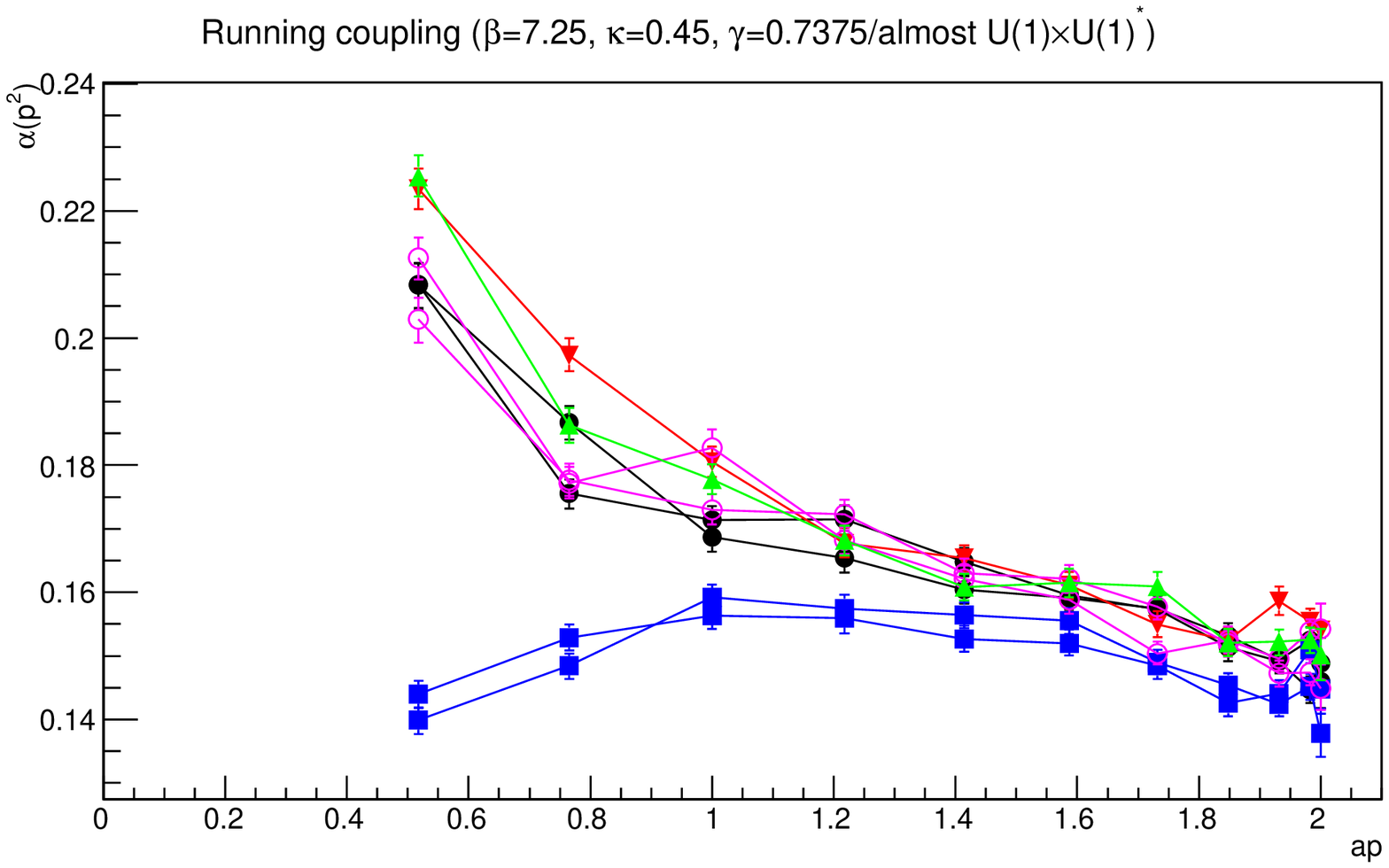} &
    \includegraphics[width=0.441\linewidth]{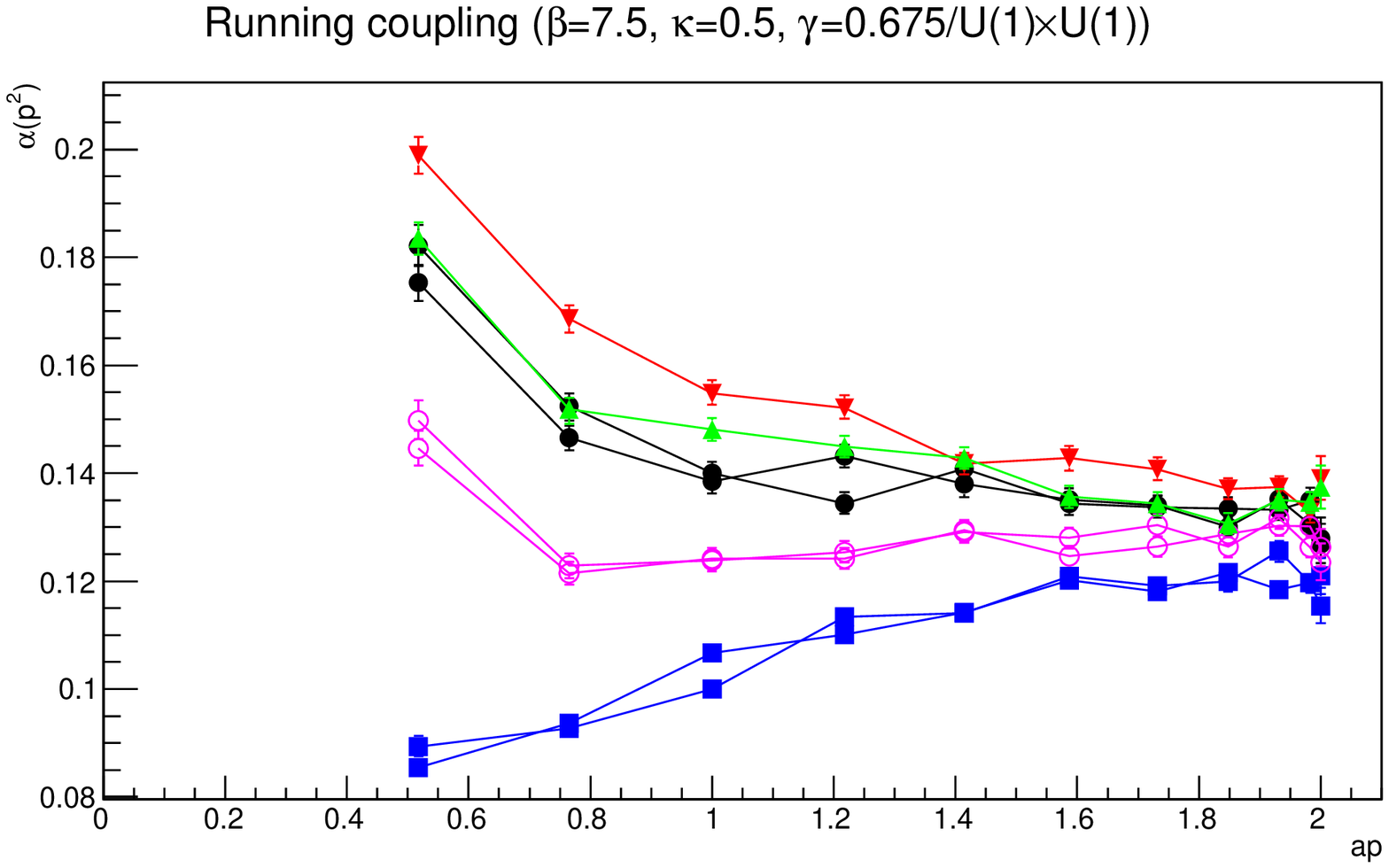}    \\
    \includegraphics[width=0.441\textwidth]{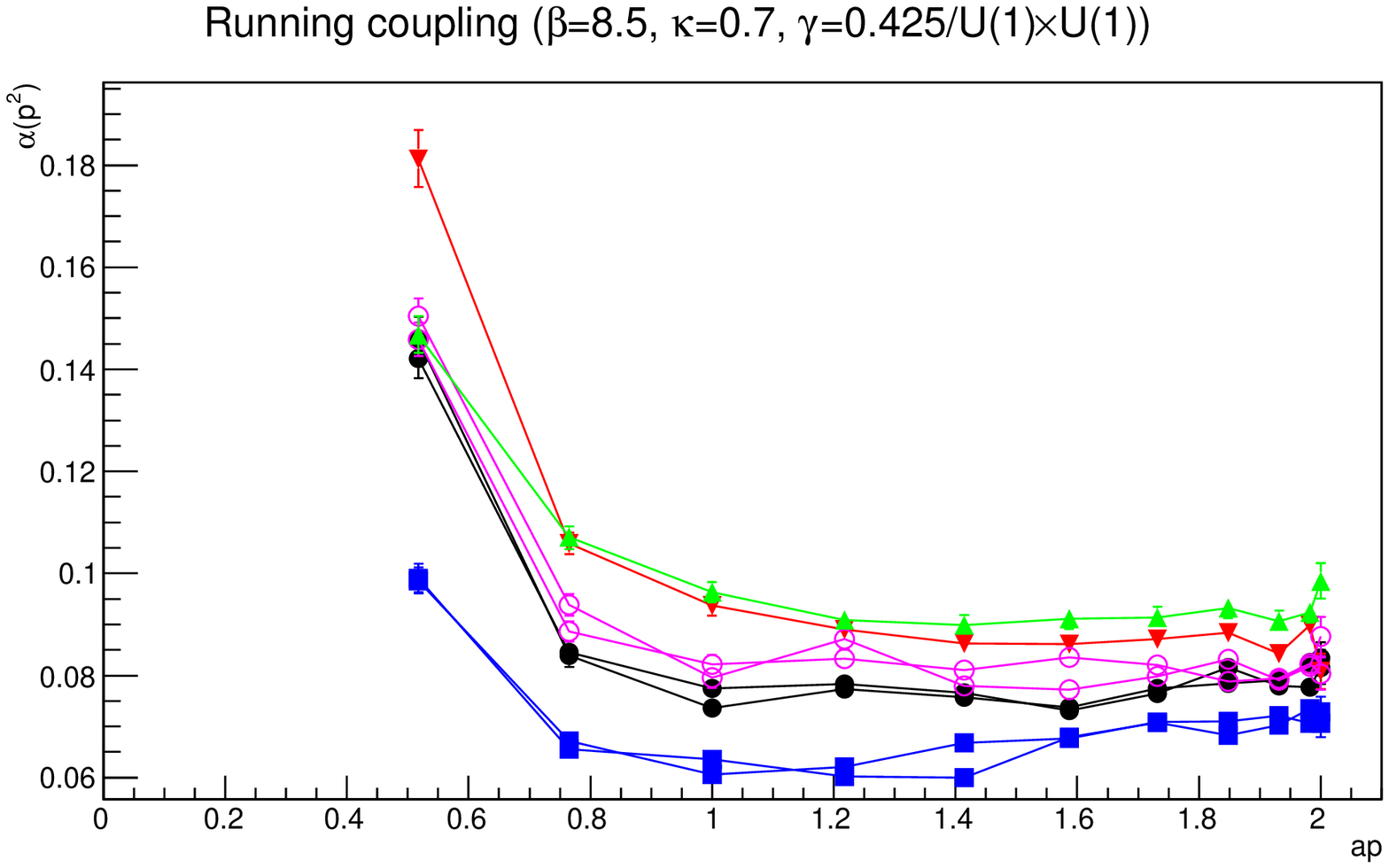}  &
    \includegraphics[width=0.441\textwidth]{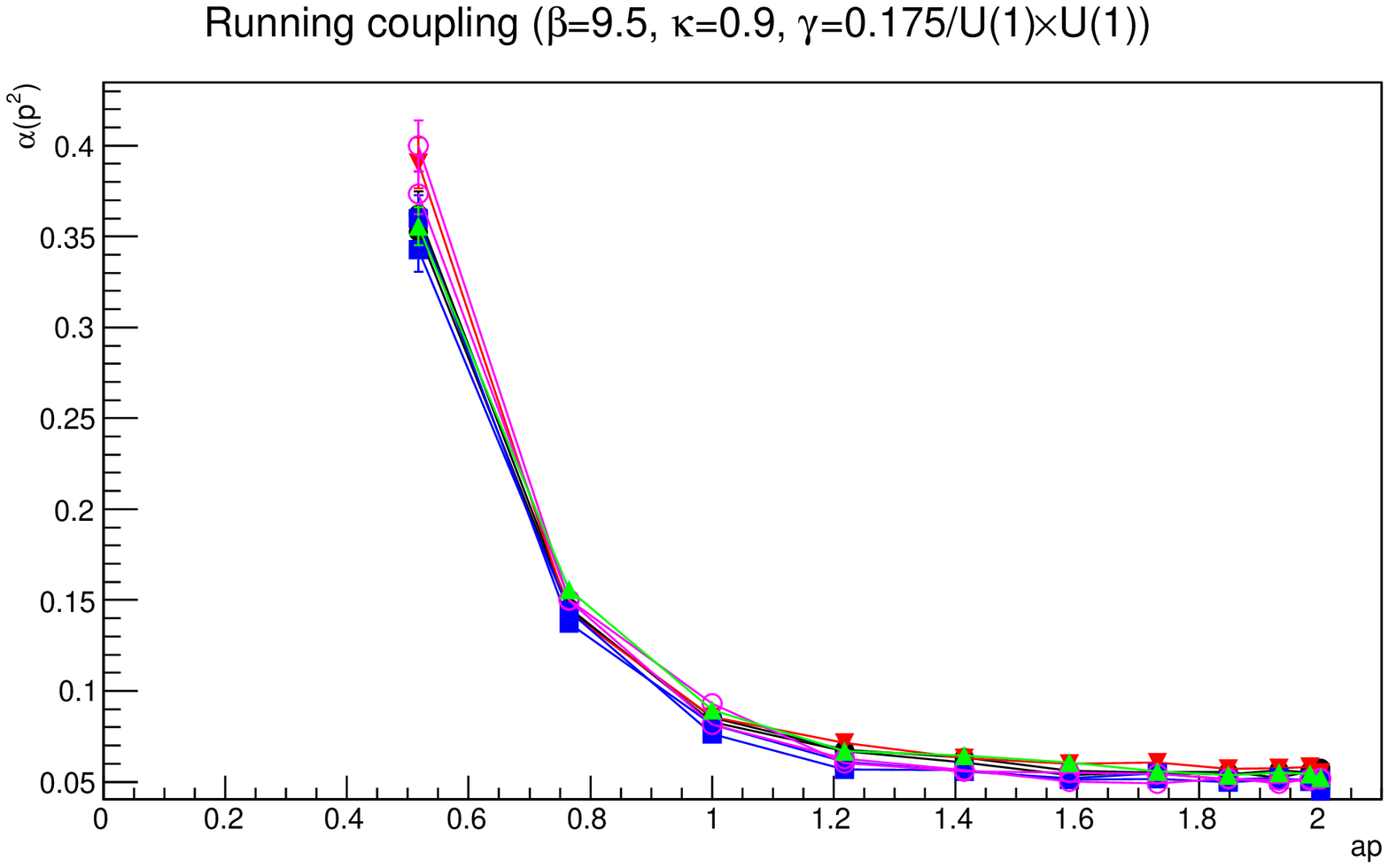}  & \\
    \includegraphics[width=0.441\textwidth]{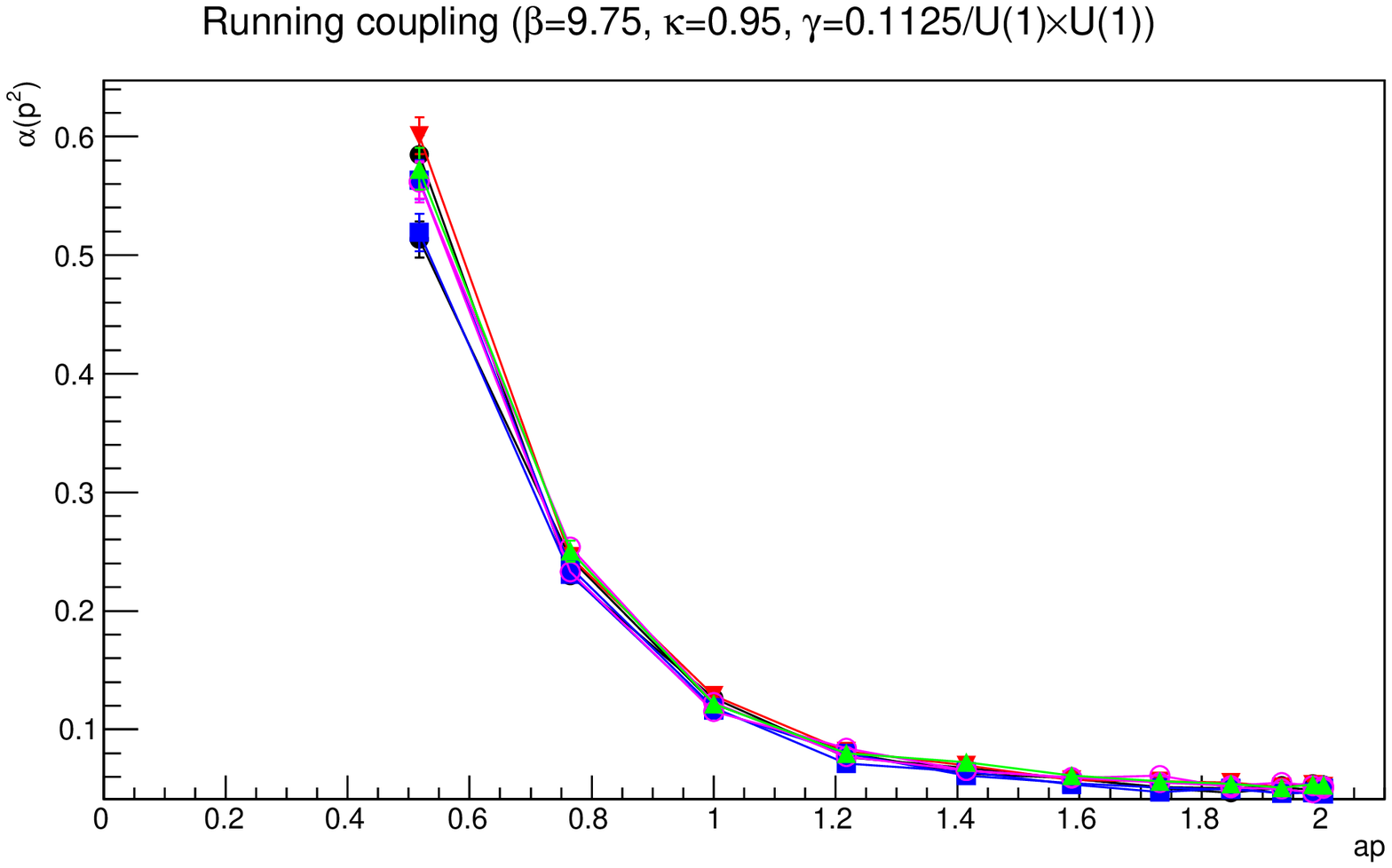} &
    \includegraphics[width=0.441\textwidth]{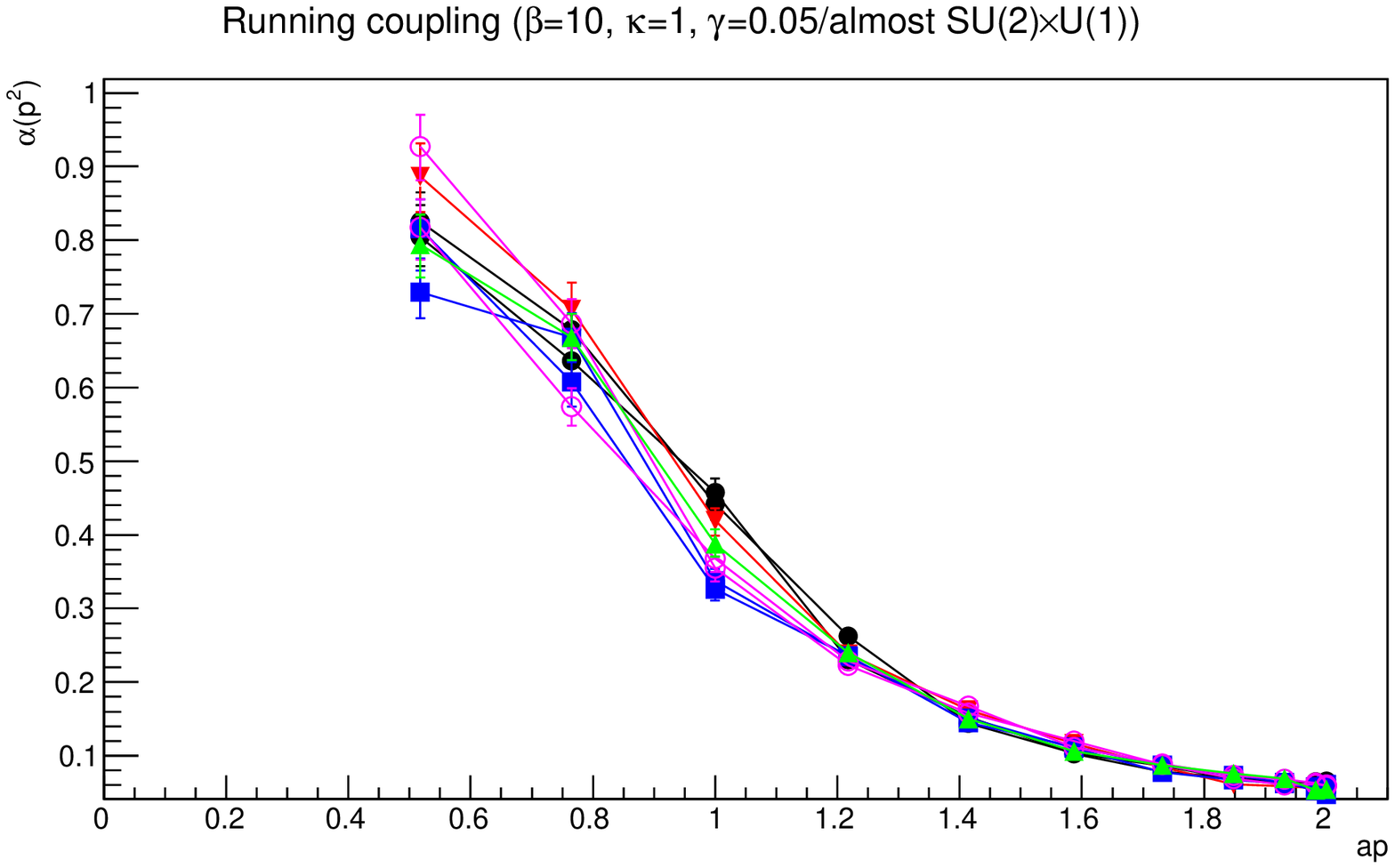}
  \end{tabular}\\
  \includegraphics[width=\textwidth]{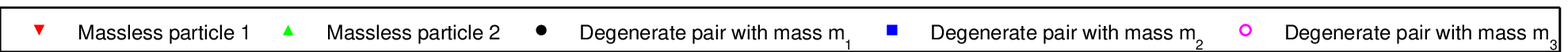}
  \caption{The running gauge coupling for the eight different charges moving from deep inside the unbroken phase ($\beta=6.5$, top right) to close to both sides of the \ZZ[2] phase transition ($\beta=7$ and $\beta=7.25$) through the broken phase ($\beta=7.5$, $\beta=8.5$, $\beta=9.5$, $\beta=9.75$) to the smallest numerically accessible value of $\gamma$ ($\beta=10$) in the lower-right panel. All results from the $24^4$ lattice, the momenta are along the $x$-axis, which minimizes finite-volume effects, and the first non-zero momentum point is suppressed due to remaining finite-volume effects.}
  \label{fig:alpha}
\end{figure*}
\begin{figure*}[t!]
  \centering
  \begin{tabular}{cccc}
    \includegraphics[width=0.441\linewidth]{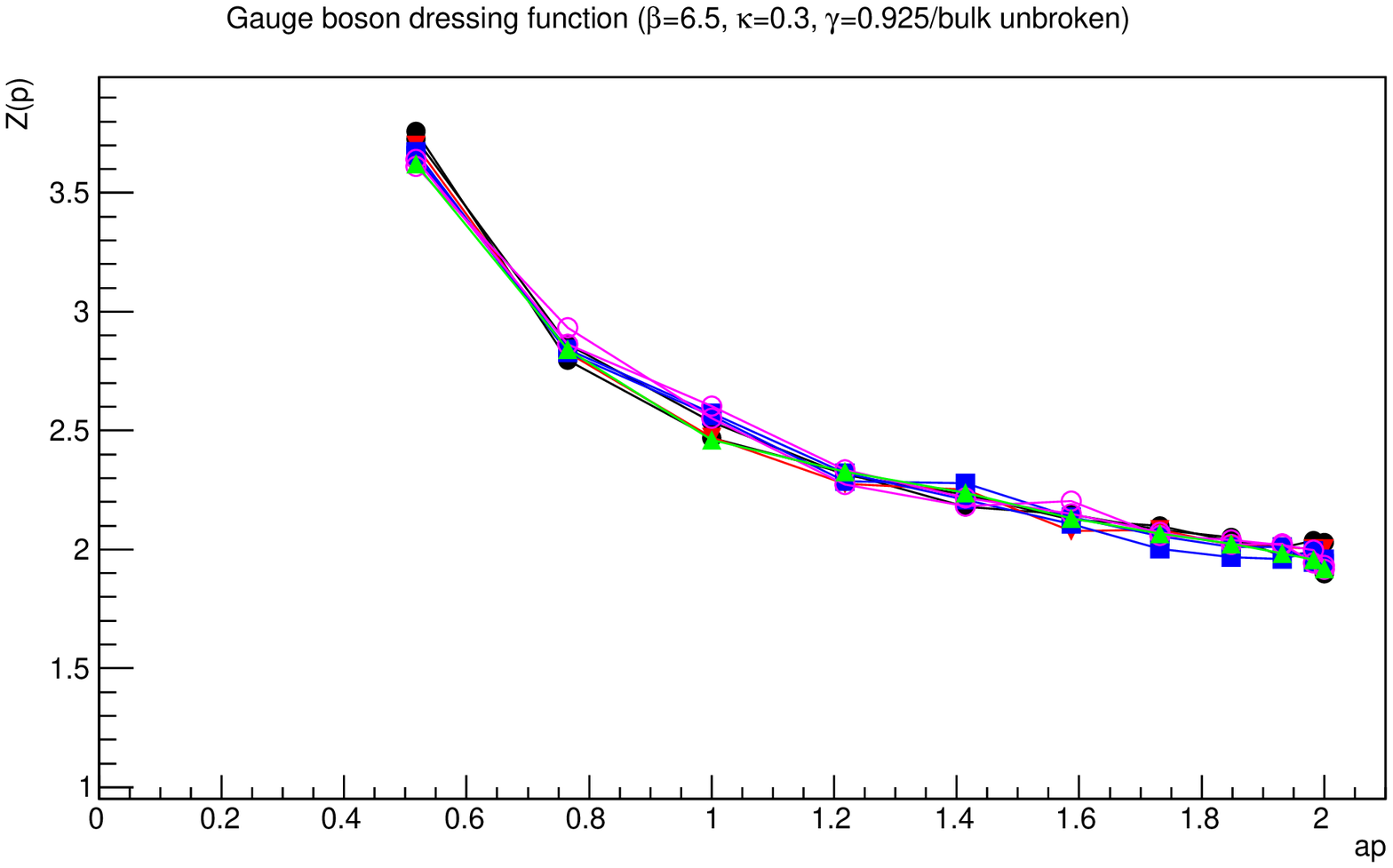}  &
    \includegraphics[width=0.441\linewidth]{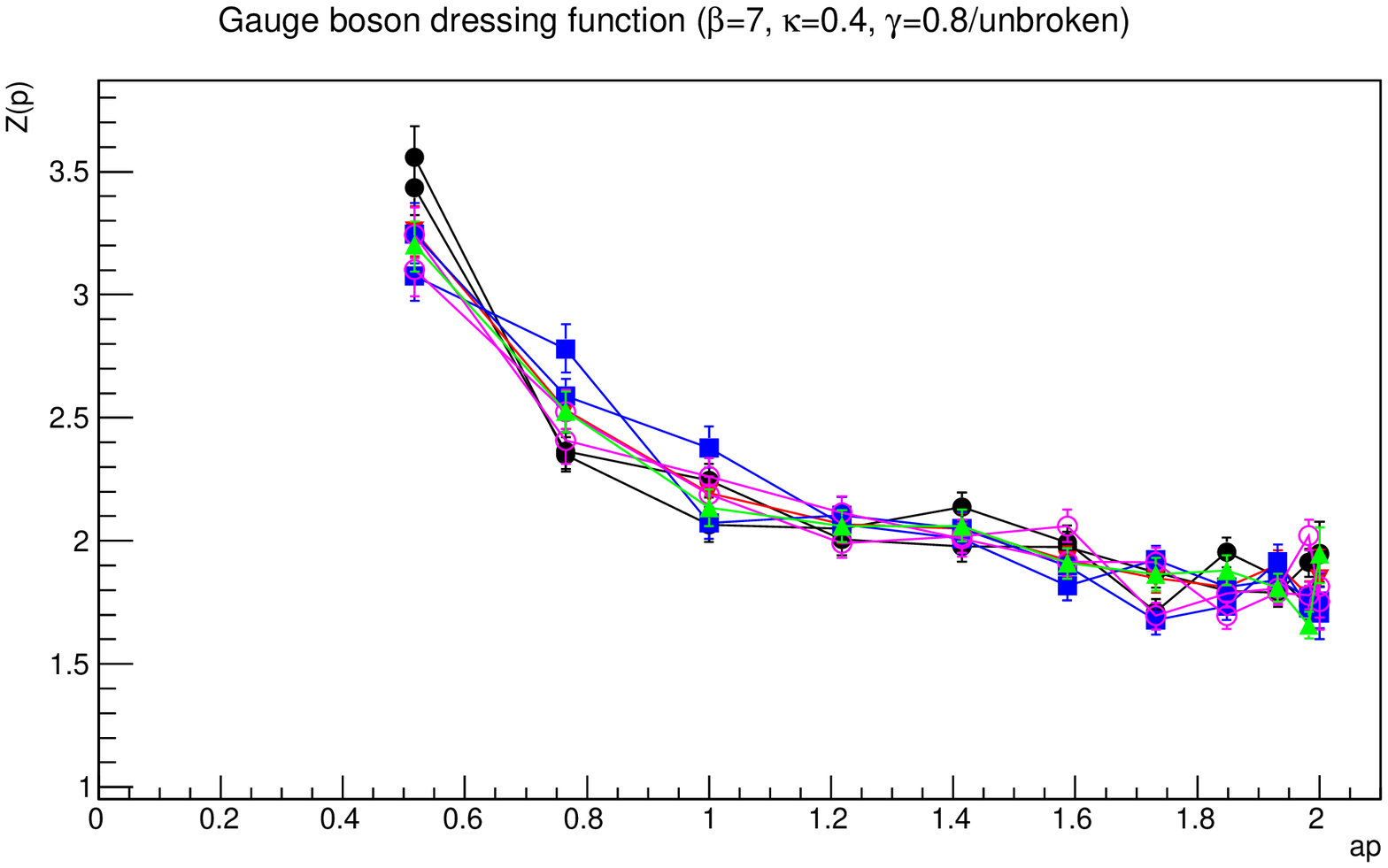}    & \\
    \includegraphics[width=0.441\linewidth]{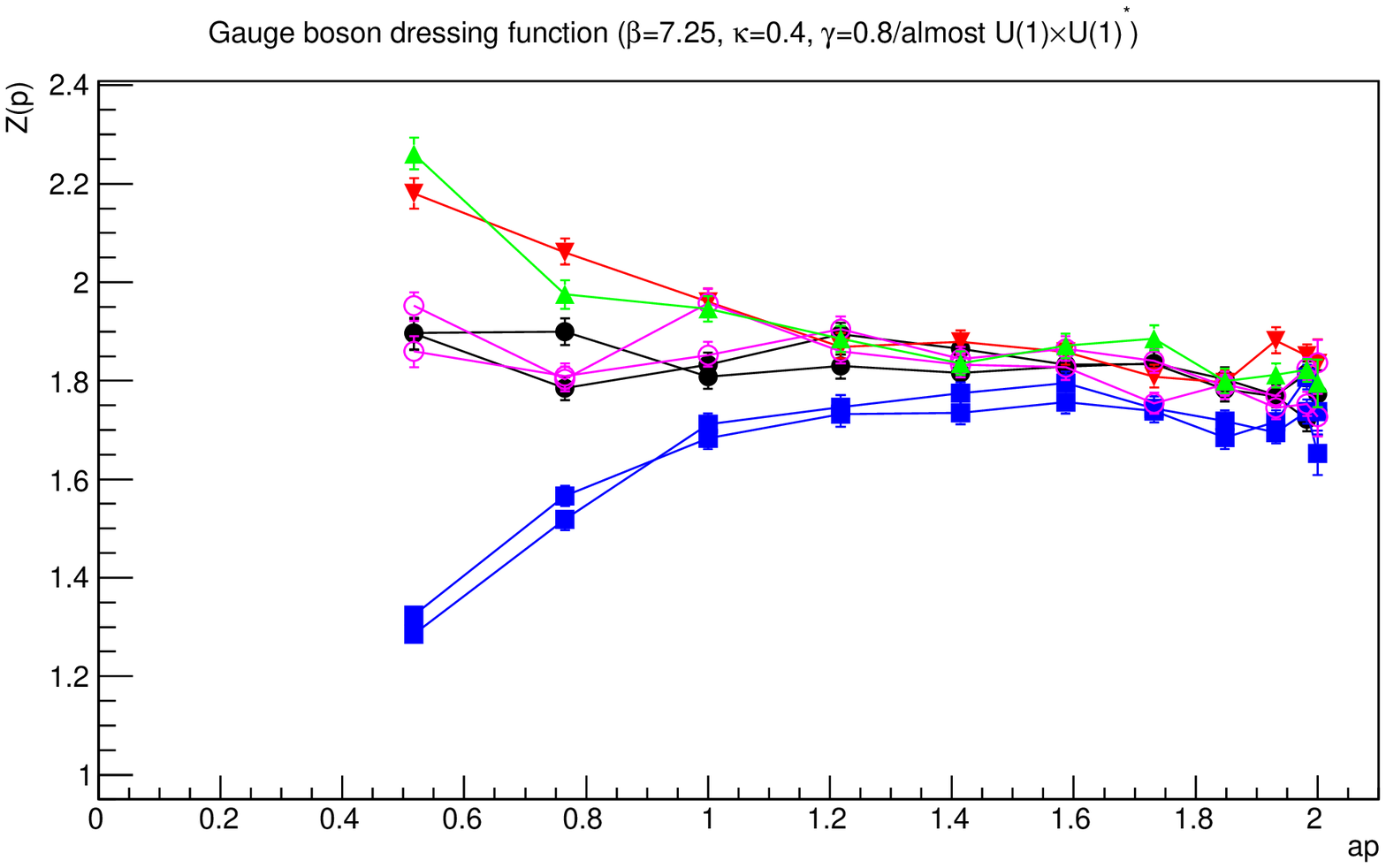} &
    \includegraphics[width=0.441\linewidth]{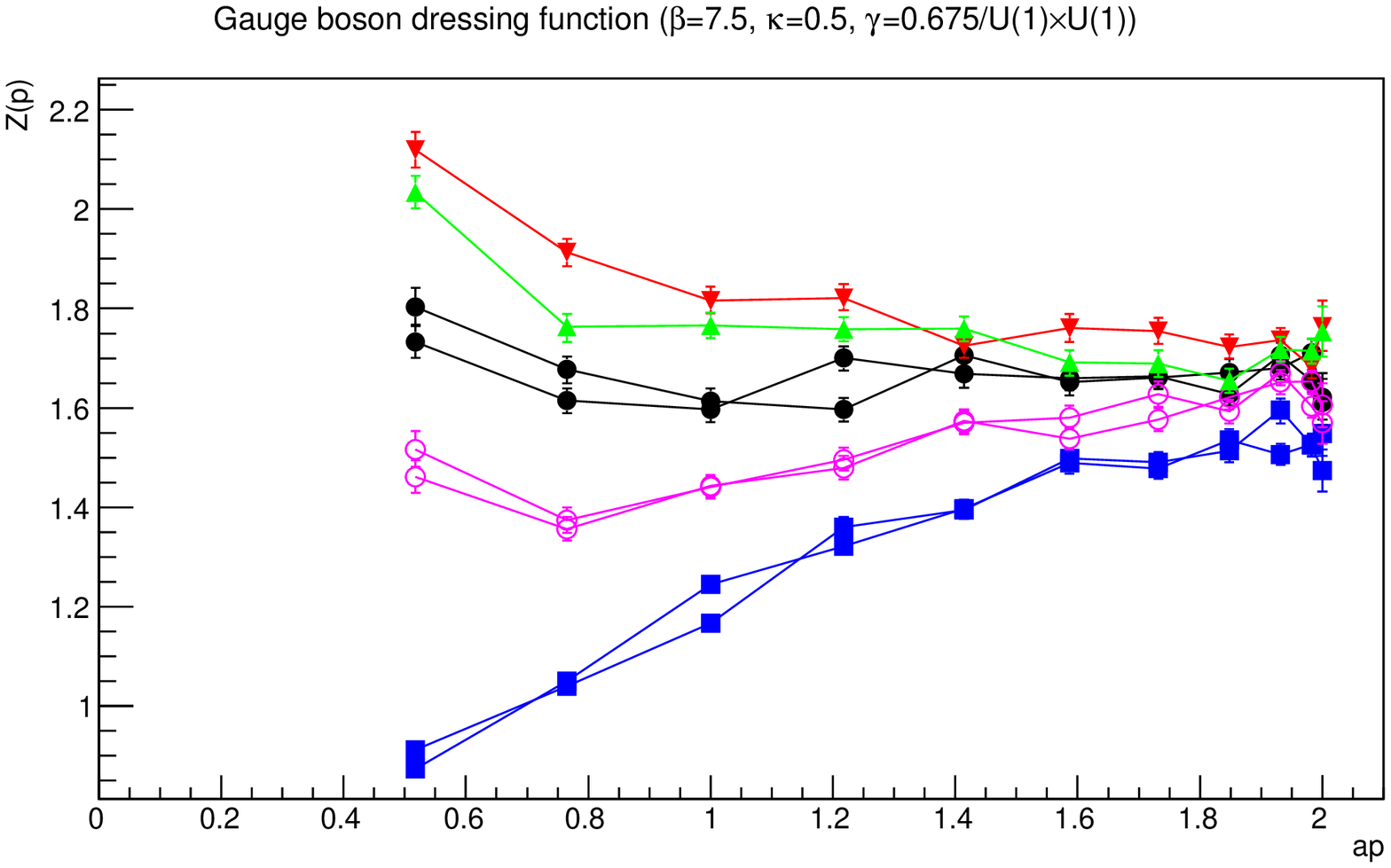}    \\
    \includegraphics[width=0.441\textwidth]{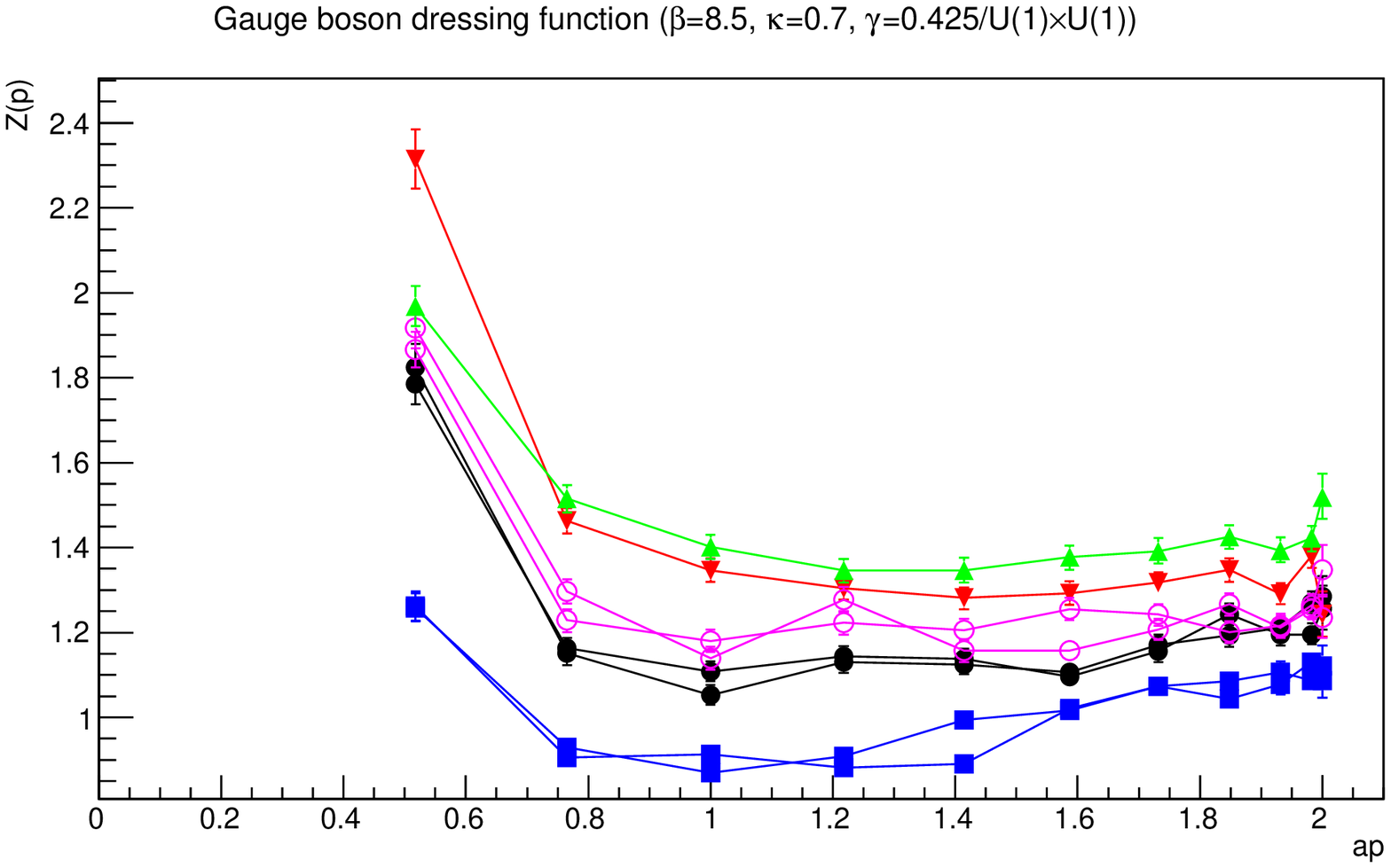}  &
    \includegraphics[width=0.441\textwidth]{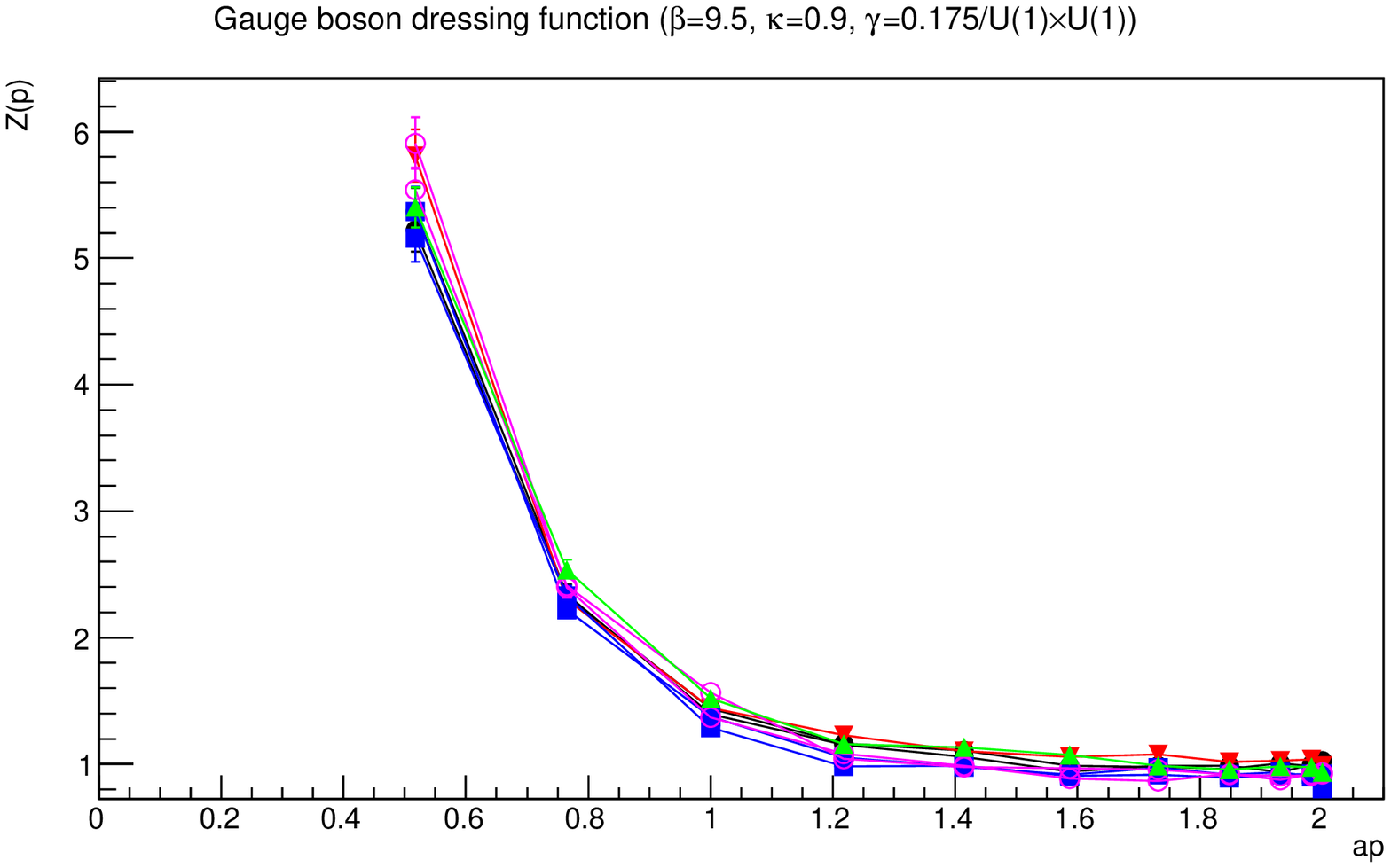}  & \\
    \includegraphics[width=0.441\textwidth]{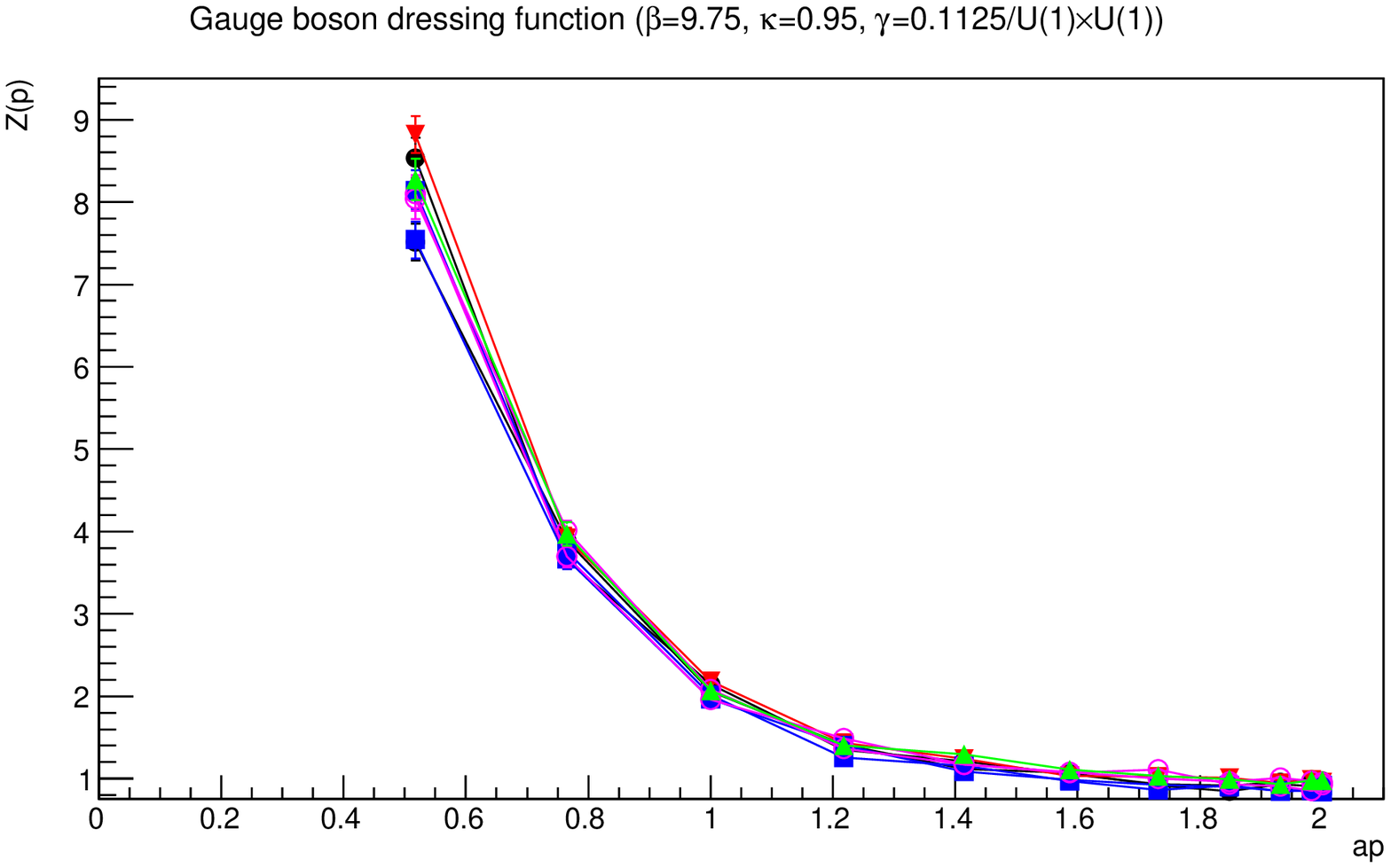} &
    \includegraphics[width=0.441\textwidth]{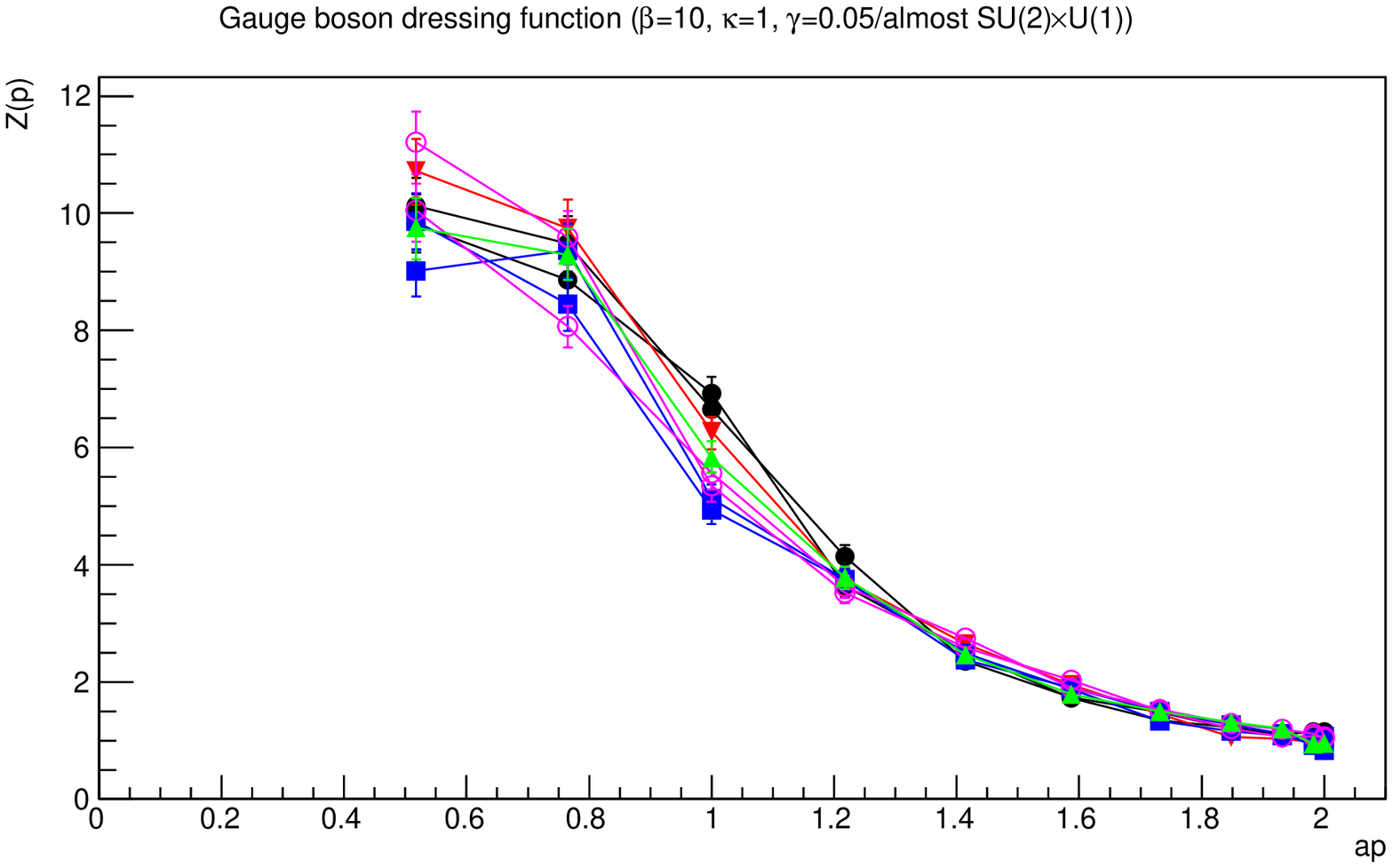}
  \end{tabular}\\
  \includegraphics[width=\textwidth]{{leg-gb}.eps}
  \caption{The gauge boson dressing function for the eight different charges moving from deep inside the unbroken phase ($\beta=6.5$, top right) to close to both sides of the \ZZ[2] phase transition ($\beta=7$ and $\beta=7.25$) through the broken phase ($\beta=7.5$, $\beta=8.5$, $\beta=9.5$, $\beta=9.75$) to the smallest numerically accessible value of $\gamma$ ($\beta=10$) in the lower-right panel. All results from the $24^4$ lattice, the momenta are along the $x$-axis, which minimizes finite-volume effects, and the zero momentum point and the first non-zero momentum point are suppressed due to remaining finite-volume effects.}
  \label{fig:gp}
\end{figure*}

Thus, what we see is a highly non-trivial averaging pattern with large cancellations, both within individual configurations and also between configurations. Nonetheless, in the end the full path integral determines the dynamics. This is best encapsulated by the quantum effective potential, displayed in figure \ref{fig:qep}. It does indeed paint a picture of such strong cancellations. In all cases it is strongly deformed from the tree-level behaviour towards a single minimum, except deep in the \ZZ[2]-unbroken phase, where the size of the induced classical fields can be considered as lattice artefacts. Thus, there are strong quantum corrections to the potential. Especially, in the cases with a $\U[1]\times\U[1]$ breaking we do not see remainder minima-like structures emerging from the strong peaking of the individual configurations at $\theta_0=\pm\pi/6$. Moreover, the classical field values saturate as a function of the sources in the \ZZ[2]-broken phase within the accessible range. This can be interpreted as the system being very stable against any attempt to perturb it out of its vacuum state. Furthermore, we observe that the minima are located at relatively small values of $\sigma_8$, while $\sigma_3$ is large in comparison, except when $\gamma$ is small. The field $\sigma_3$ is associated with the \SU[2] subgroup of \SU[3] in the Gell-Mann representation we use. If $\sigma_8$ vanishes, the system has necessarily the $\U[1]\times\U[1]^\star$ structure. The increasing value of $\sigma_8$, which would dominate in the $\SU[2]\times\U[1]$ pattern, thus corroborates our interpretation that the system moves from the $\U[1]\times\U[1]^\star$ pattern at the phase boundary to the $\SU[2]\times\U[1]$ pattern towards $\gamma\to 0$.

Hence, we could expect that the correlation functions, being differentials of the quantum-effective potential, should also markedly obey just a single pattern. That would indeed be a favourable outcome \cite{Maas:2017xzh}. The results for the running coupling (\ref{rcoupling}) and the gauge boson propagator are shown in figure \ref{fig:alpha} and \ref{fig:gp}, respectively. We note that also the gauge-dependent correlation functions are strongly affected by the critical slowing down at small $\gamma$. More details can be found in appendix \ref{a:crit}. In addition, large finite-volume effects are known to affect zero momentum and the smallest non-zero lattice momenta \cite{Maas:2011se,Afferrante:2020hqe}, and thus these are suppressed. Unfortunately, this is also the range in which the BEH effect is strongest. Finally, the ghost propagator is much more insensitive to the BEH effect, as has also been observed previously \cite{Maas:2018xxu,Afferrante:2020hqe}, and the corresponding results are therefore relegated to appendix \ref{a:ghost}.

Without an independent scale, it is not entirely trivial to interpret the results. However, by comparing to previous results \cite{Maas:2011se,Maas:2014pba,Maas:2018xxu,Afferrante:2020hqe}, it is possible to quantify the observations. Deep in the bulk phase ($\beta=6.5$), the results are typical for strongly-interacting gauge theories \cite{Maas:2011se}, and show full degeneracy, as expected. The behaviour of both the propagator and the running coupling are characteristic for a lattice spacing of the order of the characteristic scale of the theory \cite{Maas:2011se}. This remains essentially the same when moving towards the phase boundary ($\beta=7$), just with some increase of fluctuations.

After crossing the boundaries, the results indicate a markedly coarser lattice. Thus, the drop in the running coupling is therefore indeed indicative of a weakly coupled system. Moreover, the degeneracy pattern with four ($\beta=7.25$) degenerate gauge bosons, and two being markedly heavier, is precisely the expected pattern for $\U[1]\times\U[1]^\star$. While the massless modes are found at similar values as the lightest massive ones, this is likely a finite volume artefact, as has been observed previously \cite{Maas:2018xxu,Afferrante:2020hqe}. When moving further into the bulk, the lattice spacing ($\beta=7.5$, $\beta=8.5$) appears not to change drastically. However, the remaining degeneracy splits up, leaving 3 pairs of degenerate gauge bosons, being the expected pattern for $\U[1]\times\U[1]$.

However, when approaching $\gamma\to 0$ ($\beta=9.5$, $\beta=9.75$, $\beta=10$), the gauge bosons start to become degenerate again. This may be a critical slowing down effect or pure volume effect, see appendix \ref{a:crit}, though at this time this cannot be decided. At the same time the running coupling starts to increase, though still staying below one. While it appears tempting to associate this with having an unbroken non-Abelian subgroup, this cannot be alone the reason. In the case of the fundamental Higgs, there remains also an unbroken non-Abelian subgroup, but the breaking pattern remains tree-level like \cite{Maas:2016ngo,Maas:2018xxu}. Especially, the tree-level masses (\ref{eqn:boson_masses}) can never degenerate. Hence, either, despite appearances particularly at $\beta=10$, the volume is still too small and thermalization issues too strong to probe the non-degenerate regime, or the interactions are indeed strong enough to modify the behaviour substantially compared to tree-level. It would require, at the very least, an order of magnitude more computational resources to decide this.

\subsection{Implications for analytical and continuum calculations}

The, probably most important, insight from the results is that the pattern of symmetry breaking cannot be chosen at will, but only be determined at the quantum level. In this context it is important that while the vev length $w$ can be adjusted by the renormalization prescription, the angle $\theta_0$ cannot be changed by renormalization, as the possible wave-function renormalization affects only all components equally. Thus, its determination is necessarily required to be done always from the minimization of the quantum effective potential. In any expansion, this would then be required at every order, to realign the vev.

However, our investigations also suggest that the breaking angle obtained from implicit gauge-fixing may already be sufficient to get a hint of the possible breaking patterns for a given parameter set. This is an advantage for methods in which gauge fixing is expensive, like lattice simulations.

The next issue concerns the actual continuum physics. Just like any other stand-alone gauge-Higgs theory, the present one is potentially affected by a triviality issue \cite{Callaway:1988ya}. There are now two alternatives. Either, non-perturbatively there exists an interacting continuum limit at the second-order phase transition surface, or it does not. In the latter case, it is always possible to use the theory as a low-energy effective theory only \cite{Hasenfratz:1986za}. In that case, any point in the phase diagram can be used, for sufficiently low energies below the corresponding cut-off, and thus all patterns can be realized. Otherwise, a phase transition only occurs towards the boundary of the phases. Given our results, this would suggest that there is in the direction of the $\ZZ[2]$-unbroken phase only the $\U[1]\times\U[1]^\star$ pattern possible. In the $\gamma\to 0$ limit, if also manifesting a second order phase transition, apparently the $\SU[2]\times\U[1]$ pattern is possible.

Thus, for the purpose of gauge-fixed calculations beyond the lattice, being that augmented perturbation theory using the FMS mechanism, functional methods or other approaches, this implies that there are apparently none of the serious ambiguities possible considered in \cite{Maas:2017xzh}. However, it is mandatory to determine the actual breaking pattern by determining the actual value of $\theta_0$ from the quantum effective potential self-consistently.

\section{Summary}\label{s:summary}

We have presented a detailed, non-perturbative investigation of the BEH effect in a theory, in which multiple breaking patterns are possible. Our results suggest that, even if at tree-level no pattern is preferred, at the quantum level only one pattern is possible for any given set of couplings. Especially, the vev cannot be chosen by a gauge condition at will, like this is the case when one breaking pattern exists only, but needs to be determined a-posteriori from the quantum effective potential. This implies that gauge-fixing in analytical (continuum) calculation requires a simultaneous calculation and minimization of the quantum effective potential. We also show that some gauges exist, which enforce a BEH effect for any parameters, very much like gauges which forbid a BEH effect for any value of the parameters. This again demonstrates the gauge-dependence of the BEH effect, but at the same time shows it can be used nonetheless as a fruitful concept.

While this complicates issues in practice, this removes any lingering ambiguities \cite{Maas:2017xzh} for gauge-fixed calculations. Especially, this will allow manifest gauge-invariant analytical calculations using FMS-mechanism augmented perturbation theory \cite{Maas:2017xzh,Sondenheimer:2019idq,Maas:2020kda,Dudal:2020uwb} in realistic theories like GUT candidates, which are not accessible for computational costs on the lattice. Also, corresponding predictions for the spectrum of theories like the present one \cite{Maas:2017xzh,Sondenheimer:2019idq} become unambiguous, and can be tested using lattice simulations in the future.

Finally, these results imply that phenomenology in theories with multiple BEH breaking patterns cannot be done by just selecting a theory which has potentially the correct breaking pattern. It is really necessary to check whether the desired breaking pattern is indeed admitted by the quantum effective potential for the couplings necessary to fulfil phenomenological constraints, like mass spectra. Conversely, this puts such model building from a field-theoretical perspective on much more stable ground.

\acknowledgments

The computational results presented have been obtained using the Vienna Scientific Cluster (VSC) and the HPC center at the University of Graz. E.\ D.\ and B.\ R.\ have been supported by the Austrian Science Fund FWF, grant P32760. We are grateful to Vincenzo Afferrante, René Sondenheimer, and Pascal Törek for useful discussions. B.\ R.\ is specifically thankful to Fabian Zierler for sharing his autocorrelation analysis codes.

\appendix

\section{Generalization to \texorpdfstring{$N>3$}{N>3}}\label{app:sun}

The approach as outlined for $\SU[3]$ can in principle be generalized to arbitrary $\SU[N]$. As before, gauge fixing corresponds to diagonalization of the vev. In the adjoint $\SU[N]$ theory, diagonalizing in analogy to \cref{eqn:adj_breaking_angle} restricts the $\su(N)$ scalar field to the moduli space $S^{N-2}\times\RR_{\geq 0}$, parameterized by the absolute value $w$ of the vev, together with a set of $(N-2)$ spherical coordinates, corresponding to rotations in the Cartan subalgebra.

Assuming $w=1$, the $\su(N)$ matrix invariants of the scalar field define a set of $(N-2)$ real-valued functions over the unit $(N-2)$-sphere, in 1:1 correspondence with the eigenvalues. Each function (if the corresponding matrix invariant is non-zero) can be thought of as a $(N-2)$-dimensional surface, where the intersection points of all such surfaces with the unit sphere correspond to the solutions for the diagonalization of a unit $\su(N)$ matrix with different permutations of eigenvalues. Thus, in general we expect $N!$ disjoint intersections, but the intersections become degenerate for special points due to enhanced symmetries.

There are then two main questions to consider when gauge-fixing an adjoint $\SU[N]$ theory.

The first is the problem of diagonalization: what is a 1:1 map from the space of allowed values of the $\su(N)$ matrix invariants, to an (arbitrary) sector of the $(N-2)$-sphere which tiles the full sphere under the action of the $S_N$ group?
In practice, an explicit form for the eigenvalues in terms of matrix invariants becomes challenging for $N>4$, so for the purposes of lattice gauge-fixing, numerical diagonalization will need to suffice.

However, the second question is more important for analytical studies: how can the choice of `sector' on the sphere (i.e. the choice of eigenvalue ordering) affect the gauge-fixing procedure? This latter question is necessary for understanding the quantum effective potential, and is discussed in \cref{app:ssec:possible-breaking-patterns}.

It will be useful to work in a fixed coordinate system in the following, so we parameterize the diagonalized vev $\Sigma_0\in\su(N)$ via:
\begin{equation}
  \Sigma_0
  = \sum_{j=2}^{N} \sigma_{j-1}(\boldsymbol{\theta}_0) T_{j^2-1}\,,
  \label{eqn:app:defineDiagMatrix}
\end{equation}
with $\{T_a\}$ the Lie algebra of $\su(N)$ and $\boldsymbol{\theta}_0$ an angle on the $(N-2)$-sphere.
Here the coefficients $\{\sigma\}$ can be parameterized as $\sigma_i = w x_i(\{\theta_0\})$, with the coordinates $x_i(\boldsymbol{\theta})$  defined (for $N>3$) as
\begin{equation}
  \begin{aligned}
    x_{1}(\boldsymbol{\theta})   & = \cos(\theta^{(1)})\dots\cos(\theta^{(N-2)})
    \\
    x_{2}(\boldsymbol{\theta})   & = \sin(\theta^{(1)})\cos(\theta^{(2)})\dots\cos(\theta^{(N-2)})
    \\
                                 & \vdots
    \\
    x_{N-2}(\boldsymbol{\theta}) & = \sin(\theta^{(N-1)})\cos(\theta^{(N-2)})
    \\
    x_{N-1}(\boldsymbol{\theta}) & = \sin(\theta^{(N-2)})\,.
  \end{aligned}
  \label{eqn:app:polarcoords}
\end{equation}
Note that this definition deviates from the usual one for hyperspherical coordinates, but is chosen such that the $N=3$ case matches with the definition in \cref{eqn:adj_breaking_angle}. Additionally, it will turn out conveniently to have the last component depending on only one parameter in the following. To recover usual spherical coordinates one needs to replace all angles $i>1$ by $\theta^{(i)}\rightarrow -\theta^{(i)}+\pi/2$ and reverse the order of the components $x_i$.

\subsection{Possible breaking patterns for \texorpdfstring{$\SU[N]$}{SU(N)}}\label{app:ssec:possible-breaking-patterns}

We start again here from the \ZZ[2]-symmetric potential given in \cref{eqn:adj_potential} but now considering a generic scalar field in the adjoint representation of \SU[N>3]. In principle for theories with $N>3$ it would be possible to add another renormalizable term given by $\tr\qty[\Sigma^4]$ which still allows for a BEH effect to take place. However, for now we will neglect this term and thus restrict the potential to terms that have a simple correspondence between the matrix- and the vector-representation. The implications of the missing term will be discussed later.

First we need to revise the possible breaking patterns of the general \SU[N] theory. To obtain the possible patterns at tree-level we follow the steps as outlined in \cref{s:patterns}.

By inserting \cref{eqn:app:defineDiagMatrix} into the potential (\ref{eqn:adj_potential}) we see immediately that $V$ is completely independent of all angles for any $N$. Although this suggests that any angle minimizes the potential, it has been shown \cite{Ruegg:1980gf,Murphy:1983rf,O'Raifeartaigh:1986vq} that the potential has only a minimum for maximally two different non-zero eigenvalues while the remaining angles are actually saddle points. This differs from the present case in which an enhanced \O[8] symmetry of the potential allows also for three distinct eigenvalues to obtain a minimum \cite{Maas:2017xzh}. Beyond perturbation theory this restriction may however be lifted again, and an arbitrary pattern could minimize the full quantum effective action.

Nevertheless, for now we will keep the restriction to two eigenvalues. This therefore leaves us with $\floor*{\frac{N}{2}}$ possible breaking patterns of the form
\begin{align}
   & \text{S}\qty(\U[N-P]\times\U[P]) \sim \SU[N-P]\times\SU[P]\times\U[1]\nonumber \\
   & \qqtext{with} 1\le P<\frac{N}{2}\,,
\end{align}
where additional $1/\ZZ[N]$ restrictions have been dropped, as they do not play any role in the following. The corresponding normalized eigenvalues for the $\SU[M]$-subgroup are given by
\begin{equation}
  \Sigma_0^{\qty(M)} = \pm w\sqrt{\frac{1}{2}}\sqrt{\frac{1}{M}-\frac{1}{N}}
\end{equation}
with the sign of the eigenvalue such that the full $\Sigma_0$ is traceless.

The pattern with the largest remaining symmetry is always $\SU[N-1]\times\U[1]$ and corresponds to $\Sigma_0 \propto T^{N^2-1}$. Using our definition of the hyperspherical coordinates in \cref{eqn:app:polarcoords} it therefore follows that the corresponding angle is always given by \begin{equation}\label{eqn:app:sun-1_angle}
  \theta^{\qty(N-2)}=\frac{\qty(2n+1)\pi}{2}=\arcsin((-1)^{n})\,,
\end{equation}
while all other angles can be chosen arbitrarily.

To find the corresponding angles for the other breaking patterns characterized by $P$ we can always start with the last angle $\theta^{\qty(N-2)}$ by solving the equation
\begin{equation}
  \sin(\theta^{(N-2)})=\frac{\Sigma_0^{\qty(P)}}{\Sigma_0^{\qty(1)}}.
\end{equation}
Plugging this back into the vev one finds iteratively that for a specific breaking pattern $P$ it is necessary to fix exactly the last $P$ angles. These are then given by
\begin{equation}\label{eqn:app:sun_angles}
  \theta^{(N-2-j)}=\arcsin((-1)^{n}\sqrt{\frac{N-P}{\qty(P-j)\qty(N-1-j)}})
\end{equation}
for $0\le j < P$, which reduces for $P=1$ to \cref{eqn:app:sun-1_angle}. The factor $n\in\mathbb{N}$ defines the overall sign of the vev and is chosen such that for $n=0$ or $n$ even the first eigenvalues are positive.

In principle, it is possible to extend this construction to the other breaking patterns that are not minima of the tree-level potential. This can be done iteratively by realizing that for a \SU[N] theory the breaking patterns of the smaller groups are also included with 0-padded eigenvalues. The patterns of the group \SU[n] with $n<N$ can always be located at the angles $\theta^{(N-2)}=\dots=\theta^{(n-1)}=0$ and the remaining angles as calculated for the \SU[n] group. To know at least one angle for all possible breaking patterns of the \SU[N] group needs the knowledge of all angles for the \SU[N-1] group.

Also, it should be kept in mind that there are again multiple combinations of angles corresponding to the same breaking patterns due to the possible permutations of the eigenvalues. This is discussed in the last section of this appendix but can be ignored for now.

\subsection{Possible phases and a general potential}

We are now in the position to try and expand our results for the \SU[3] case to arbitrary $N$. The first thing to be noticed is that for odd $N>3$ there exists a major difference to our system. In our case it was possible, due to the enhanced symmetry of the potential, that there exists one breaking pattern that still preserves the global \ZZ[2]-symmetry in a diagonal fashion, namely the special $\U[1]\times\U[1]^\star$. This is not possible for odd $N$ due to the restriction of needing exactly 2 distinct eigenvalues. Hence, the eigenvalues cannot be equally distributed anymore while still preserving the traceless condition. Therefore, the \ZZ[2]-order parameter from \cref{eqn:z2order} will never vanish for a gauge-fixed vev and thus BEH-breaking can only happen perturbatively when \ZZ[2] is broken.

For even $N>3$ however there still exists the pattern $\SU[N/2]\times\SU[N/2]\times\U[1]$ which has exactly $N/2$ times the eigenvalues $\pm\frac{1}{\sqrt{2N}}$. Here again a simple permutation of the eigenvalues can be used to perform a \ZZ[2]-transformation and therefore being a subgroup of the gauge-group. It is thus the equivalent of the $\U[1]\times\U[1]^\star$ pattern in the \SU[3] theory and in this case the \ZZ[2]-order parameter can vanish.

Additionally, in the non-perturbative case different combinations of $\SU[n]\times\SU[m]\times\dots$ with $n \text{ and } m$ even, can lead to the same behaviour. These patterns, with an additional suitable \U[1], can also show up for odd $N>3$ with zero-padded or non-zero eigenvalues. All these patterns would classify as `special' patterns, due to the fact of \ZZ[2]-breaking being possible.

Assuming that our results are generalizable, and that for these theories also distinct \ZZ[2] phases do exist, this would predict that for all $N$ the same complications as discussed in the main text can arise. In particular, it would be possible to find a gauge, i.e.\ unitary gauge with a-priori strict ordering, that can be used to obtain BEH-like physics with the $\SU[N/2]\times\SU[N/2]\times\U[1]$ (sub-)pattern in the \ZZ[2]-unbroken phase and close to the phase transition. For the odd cases the only prediction that can be made is that \ZZ[2] breaking also separates regions where a BEH-effect may be possible and those where it is not possible at all, except for these special patterns.

Another point that should be mentioned is that in theories with larger gauge groups there is perturbatively no continuous angle connecting the different breaking patterns, because only certain angle combinations minimize the potential at tree-level. This has to be taken into account when averaging over configurations. While in the present \SU[3] case the distributions of (local or global) angles still behaved as Gaussian in some sense (see \cref{fig:theta}), this is not the case anymore for more than one angle. This implies that averaging can produce either specific breaking patterns in distinct parts of the phase diagram if one direction is indeed singled out, or may average over multiple breaking patterns for individual configurations. The second option would be fairly dramatic for perturbation theory since this would mean that the average field is not at a minimum of the potential, but instead located at an arbitrary combination of angles. Therefore, it is extremely important to monitor the histograms of angles when going to higher $N$. Although, being a problem for perturbation theory, the scenario would still be similar to what we found for the $\SU[3]$ case, where independent configurations of an (almost) $\SU[2]\times\U[1]$ type averaged to a generic $\U[1]\times\U[1]$ pattern. The difference is though that in the present case this is still a minimum of the potential. Nevertheless, in the end the dynamics are really governed again by the (minimum of) the quantum effective potential, which may lie at an arbitrary point on the $S^{N-2}$-sphere.

Although all of this seems quite discouraging for usual phenomenology there may be a way out. So far we have only discussed the properties of a very simple tree-level potential. However, it is already known in the literature \cite{Ruegg:1980gf,Murphy:1983rf} that by explicitly adding a $\tr[\Sigma^3]$ or a $\tr[\Sigma^4]$ to the potential, the minimum structure is drastically changed and therefore certain patterns are singled out; see \cite[Table 1]{Murphy:1983rf} for a comprehensive list. Also since $\tr[\Sigma^4]$ is not protected by any symmetry the term is expected to be generated \cite{Maas:2017xzh} even without explicitly adding it. To generalize this to our results we should remember that we found in the region of $\gamma\to 0$ hints that the realized breaking pattern indeed becomes of the form $\SU[2]\times\U[1]$ independent of the chosen gauge. This is however the same region where \ZZ[2]-breaking becomes strong in the sense of the order parameter rising strongly. Also, we saw that in the \ZZ[2]-unbroken region and close to the phase transition the $\U[1]\times\U[1]^\star$ pattern became favoured. In fact using the implicitly gauge-fixed breaking angle, as discussed in \cref{sec:results_implicit} turned out to be a good precursor to identify possible breaking patterns, if a BEH effect is allowed by the chosen gauge. That this happens may be expected since these are gauge-invariant quantities which are the ones really governing the dynamics, and therefore will also have an influence on the quantum effective potential.

For larger $\SU[N]$ gauge groups there do exist additional matrix invariants built up from $\tr\qty[\Sigma^p]$ with $p\le N$ and $\det\qty[\Sigma]$. These quantities are again connected to the breaking-angles in \cref{eqn:app:sun_angles}, however the corresponding relations between invariants and angles become quite involved and analytically not explicitly solvable. Nevertheless, it seems that already the two invariants $\det\qty[\Sigma]$ and $\tr\qty[\Sigma^3]$ may be sufficient to restrict the possible patterns for a given $\Sigma$ substantially. This is because \SU[4] is fully specified by these two invariants, as will be seen in \cref{app:ssec:su4}, and again an iterative procedure results in information for the \SU[N] groups. In addition, all patterns with non-zero eigenvalues can be identified uniquely using the matrix invariant $\tr\qty[\Sigma^N]$ and the corresponding powers of the involved subgroups.

So while also for larger $N$ it has to be expected that choosing a specific breaking pattern a-priori is also not possible, it still may be possible to at least restrict the possible patterns for certain parameter sets to some extent. The two possibilities that seem to be realizable are modifying the potential or checking the possible angles for certain parameter sets. The first option can be done by adding some terms explicitly to the tree-level potential\footnote{Gauge conditions involving scalars in different e.g.\ non-linear ways may also be an option \cite{Dolan:1974gu}, but this is essentially unexplored.} and thus reducing the set of possible patterns. However, beyond \SU[4] the potential cannot keep increasing in complexity, as matrix invariants $\sim \tr\qty[\Sigma^n] \text{ for } n>4$ are non-renormalizable. The other option is to simply calculate/measure the invariant quantities and obtain the corresponding angles. Since this is a general extension of the implicit gauge-fixing condition it may therefore be used to identify parts in the phase-diagram that produce a quantum effective potential close to the breaking pattern(s) in question. Nevertheless, one should always keep in mind that the quantum effective potential is not limited to the tree-level behaviour and therefore the absolute minimum may be at an arbitrary combination of angles and therefore simply be of a $\U[1]^{N-1}$ type. Thus, a careful analysis of the quantum effective potential cannot be avoided.

\subsection{Dealing with breaking pattern degeneracies}

On top of the previous discussion the larger gauge groups also suffer again from degeneracies in terms of breaking angles due to possible permutations of the eigenvalues. A closed form expression for the angle degeneracies is not yet known, but may be deduced for specific breaking patterns from \cref{eqn:app:sun_angles}. Since this turns out to be very cumbersome we will now shortly sketch also another way to identify/deal with these degeneracies.

\subsubsection{The sectors for \texorpdfstring{$\SU[4]$}{SU(4)}}\label{app:ssec:su4}

\begin{figure}[ht!]
  \centering
  \includegraphics[width=0.49\linewidth]{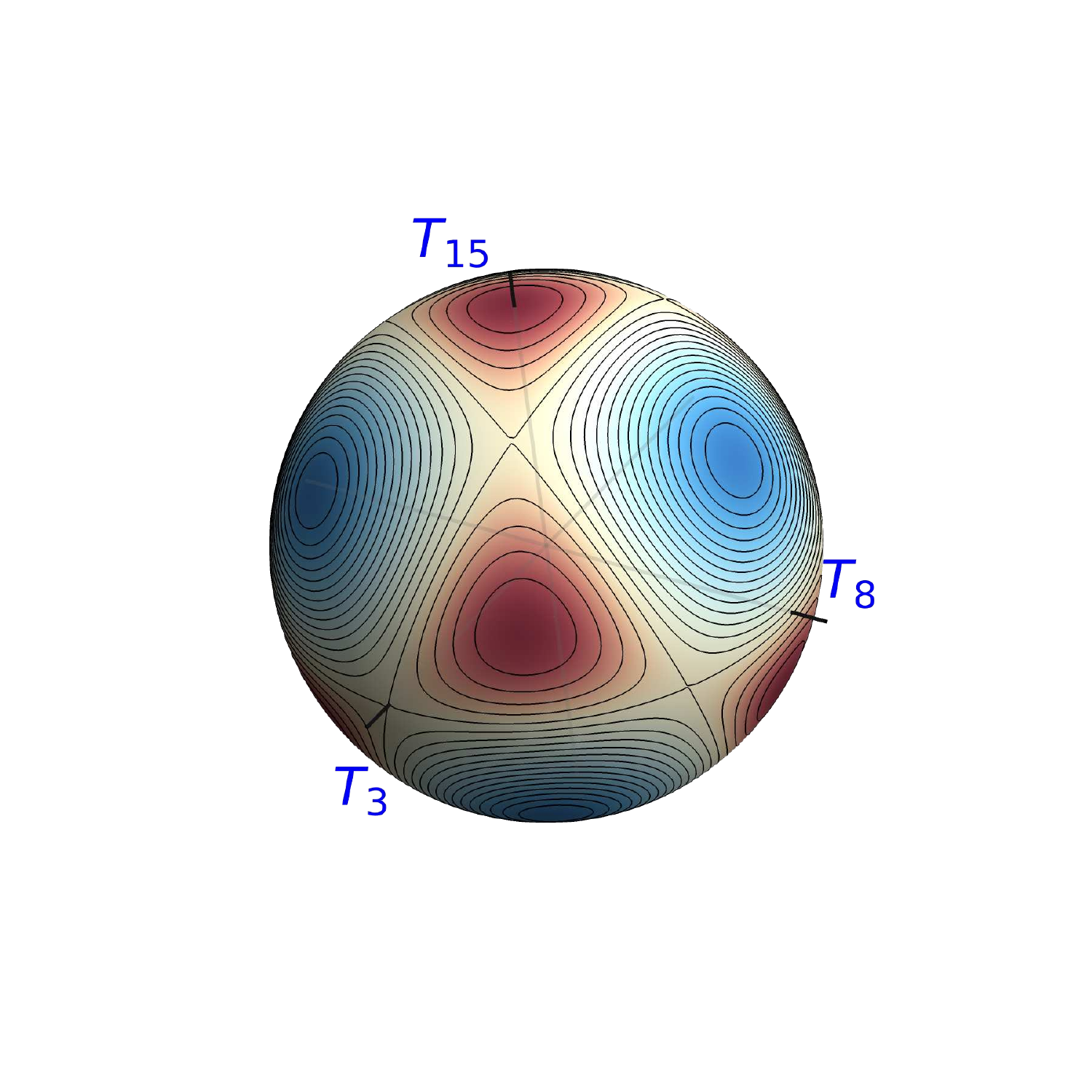}\hfill
  \includegraphics[width=0.49\linewidth]{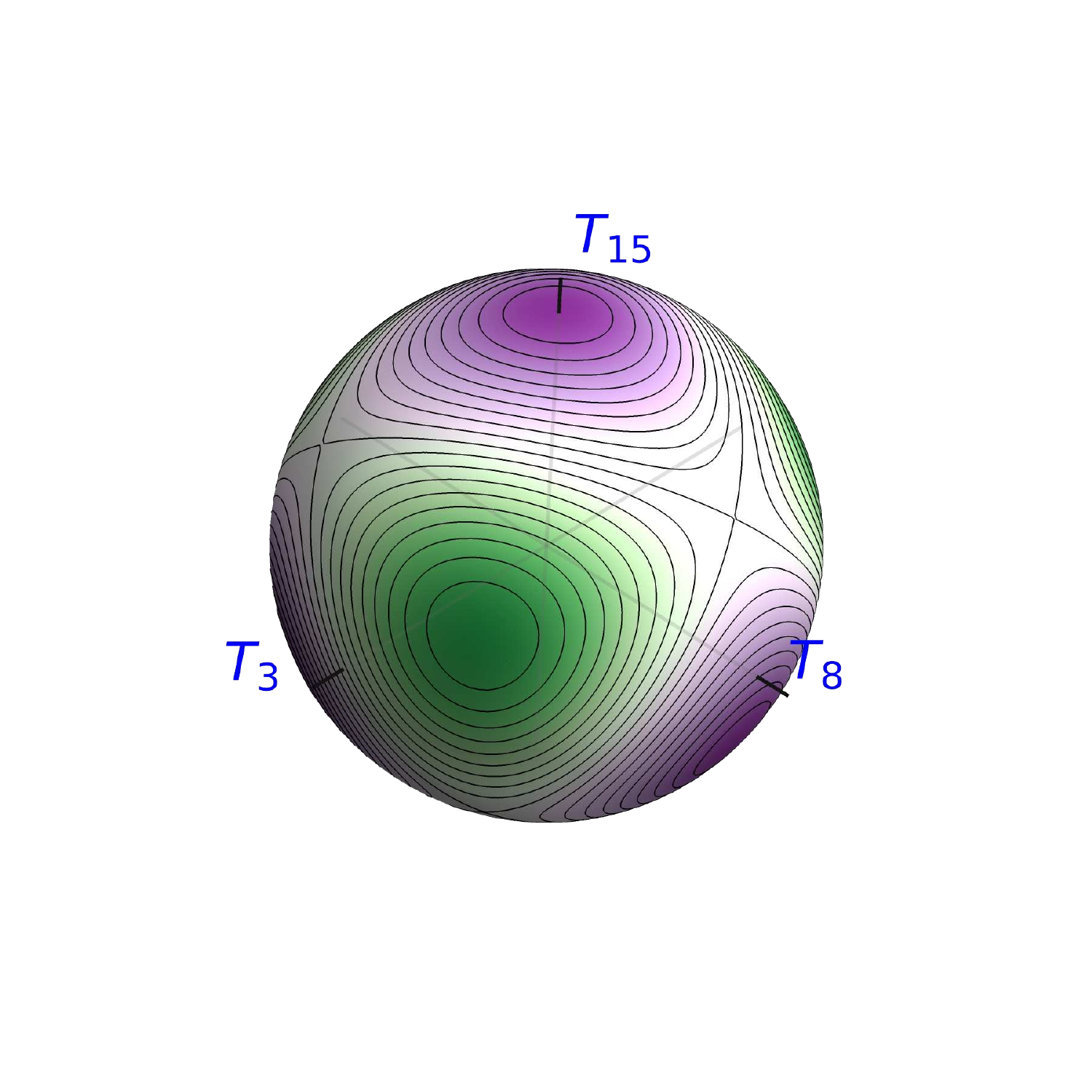}
  \\
  \includegraphics[width=0.35\linewidth]{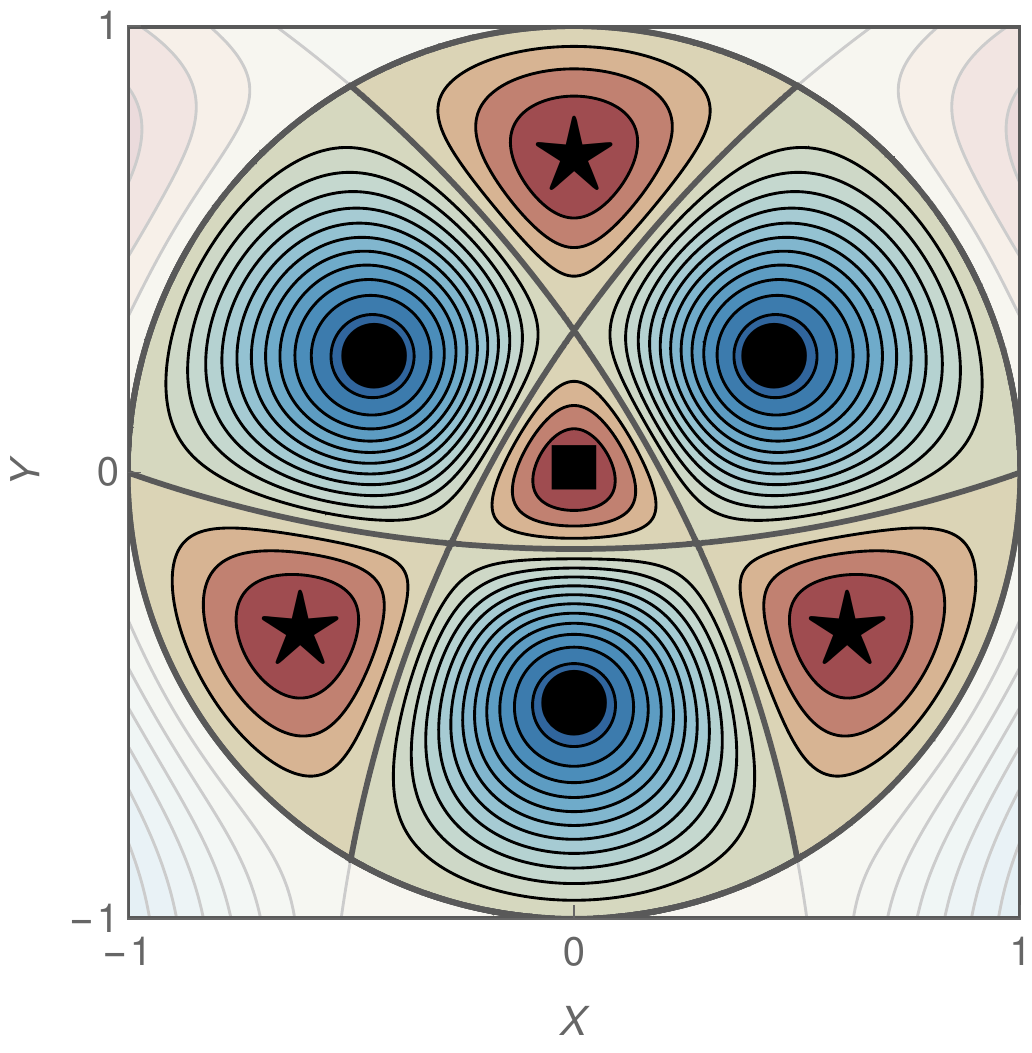}
  \hspace*{0.15\linewidth}
  \includegraphics[width=0.35\linewidth]{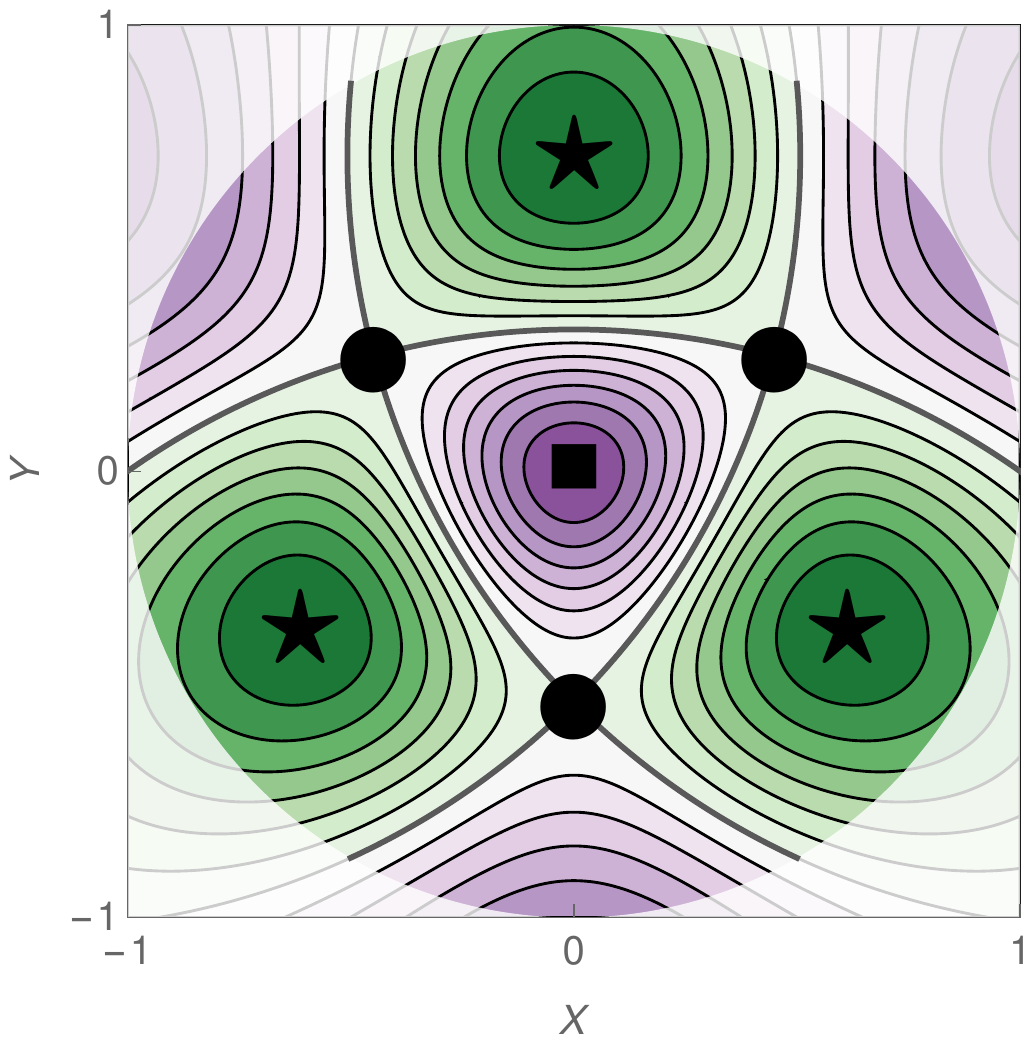}
  \caption{Plots of $\alpha$ (maxima in blue, minima in red) and $\beta$ (maxima in purple, minima in green) over the sphere together with the stereographic projection of the lower hemisphere. Extrema of the determinant $\alpha$ are highlighted: the square and star-shaped points also correspond to the extrema of the trace-cubed and are essentially the same as the special points for the $\SU[3]$ case. The round points, maximizing the determinant, are only relevant for $N\geq 4$. Any choice of three adjacent points with each of the symbols ({\small$\blacksquare$}, $\bigstar$, {\large$\bullet$}) defines a fundamental region on the sphere, which tiles the full sphere under the symmetries generated by permuting eigenvalues (i.e. the cubic group).
  }
  \label{fig:spherecontours}
\end{figure}

We describe here the situation for $\SU[4]$, as it is the simplest non-trivial example after $\SU[3]$. Working in standard spherical coordinates $(\theta,\phi)$ we can parameterize a diagonal matrix with unit norm, $M(\theta,\phi)\in\su(4)$, by $M(\theta,\phi) = \cos(\theta)T_{15}+\sin(\theta)\cos(\phi)T_3 + \sin(\theta)\sin(\phi) T_8$.
The matrix invariants are given in these coordinates by
\begin{multline}
  \alpha(\theta,\phi) = \det\qty[ M(\theta,\phi) ] \\
  = -\frac{1}{192}\Big[1+\sin(\theta)^2(\sqrt{8}\sin(2\theta)\sin(3\phi)+7\sin(\theta)^2-8)\Big]
\end{multline}
\begin{multline}
  \beta(\theta,\phi)= -\frac{1}{3}\tr\qty[M(\theta,\phi)^3] \\
  =\frac{1}{48\sqrt{6}}(3\cos(\theta)+5\cos(3\theta)-4\sqrt{2}\sin(\theta)^3\sin(3\phi))\,,
\end{multline}
which are plotted over the sphere in \cref{fig:spherecontours}.

\begin{figure}[t!]
  \centering
  \includegraphics[width=.75\linewidth]{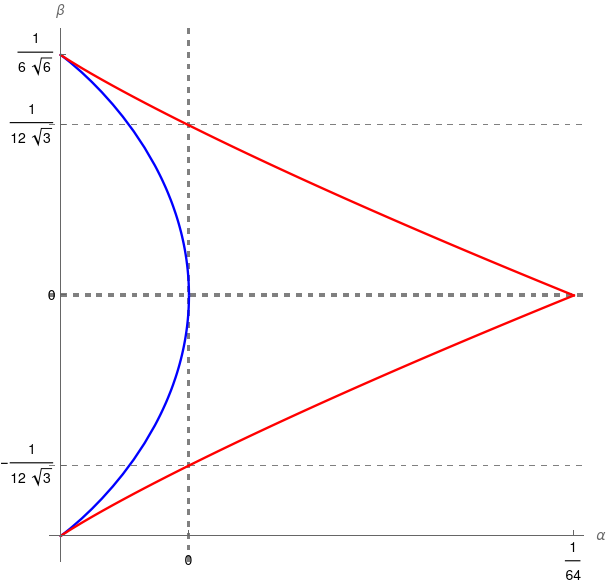}
  \caption{Possible values of matrix invariants $\alpha=\det\qty[\Sigma/w]$ and $\beta=-\tfrac{1}{3}\tr[(\Sigma/w)^3]$ for $\Sigma\in\su(4)$. Geometrically, the edges correspond to regions of enhanced symmetry, where the `surfaces' corresponding to $\alpha$ and $\beta$ merely touch once on the unit sphere, rather than intersecting twice, leading to 12 degenerate points on the sphere each with multiplicity 2. Algebraically, the bounding lines can be found by setting discriminant of the $\su(4)$ eigenvalue equation to zero.}
  \label{fig:arrow}
\end{figure}

\begin{figure}[t!]
  \centering
  \includegraphics[width=.6\linewidth]{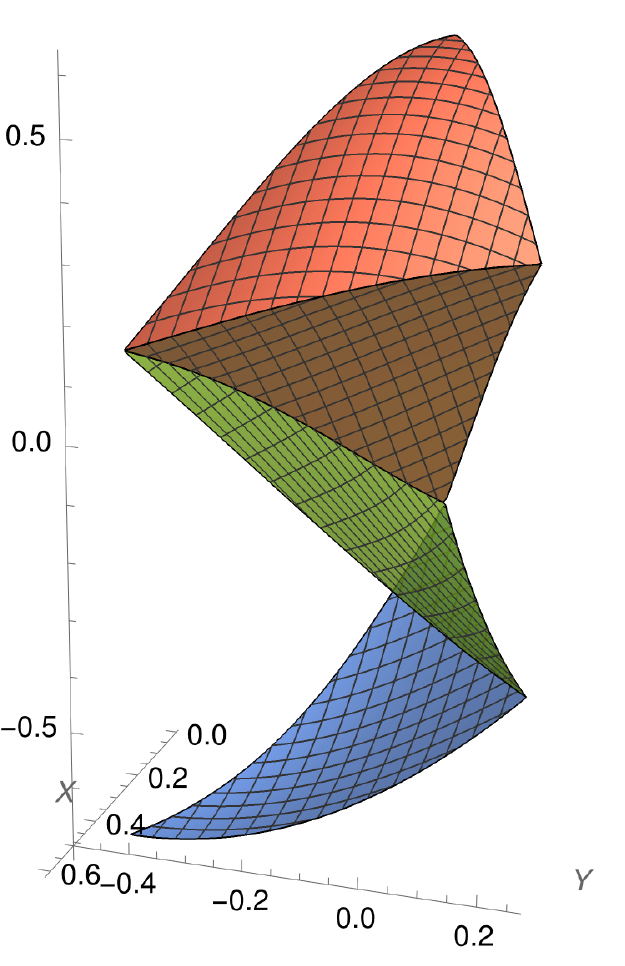}
  \\
  \includegraphics[width=.6\linewidth]{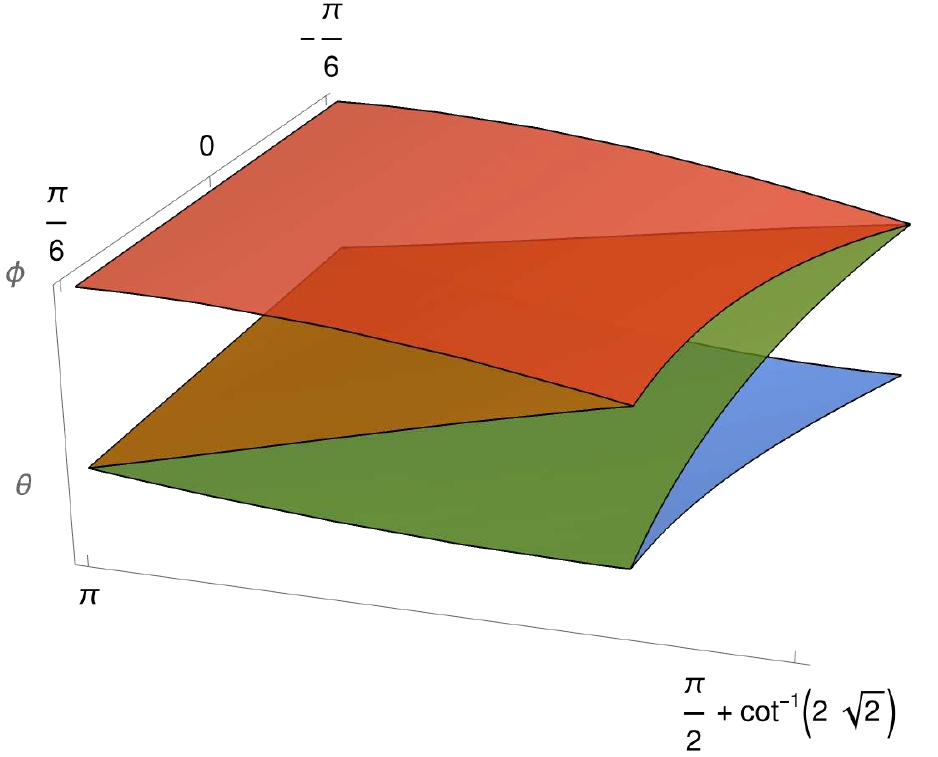}
  \caption{(Top) The sector (in the stereographic plane $(X,Y)$) joining three extrema of the determinant corresponds to a fixed size-ordering for the eigenvalues of a diagonal $\su(4)$ matrix, analogous to the fundamental sector in \cref{fig:comp}.
    This triangle on $S^2$ can be parameterized by $(\theta,\phi)$ in usual spherical coordinates, with $\phi\in[-\pi/6,\pi/6]$ as in the $\SU[3]$ case, and $\theta\in[\tfrac{\pi}{2}+\cot^{-1}(\sqrt{2}\sec(\phi-\tfrac{\pi}{6})),\pi]$.
    The other possible sectors can be found by acting on the sphere with the cubic group.
    (Bottom) The same sector in standard spherical coordinates.}
  \label{fig:su4eigs}
\end{figure}

The solutions $(\theta_0,\phi_0)$ for the diagonalization of the matrix $(\Sigma_0/w)\in\su(4)$ correspond to the $4!$ points where these surfaces intersect with each other and the unit sphere.\footnote{These coordinates correspond to different permutations of the solutions of the eigenvalue equation $x^4 -\tfrac{1}{4}x^2 + x \beta + \alpha = 0$.}
The choice of solution affects the gauge-fixing procedure as before, and must be chosen to preserve the relevant global symmetries. We then seek a `fundamental region' with a specific eigenvalue ordering, which tiles the full sphere under the action of the cubic group and has a 1:1 mapping to the space of allowed values for $(\alpha,\beta)$, as given in \cref{fig:arrow}.

Analogous to the approach for $\SU[3]$, we consider a stereographic projection of the sphere to $\RR^2$, also plotted in \cref{fig:spherecontours} for the hemisphere with $\theta\geq\pi/2$.
Note that, for each `slice' of the 3-sphere along the $T_{15}$ axis, the problem is of course reduced to that of the  $\SU[3]$ case with a different overall $w$-scaling, so the system is therefore symmetric under the same rotational symmetries.
Recall that for $\SU[3]$, the choice of sector corresponded to an arc which spanned the region between the maximum and minimum of its matrix invariant: here this corresponds to a choice of any of the six lines on the sphere which join adjacent pairs of maxima and minima for $\beta$.
We can therefore map the blue curve spanning the range of $\beta$ in \cref{fig:arrow}, to the line on the disc in \cref{fig:spherecontours} that joins any neighbouring pair of square and star-shaped points.
Along this line, the symmetries of the $\SU[3]$ case extend naturally: the diagonalized form of the $\Sigma$ matrix would be invariant under a transformation $\operatorname{S}(\U[3]\times\U[1])\sim\SU[3]\times\U[1]$ for the extrema of $\beta$ (where the eigenvalues are proportional to a permutation of $T_{15}$, and $\alpha$ is also minimized); and $\operatorname{S}(\U[2]\times\U[2]\times\ZZ[2])\sim\SU[2]^2\times\U[1]\times\ZZ[2]$ at $(\beta=0,\alpha=0)$ (where the eigenvalues correspond to those of $T_3$, and $\alpha$ has a saddle point), and $\operatorname{S}(\U[2]\times\U[2])\sim\SU[2]^2\times\U[1]$ otherwise.

The sector for $\SU[4]$ is now augmented by an additional coordinate corresponding to the maximum of the determinant, i.e.\ the point $(\alpha=1/64,\beta=0)$ in \cref{fig:arrow} and the round points in \cref{fig:spherecontours}.
The $\beta=0$ axis of \cref{fig:arrow} of course maps to the line joining the previous saddle point of $\alpha$ to an adjacent maximum of $\alpha$, and has the symmetry $\SU[2]^2\times\U[1]\times\ZZ[2]$.
Finally, the red curves of \cref{fig:arrow} map to the corresponding geodesics on the sphere that join the minima and maxima of $\alpha$, and these correspond to $\SU[2]\times\U[1]^2$ symmetries.
All remaining points correspond to only a $\U[1]^3$ symmetry.

This sector is a valid choice for Landau--'t Hooft gauge, as it provides a fixed ordering for the eigenvalues and is 1:1 over the space of possible $(\alpha,\beta)$ (see \cref{fig:su4eigs}).

As we saw in the $\SU[3]$ case, the choice of diagonalization for the vev plays an important role in the gauge-dependent physics, and depending on the gauge-fixing procedure it can artificially introduce or destroy symmetries arising from the BEH effect.
We therefore need to understand how the choice of ordering of the eigenvalues, together with the gauge-fixing strategy, affects the overall breaking pattern.
Note that a choice of sector as described above, joining three adjacent extrema of the determinant, would suffice for Landau--'t Hooft gauge but not for unitary gauge, since a global $\ZZ[2]$ reflection $\Sigma\mapsto-\Sigma$ would require both $\phi\mapsto\phi+\pi$ and $\theta\mapsto\pi-\theta$.
A suitable choice obeying the global symmetry would then be to divide the sector into two disjoint parts on opposite hemispheres according to the value of $\beta$ in analogy to \cref{eqn:app:unitary_gauge_z2}.

\subsubsection{The sectors for general \texorpdfstring{$N$}{N}}

Although for larger $N$ the explicit determination of the eigenvalues in terms of matrix invariants is more cumbersome or (analytically) impossible, the generalization of the sectors from $\SU[N-1]$ to $\SU[N]$ is straightforward.
One can project stereographically from $S^{N-2}$ to $\RR^{N-2}$, or from the lower hemisphere to a unit $(N-2)$-dimensional ball.
Since at each value of the radius, the same symmetries hold as for the $\SU[N-1]$ case, the problem reduces to that of diagonalizing a $\SU[N-1]$ matrix with variable norm.
An expression for a `fundamental sector' can thus be determined by choosing one of the $(N-1)!$ possible regions corresponding to the $\SU[N-1]$ sector, and subdividing it into $N$ parts, which produces a $(N-2)$-simplex that tiles the sphere under the rotations and reflections that arise from permuting the eigenvalues. As the choice of $\SU[N-1]$-sector singles out a specific ordering for the eigenvalues $(\lambda_1,\dots,\lambda_{N-1})$, the subdivision into N further sectors can thus be reduced to finding the radii where $\lambda_N=\lambda_i$ for each $i$ respectively.
Since maximal (i.e. $\SU[N-1]\times\U[1]$) symmetry is obtained when the eigenvalues are one of the $2N$ permutations of $\pm T_{N^2-1}$, we can choose, for simplicity, to place one of the vertices of the sector as the origin of the projected space.
If the sector is further defined so that the eigenvalues are ordered from smallest to largest, the remaining $(N-2)$ vertices are defined from the intersection of the corresponding $\SU[N-1]$-sector. This leaves us with the iterative equation for the bounds of a `fundamental domain'
\begin{gather}
  \frac{\pi}{6} \le \theta^{(1)} \le \frac{\pi}{2} \notag\\
  \arctan(\sqrt{\tfrac{N-2}{N}}\sin(\theta^{(N-3)})) \le \theta^{(N-2)} \le \frac{\pi}{2}
\end{gather}
for $\SU[N\geq 4]$\footnote{Note that this is of course a different sector from that of the previous section.}.
This is a direct consequence of the definitions in \cref{app:ssec:possible-breaking-patterns}.

Explicit expressions for the boundaries of the $\SU[N]$-sectors can be found in terms of intersections of the $\SU[N-1]$-sectors, together with the set of $(N-1)$ further $(N-3)$-spheres with origins at the points corresponding to permutations of the eigenvalues of $T_{N^2-1}$, and radii $\sqrt{\tfrac{2(N-1)}{N}}$.

Using the degeneracies of the points where different sectors meet, together with the symmetries from preserving $T_{N^2-1}$ or the $\SU[N-1]$-sector, the symmetry breaking patterns outlined for $\SU[4]$ can be easily generalized to arbitrary $N$.

Note that within this sector there is a well-defined ordering of the eigenvalues, and symmetries are enhanced on the boundaries.\footnote{A more detailed discussion using Morse theory can be found in \cite{Houston1983am}.}
As before, to adapt the sector so that it preserves the global-$\ZZ[2]$ symmetry, as needed for unitary gauge, one should split this into two parts depending on the sign of $\tr\qty[\Sigma^3]$, and reflect one part to the opposite sector on the sphere.

\section{Algorithms}\label{app:algorithms}

For our simulations we always used different versions and modifications of generic heatbath algorithms \cite{Creutz:1980zw,Cabibbo:1982zn,Bunk:1994xs,Knechtli:1999tw}, consisting of local updates of the individual gauge-fields and scalar-fields. These will be described in more detail below. For consistency, we also cross-checked our results with a standard multi-hit-metropolis (MHM) \cite{Maas:2017xzh,Maas:2016ngo}, and they agreed. However, the heatbath algorithms outperformed the MHM algorithms in terms of equilibration time and decorrelation of configurations, as expected. The reason for not choosing a Hybrid Monte Carlo algorithm (HMC) like in \cite{Wellegehausen:2011sc} is mainly due to the problems encountered in a previous study of a $\SU[2]$ theory with an adjoint scalar field \cite{Afferrante:2020hqe}. This system has some properties in common with the theory studied in this paper and therefore similar problems were expected to occur.

\subsection{Link updates}

A usual heatbath algorithm exploits the fact that the local probability distribution of a single link $U_\mu\qty(x)$ can be written as \cite{Gattringer:2010zz}
\begin{equation}
  \dd{P\qty(U_\mu\qty(x))} = \dd{U_\mu\qty(x)}\exp\qty(\frac{\beta}{3}\Re\qty[\tr\qty[U_\mu\qty(x)A_\mu\qty(x)]])
\end{equation}
with $A_\mu\qty(x)$ being the so-called staples and $\dd{U_\mu\qty(x)}$ the Haar measure. As long as all parts of the action of the theory are linear in $U_\mu\qty(x)$ the staples can be modified accordingly and a usual heatbath update as outlined in e.g.\ \cite{DeGrand:2006zz,Gattringer:2010zz} can be performed. Otherwise, heatbath algorithms cannot be applied to probability distributions that are of a Gaussian-type or even higher orders in the updated field. Indeed, in our theory the interaction term (\ref{eqn:adj_lattice_lagrangian_int}) is quadratic in the local link and therefore makes the usual approach impossible. To circumvent this problem there are now two different approaches: Either using a different representation for this term or modifying the update strategy.\footnote{Another approach by using additional auxiliary fields to linearize the action as in \cite{Vairinhos:2010ha} is currently under investigation \cite{newhb:2023}.}

The first approach of using a different representation, would be to use the adjoint representation of the links
\begin{equation}\label{eqn:adjoint_link}
  U_{\mu,ab}^A\qty(x)=2\tr\qty[T_aU_\mu\qty(x)T_bU_\mu^{\dagger}\qty(x)]\,.
\end{equation}
This allows the interaction term to be written in the usual form $\Sigma_a\qty(x)U_{\mu,ab}^A\qty(x)\Sigma_b\qty(x+\hat{\mu})$ which is linear in the adjoint link. However, to the knowledge of the authors, it is not possible to rewrite the pure gauge part of the action (\ref{eqn:adj_lattice_lagrangian_gauge}) in linear terms of the adjoint link variable. Therefore, this approach simply shifts the problem and does not really solve it.

This leaves us with the second approach, which is the one used to obtain our configurations; modifying the update strategy. To motivate the used update strategy one should remember that a heatbath update is actually a usual Monte Carlo update where the new variable is chosen such that the accept-reject step can be skipped. So we instead follow a `hybrid' approach of an approximate heatbath update (see \citet[7.2]{DeGrand:2006zz}) for modified staples followed by an accept-reject step. The procedure is thus as follows:
\begin{enumerate}
  \item Sampling a new local link variable $U_\mu\qty(x)'$ from a heatbath of the linear part of the action
  \item Accept-reject it with the acceptance probability $\min\qty(1,\exp\qty(-\Delta S_{\text{int}}))$ using the interaction part of the action (\ref{eqn:adj_lattice_lagrangian_int})
\end{enumerate}
This algorithm is still expected to produce better candidate configurations than the simple MHM algorithm in most parts of the phase diagram due to taking into account the linear change in the action. Only in parts where scalar dynamics have a very large influence on the system, i.e. when $\gamma\to 0$ or $\kappa\gg 1$, this algorithm is expected to equilibrate slowly. This has indeed been observed as discussed in \cref{a:crit}. However, note that an MHM algorithm is expected to perform similarly poorly in such regions.

Apart from the approximate heatbath updates we also performed overrelaxation sweeps in between measurements to increase the step-size in configuration space. The main idea of an overrelaxation update is to propose a new variable such that the action stays unchanged. The usual procedure is to update the local links by $U_\mu\qty(x)\to \qty(V_\mu\qty(x)U_\mu\qty(x)V_\mu\qty(x))^{\dagger}$ where $V_\mu\qty(x)=A_\mu\qty(x)/\sqrt{\det\qty[A_\mu\qty(x)]}$. This again faces the same issue as the usual heatbath algorithm and will only work when the action can be reformulated in terms of (modified) staples. So again we choose here to do an `approximate overrelaxation' update with an additional accept-reject step. It should be mentioned that this is strictly speaking not an overrelaxation update since it modifies the overall action. However, it is still a valid update since it respects the usual criteria of a Monte Carlo step and only changes the interaction part of the action.

\subsection{Scalar updates}

For the update of the scalar fields we immediately run into the same problem as for the links since the action is up to quartic order in the local fields. However, in \cite{Bunk:1994xs,Knechtli:1999tw} an algorithm for an approximate heatbath for a scalar field in the fundamental representation of $\SU[2]$ has been proposed and used. Here we will adapt the steps outlined in Appendix C.3 of \cite{Knechtli:1999tw} to fit our needs. Also, without loss of generality, we will sketch the steps for an arbitrary $\SU[N]$ theory with a scalar in the adjoint representation but keeping the same potential as in \cref{eqn:adj_lattice_lagrangian}.

To start with one needs to reformulate the action parts that depend on the scalar-field $\Sigma\qty(x)$ in terms of a real vector $\vec{\Sigma}:=\Sigma_a\qty(x)=2\tr\qty[\Sigma\qty(x)T_a]$ of dimension $d=N^2-1$ according to \cref{eqn:vec}. This simply replaces $\tr\qty[\Sigma\qty(x)^2]$ by $\vec{\Sigma}^2/2$ in \cref{eqn:adj_lattice_lagrangian_scalar} and again uses the adjoint link in \cref{eqn:adjoint_link} for the interaction term (\ref{eqn:adj_lattice_lagrangian_int}). This leaves us after some reshuffling with the following terms in the action that are dependent on a local scalar field $\vec{\Sigma}$
\begin{align}\label{eqn:app:act_scalar}
  S_{\text{scalar}} & = \qty(\vec{\Sigma} - \vec{k})^2 + \gamma\qty(\vec{\Sigma}^2-1)^2 - \vec{k}^2                                    \\
  \vec{k}           & := \frac{\kappa}{2} \sum_{\mu=\pm 1}^{\pm 4}U_{\mu,ab}^A\qty(x)\Sigma_b\qty(x+\hat{\mu}) \label{eqn:app:int_vec}
\end{align}
where we used that $U_{\mu,ab}^A=U_{-\mu,ab}^{A\dagger}$ and have introduced the interaction vector $\vec{k}$\footnote{Note that also for simulations with scalar fields in different representations an equivalent expression may be obtained with corresponding modifications to $\vec{k}$.}. Now we want to generate a configuration following the distribution
\begin{align}
  \dd{P\qty(\vec{\Sigma})} & = \frac{1}{Z}\dd{\vec{\Sigma}}\exp\qty(-V_{\text{scalar}})\notag \\
  V_{\text{scalar}}        & =\qty(\vec{\Sigma} - \vec{k})^2 + \gamma\qty(\vec{\Sigma}^2-1)^2
\end{align}
with $Z$ being a suitable normalization. To obtain the potential $V_{\text{scalar}}$ the constant term from the action (\ref{eqn:app:act_scalar}) has been dropped.

Due to the quartic dependence of $V_{\text{scalar}}$ on the scalar field we cannot simply produce new configurations from a Gaussian heatbath. Following now \cite{Knechtli:1999tw} we will therefore approximate the potential by its quadratic part and add a Metropolis-step to correct for the quartic part. To improve the acceptance rate for this procedure we introduce an additional parameter $\alpha$ such that the split between the individual parts is optimized
\begin{gather}
  V_{\text{scalar}}^{\alpha} = \alpha\qty(\vec{\Sigma} - \frac{\vec{k}}{\alpha})^2 + \gamma\qty(\vec{\Sigma}^2-v_{\alpha}^2)^2 - c_\alpha \notag\\
  v_{\alpha}^2 = 1 + \frac{\alpha-1}{2\gamma}\,,\: c_{\alpha}=\gamma\qty(v_{\alpha}^4 - 1) + \qty(\frac{1}{\alpha}-1)\vec{k}^2\,. \label{eqn:app:scalar_pot}
\end{gather}

A new trial scalar field can now be generated from a Gaussian distribution
\begin{equation}\label{eqn:app:p_trial}
  P_{\text{trial}}\qty(\vec{\Sigma}') = \qty(\frac{\alpha}{\pi})^{\frac{d}{2}}\exp\qty(-\alpha\qty(\vec{\Sigma}' - \frac{\vec{k}}{\alpha})^2)
\end{equation}
and accepted with the probability
\begin{equation}\label{eqn:app:p_accept}
  P_{\text{accept}}\qty(\vec{\Sigma}') = \exp\qty(-\gamma\qty(\vec{\Sigma}'^2-v_{\alpha}^2)^2)\,.
\end{equation}
The total acceptance rate is then given by
\begin{align}
  A(\alpha) & =\int_{\RR^{d}}\dd{\vec{\Sigma}'} e^{-c_{\alpha}}P_{\text{trial}}\qty(\vec{\Sigma}')P_{\text{accept}}\qty(\vec{\Sigma}') \notag \\
            & = \qty(\frac{\alpha}{\pi})^{\frac{d}{2}}e^{-c_{\alpha}}\dd{P\qty(\vec{\Sigma})}\,,
\end{align}
with $d$ being here the dimension of coefficient vector, i.e.\ $d=N^2-1$ for the adjoint representation.
To maximize the acceptance we take the derivative with respect to $\alpha$ and obtain the following cubic equation
\begin{equation}\label{eqn:app:cubic}
  \dv{A(\alpha)}{\alpha} = 0 \Rightarrow \alpha\qty(\alpha^2 + \qty(2\gamma-1)\alpha-d\gamma)=2\gamma\vec{k}^2\,.
\end{equation}
Since we are interested in large parts of the phase diagram we decided to solve the cubic equation for every update individually using Cardano's method. Especially, it can be shown that the $\alpha$ which minimizes the acceptance rate has to be positive and real, thus it follows that we can always use the first solution of Cardano's method. In case of three real solutions the first solution is always the one that maximizes the acceptance step.

Following these considerations we are finally at the point to state the algorithm for obtaining a new configuration:
\begin{enumerate}
  \item Compute the local interaction vector $\vec{k}$ in \cref{eqn:app:int_vec}
  \item Calculate the optimal value of $\alpha$ by solving \cref{eqn:app:cubic}
  \item Create a new Gaussian field $\vec{\Sigma}'$ according to the distribution in \cref{eqn:app:p_trial}
  \item Calculate $v_{\alpha}^2$ from \cref{eqn:app:scalar_pot}
  \item Accept-reject the new configuration according to the probability $\min\qty(1,P_{\text{accept}})$
\end{enumerate}
This strategy performs quite well in most parts of the phase diagram. Especially large $\kappa$ values are also accessible which would not be the case when using the approximate solution for the cubic equation as was done in \cite{Knechtli:1999tw}.

To improve the updates we also employed an overrelaxation update for the scalar field. Consider again the scalar part of the action in \cref{eqn:app:int_vec} but expanding the quadratic part explicitly.
\begin{equation}\label{eqn:app:s_scalar}
  S_{\text{scalar}} = \vec{\Sigma}^2 - 2\vec{\Sigma}\vec{k} + \gamma\qty(\vec{\Sigma}^2-1)^2\,.
\end{equation}
We now look for a transformation of the scalar that leaves this invariant, i.e.\ a rotation about the $\vec{k}$-axis and norm preserving.
This is achieved by choosing a random vector $\vec{\rho}_\perp$ in the space orthogonal to $\vec{k}$, and using this to reflect $\vec{\Sigma}$ in the plane spanned by $\{\vec{\rho}_\perp,\vec{k}\}$:
\begin{equation}
  \vec{\rho}_\perp=\vec{\rho}-\vec{k}\frac{\vec{\rho}\cdot\vec{k}}{|k|^2}\quad\text{with $\vec{\rho}\in\RR^{d}$ a random vector},
\end{equation}
The new scalar field is then obtained from the reflection
\begin{equation}
  \vec{\Sigma}' = \vec{\Sigma}-2\frac{\vec{\Sigma}\vec{\rho}_\perp}{|\vec{\rho}_\perp|^2}\vec{\rho}_\perp\,,
\end{equation}
which indeed satisfies the condition of keeping (\ref{eqn:app:s_scalar}) and thus the full action invariant.

\subsection{Gauge-fixing}

For gauge-fixing it was partly necessary to develop new methods as has already been described throughout the main text. The two gauges that have been used are unitary gauge and Landau--'t Hooft gauge.

\subsubsection{Unitary gauge}

The main aim of unitary gauge is to diagonalize the scalar fields locally and apply the corresponding gauge transformation also to the link fields. Therefore, one needs to calculate the gauge transformation $G\qty(x)$ such that
\begin{equation}
  \Sigma\qty(x) \rightarrow \Sigma_\text{gf}\qty(x) =  G(x)\Sigma\qty(x)G^\dagger(x) =  \diag{\qty[\lambda_1,\lambda_2,\lambda_3]}\,.
\end{equation}
Subsequently, the transformation is applied to the links according to
\begin{equation}
  U_{\mu\qty(x)}\rightarrow G(x)U_{\mu\qty(x)}G^\dagger(x+\hat{\mu})
\end{equation}
To obtain $G(x)$ we therefore need to find the eigenvector matrix. This has been done using the `direct analytical calculation of the eigenvectors' algorithm described in \cite{Kopp:2006wp}. As an input for this algorithm the eigenvalues need to be obtained first. This can be done by calculating a local breaking angle $\theta\qty(x)$ according to \cref{eqn:theta_gauge_invariant} and reconstruct the matrix using \cref{eqn:adj_breaking_angle}. However, as already mentioned in the main text the sorting of the eigenvalues, and correspondingly the ordering of the columns of $G(x)$, has a significant impact on the gauge-fixed observables. To preserve the \ZZ[2] alignment of the scalar fields it is necessary to make sure that after diagonalization fields of opposing signs still correspond to oppositely aligned fields. This can be achieved by using the sign of the obtained breaking angle to sort the eigenvalues accordingly. The problem that appears is that the $\arcsin$-function used to obtain the local angle from \cref{eqn:theta_gauge_invariant} will always yield angles in the range $[-\pi/6,\pi/6]$. Therefore, when performing the reconstruction of the matrix this will always destroy the \ZZ[2]-alignment. The easiest way to compensate this is to modify the reconstruction \cref{eqn:adj_breaking_angle} to
\begin{align}\label{eqn:app:unitary_gauge_z2}
  \Sigma_\text{gf}\qty(x) = & \cos\qty(\abs{\theta\qty(x)}+\frac{\qty(2k+3\sgn\qty(\theta\qty(x))-3)\pi}{6}) T_3 \notag \\
  +                         & \sin\qty(\abs{\theta\qty(x)}+\frac{\qty(2k+3\sgn\qty(\theta\qty(x))-3)\pi}{6})T_8\,,
\end{align}
where $k=0,1,\dots,6$ can be used to define the sorting of the eigenvalues for the \ZZ[2]-positive fields. In our code we used $k=0$ which returns for the special angles $0$ and $\pm\pi/6$ the matrices $T_3$ and a permutation of $\pm T_8$ respectively, while $k=1$ corresponds to descending/ascending sorted eigenvalues for positive/negative angles respectively. Any of these reconstruction methods does guarantee the maximal preservation of the (local) \ZZ[2]-symmetry and thus keeps the alignment of neighbours intact.

\subsubsection{Landau--'t Hooft gauge}

In Landau--'t Hooft gauge these problems do not show up since the local transformation is fully specified by the (minimal) Landau gauge condition. In our simulations we therefore calculated first the local gauge transformations $G(x)$ in minimal Landau gauge as described in \cite{Ilgenfritz:2010gu,Maas:2011se} using the `Stochastic Overrelaxation Method' as described in \cite{Cucchieri:1995pn}. The corresponding code for \SU[3] uses again a Cabibbo-Marinari trick \cite{Cabibbo:1982zn}, like the heatbath methods, and has been extensively tested already in simulations with a fundamental Higgs \cite{Maas:2018xxu}.

For the 't Hooft part \cite{Bohm:2001yx} of the gauge condition we still got a remaining global degree of freedom to fix. In analogy to the fundamental case we therefore fix the gauge by restricting the global direction of the space-time averaged scalar field with the condition
\begin{equation}
  \tilde{\Sigma}=\frac{1}{V}\sum_x \Sigma\qty(x) \rightarrow \tilde{\Sigma}_\text{gf} =  G\tilde{\Sigma}G^\dagger =  \diag{\qty[\lambda_1,\lambda_2,\lambda_3]}\,.
\end{equation}
The main complication that arises again is that it is impossible to restrict this condition further to a specified direction of the vev, i.e.\ to a certain pattern. Instead, we can only fix it to the Cartan space which contains all breaking patterns. This mirrors again our result that beyond perturbation theory it is not possible to a-priori choose a pattern.

Form the algorithmic point of view, we used the same algorithm as for the unitary gauge update, without any additional \ZZ[2]-specific sorting. Although it is possible to do this here too, and has also been tried, it is not necessary to add this additional overhead since this condition does not alter subsequent \ZZ[2]-alignments. Therefore, the \ZZ[2]-symmetry is not broken explicitly by the gauge-fixing condition nor the algorithm.

\subsection{Critical slowing down}\label{a:crit}

\begin{figure}[t!]
  \centering
  \includesvg[width=\linewidth]{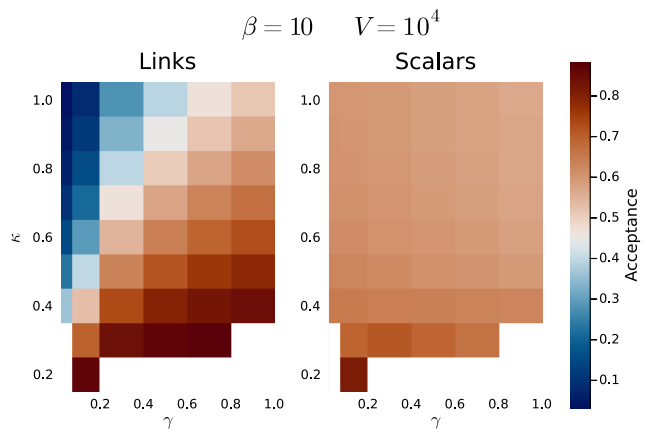}
  \caption{Acceptance rates of the link- (left) and scalar-updates (right).}
  \label{fig:app:acceptance}
\end{figure}

As already mentioned in the main text we encountered regions in the phase diagram with critical slowing down of our simulations. The effect becomes especially severe with increasing values of $\kappa$ and $\gamma$ becoming small. As an example \cref{fig:app:acceptance} shows the obtained acceptance rates as a function of $\kappa$ and $\gamma$ for a fixed value of $\beta=10$.

We see that in the region where $\gamma\to 0$ the acceptance rate of the link updates using the approximate heatbath update dropped to about \SI{5}{\percent} compared to the usual rates of approximately \SIrange{50}{70}{\percent}. The scalar updates using the approximate heatbath method stayed at approximately \SI{60}{\percent} throughout the scanned phase-diagram.

To reduce the impact of this effect we therefore already drop several configurations between consecutive measurements to decorrelate our observables and also skip $\order{5000}$ configurations in the beginning for equilibration, for all parts of the phase-diagram. Nevertheless, for gauge-invariant observables an additional post-analysis of our data can still not be avoided in most parts. To further decorrelate our measurements we calculate the exponential autocorrelation time $\tau$ of the plaquette according to \cite{Gattringer:2010zz} and skip additional measurements such that the autocorrelation time becomes approximately 1. To obtain gauge-fixed configurations we used only every tenth configuration for gauge-fixing which always yields sufficient decorrelation according to the autocorrelation time.

\begin{figure*}[ht]
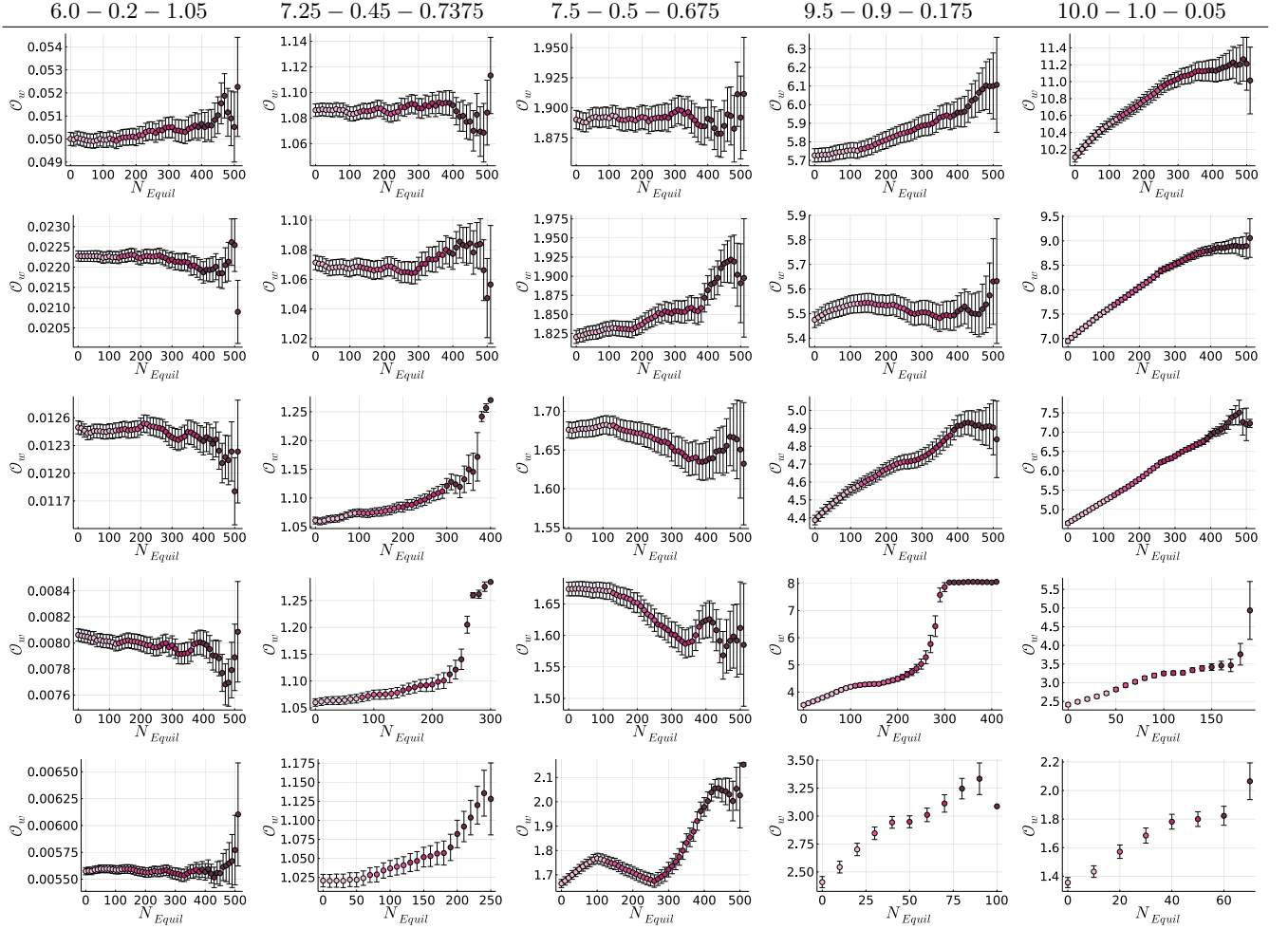

  \centering
  \begin{tabular}{ccccc}
    $6.0-0.2-1.05$                                                               &
    $7.25-0.45-0.7375$                                                           &
    $7.5-0.5-0.675$                                                              &
    $9.5-0.9-0.175$                                                              &
    $10.0-1.0-0.05$                                                                \\\hline
    \includesvg[width=0.19\textwidth]{{vev_equil_8-6.000000-0.200000-1.050000}}  &
    \includesvg[width=0.19\textwidth]{{vev_equil_8-7.250000-0.450000-0.737500}}  &
    \includesvg[width=0.19\textwidth]{{vev_equil_8-7.500000-0.500000-0.675000}}  &
    \includesvg[width=0.19\textwidth]{{vev_equil_8-9.500000-0.900000-0.175000}}  &
    \includesvg[width=0.19\textwidth]{{vev_equil_8-10.000000-1.000000-0.050000}}   \\
    \includesvg[width=0.19\textwidth]{{vev_equil_12-6.000000-0.200000-1.050000}} &
    \includesvg[width=0.19\textwidth]{{vev_equil_12-7.250000-0.450000-0.737500}} &
    \includesvg[width=0.19\textwidth]{{vev_equil_12-7.500000-0.500000-0.675000}} &
    \includesvg[width=0.19\textwidth]{{vev_equil_12-9.500000-0.900000-0.175000}} &
    \includesvg[width=0.19\textwidth]{{vev_equil_12-10.000000-1.000000-0.050000}}  \\
    \includesvg[width=0.19\textwidth]{{vev_equil_16-6.000000-0.200000-1.050000}} &
    \includesvg[width=0.19\textwidth]{{vev_equil_16-7.250000-0.450000-0.737500}} &
    \includesvg[width=0.19\textwidth]{{vev_equil_16-7.500000-0.500000-0.675000}} &
    \includesvg[width=0.19\textwidth]{{vev_equil_16-9.500000-0.900000-0.175000}} &
    \includesvg[width=0.19\textwidth]{{vev_equil_16-10.000000-1.000000-0.050000}}  \\
    \includesvg[width=0.19\textwidth]{{vev_equil_20-6.000000-0.200000-1.050000}} &
    \includesvg[width=0.19\textwidth]{{vev_equil_20-7.250000-0.450000-0.737500}} &
    \includesvg[width=0.19\textwidth]{{vev_equil_20-7.500000-0.500000-0.675000}} &
    \includesvg[width=0.19\textwidth]{{vev_equil_20-9.500000-0.900000-0.175000}} &
    \includesvg[width=0.19\textwidth]{{vev_equil_20-10.000000-1.000000-0.050000}}  \\
    \includesvg[width=0.19\textwidth]{{vev_equil_24-6.000000-0.200000-1.050000}} &
    \includesvg[width=0.19\textwidth]{{vev_equil_24-7.250000-0.450000-0.737500}} &
    \includesvg[width=0.19\textwidth]{{vev_equil_24-7.500000-0.500000-0.675000}} &
    \includesvg[width=0.19\textwidth]{{vev_equil_24-9.500000-0.900000-0.175000}} &
    \includesvg[width=0.19\textwidth]{{vev_equil_24-10.000000-1.000000-0.050000}}
  \end{tabular}
  \caption{The $\mathcal{O}_w$ obtained from binned data on individual configurations from deep in the $\ZZ[2]$ unbroken region across the phase boundary down to $\gamma\approx 0$. The header of the column gives the respective simulation parameters $\beta-\kappa-\gamma$. $N_{Equil}$ gives the number of additionally skipped initial configurations for calculating the expectation value. Darker points therefore contain only configurations obtained later in the MC-history.}
  \label{fig:equilibration_vev}
\end{figure*}

This works very well for both types of configurations as long as the data is sufficiently equilibrated, which is always the case in the \ZZ[2]-unbroken area (blue points in \cref{fig:z2}) and in the broken region close to the phase transition. Deep in the Higgs-like region our simulations for $L>16^4$ do not fully equilibrate at all, but the slope of the action becomes smaller than the statistical fluctuations and should therefore still yield a reliable estimate within the obtained errors. As an example we see in \cref{fig:equilibration_vev} the gauge-fixed scalar length obtained from \cref{eqn:adj_gauge_order_lat} in Landau--'t Hooft gauge as a function of skipped initial measurements for several points in the phase-diagram and volumes. What can be clearly seen is that in the regions where the data is not well equilibrated the vev becomes suppressed. This needs to be carefully taken into account when analysing the data, since this may otherwise be wrongly interpreted as a vanishing vev in the $V\to\infty$ limit, which is not the case. The second problem that turns up is that equilibration takes much longer for configurations that are close to the $\SU[2]\times\U[1]$ type. To see this we look again at the scalar length distributions shown in \cref{fig:vev} and the angle against length plot in \cref{fig:theta_vs_vev}. In these plots a darker colour indicates in general that the corresponding configuration has been obtained later in the respective MC-history. First we see that deep in the broken region, i.e. close to $\gamma\approx 0$, darker bins in \cref{fig:vev} usually tend to higher values of the vev $w$. In the other plot \cref{fig:theta_vs_vev} we however also see that darker colours and respectively higher vev values tend towards angles of $\pm\pi/6$. Combining these two observations we can conclude that equilibration for $\SU[2]\times\U[1]$-like configurations, in this gauge, are the ones that are approached in equilibrium. This can also be seen by realizing that darker points are points that are obtained later on in the Monte-Carlo history. Finally, this effects can also be seen when comparing \cref{fig:conf_bulk}, which is well equilibrated, with \cref{fig:conf_deep_broken}. In \cref{fig:conf_deep_broken} we see, that towards the boundary of the cube there are still quite a lot of configurations left that are not as dark, i.e. strongly aligned, as one would expect. In regions where a sufficient compensation by removing data could not be achieved we rescaled our errors accordingly to
\begin{equation}
  \sigma \rightarrow 2\sqrt{\tau}\sigma
\end{equation}
to compensate the effect. The factor two is here larger than the usual factor of $\sqrt{2}$ to be a bit more conservative with error estimation in these parts.

\begin{figure*}[t!]
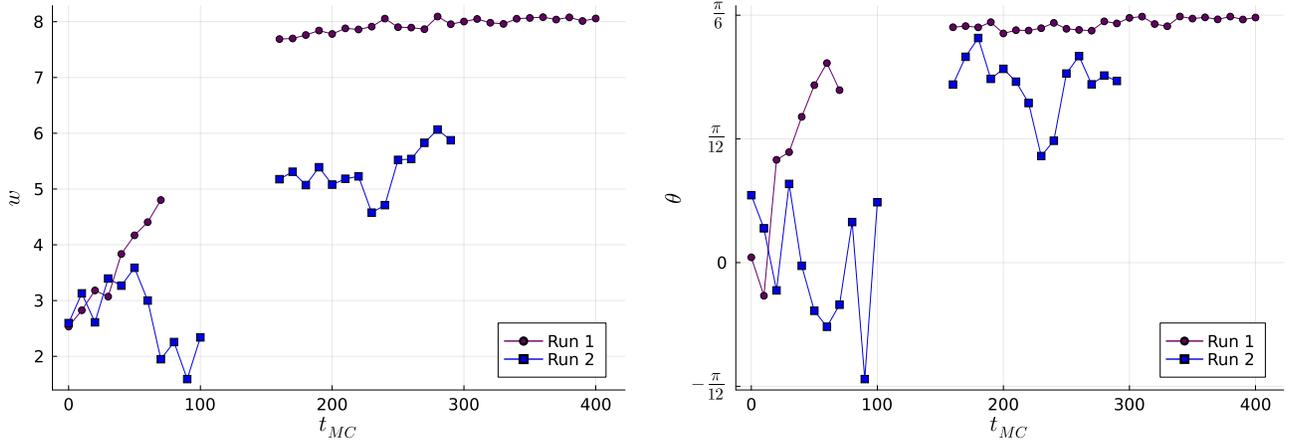

  \centering
  \includesvg[width=\columnwidth]{{w_individual_mc_history_20-9.500000-0.900000-0.175000}}
  \includesvg[width=\columnwidth]{{theta_individual_mc_history_20-9.500000-0.900000-0.175000}}
  \caption{$\mathcal{O}_w$ (top panel) and $\theta$ (bottom panel) along two different Monte Carlo trajectories, where Run 1 equilibrated fast, but is a rare case. Run 2 shows a more typical behaviour of very slow equilibration. Note that $\theta=\pm\pi/6$ is the $\SU[2]\times\U[1]$ pattern.}
  \label{fig:app:trajectories}
\end{figure*}

\begin{figure*}[ht!]
  \centering
  \begin{subfigure}[b]{0.19\textwidth}
    \includegraphics[width=\linewidth]{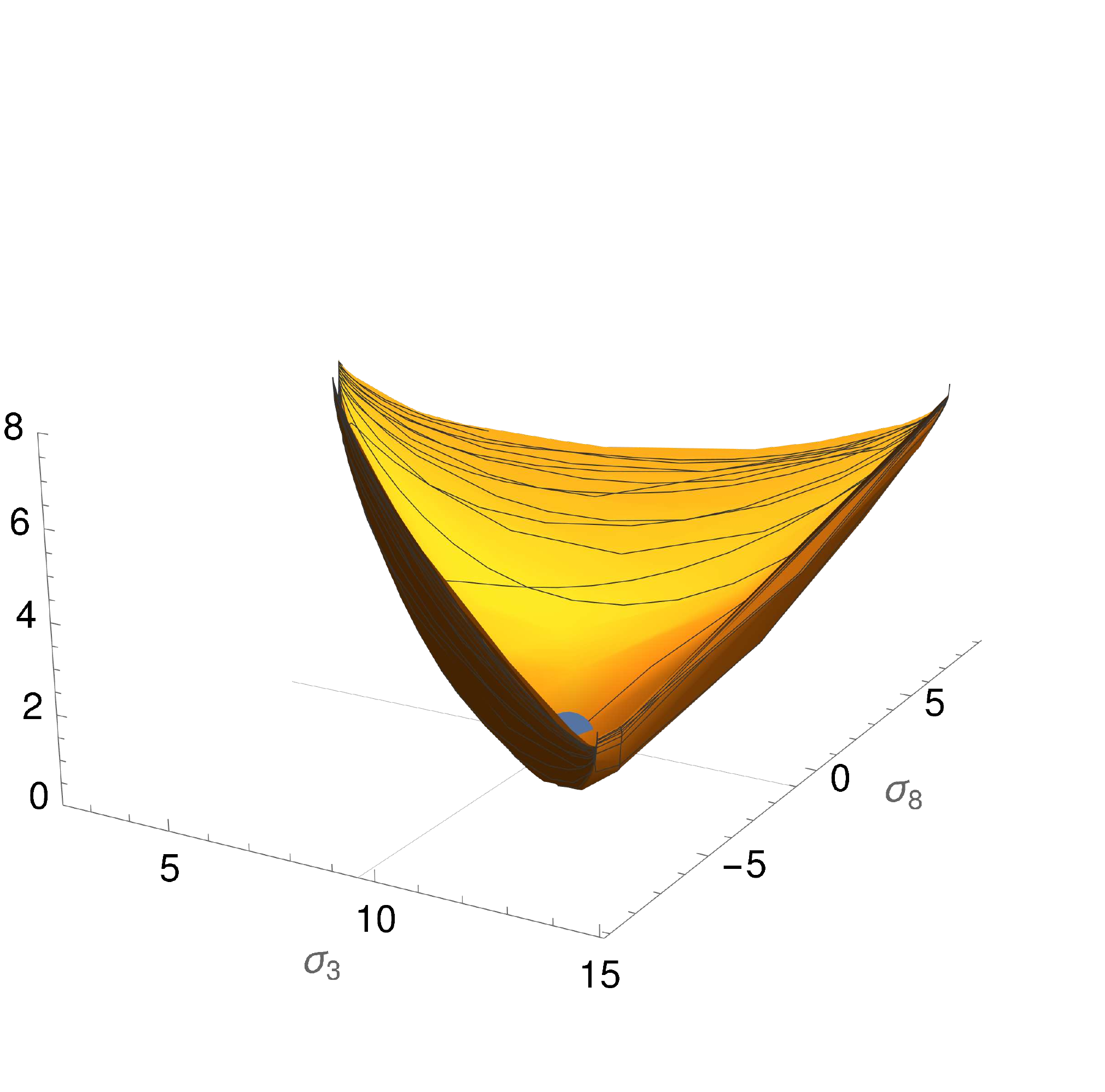}
    \caption{$V=8^4$}
  \end{subfigure}
  \begin{subfigure}[b]{0.19\textwidth}
    \includegraphics[width=\linewidth]{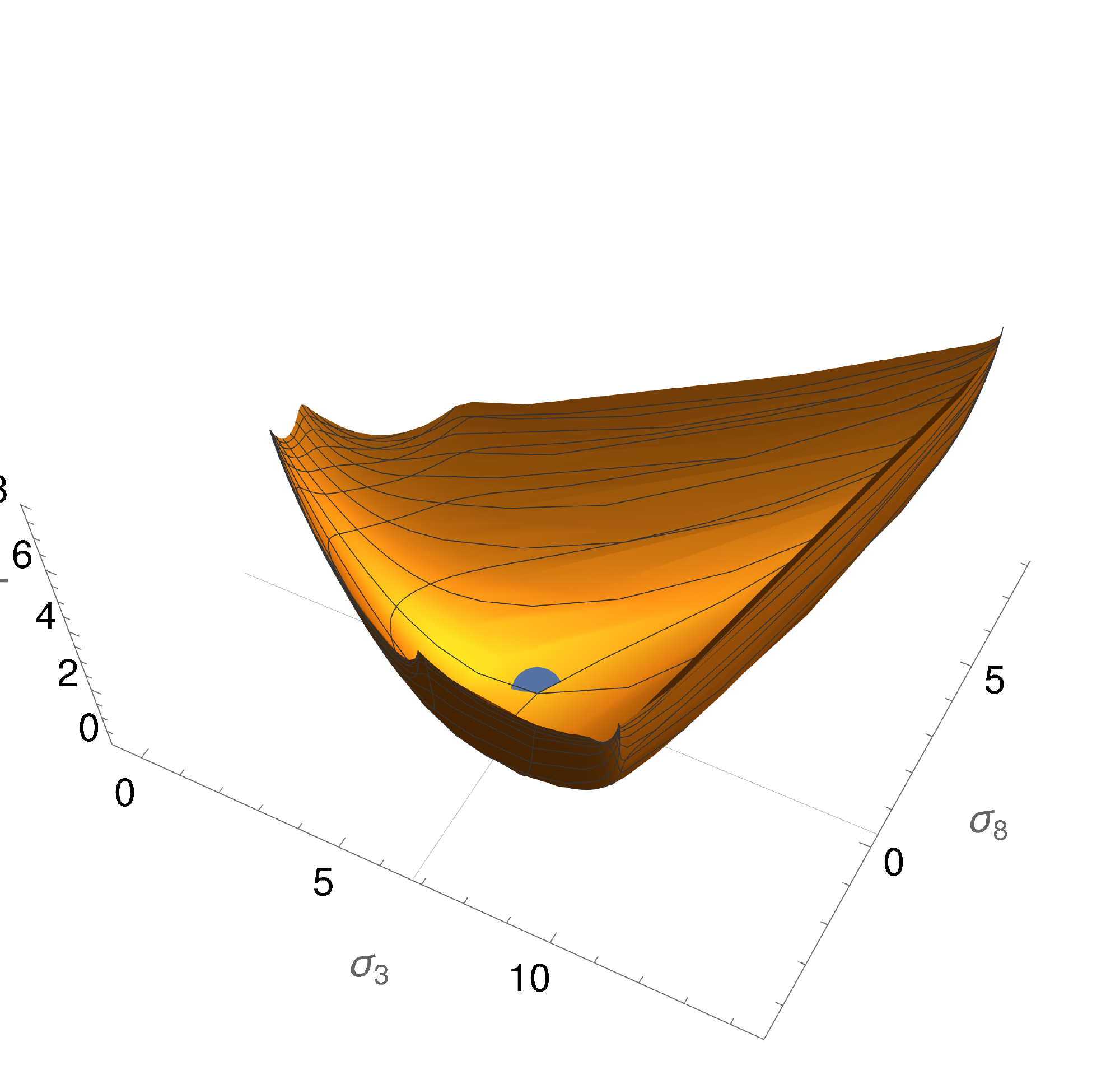}
    \caption{$V=12^4$}
  \end{subfigure}
  \begin{subfigure}[b]{0.19\textwidth}
    \includegraphics[width=\linewidth]{{effective_potential_16-10.000000-1.000000-0.050000}.pdf}
    \caption{$V=16^4$}
  \end{subfigure}
  \begin{subfigure}[b]{0.19\textwidth}
    \includegraphics[width=\linewidth]{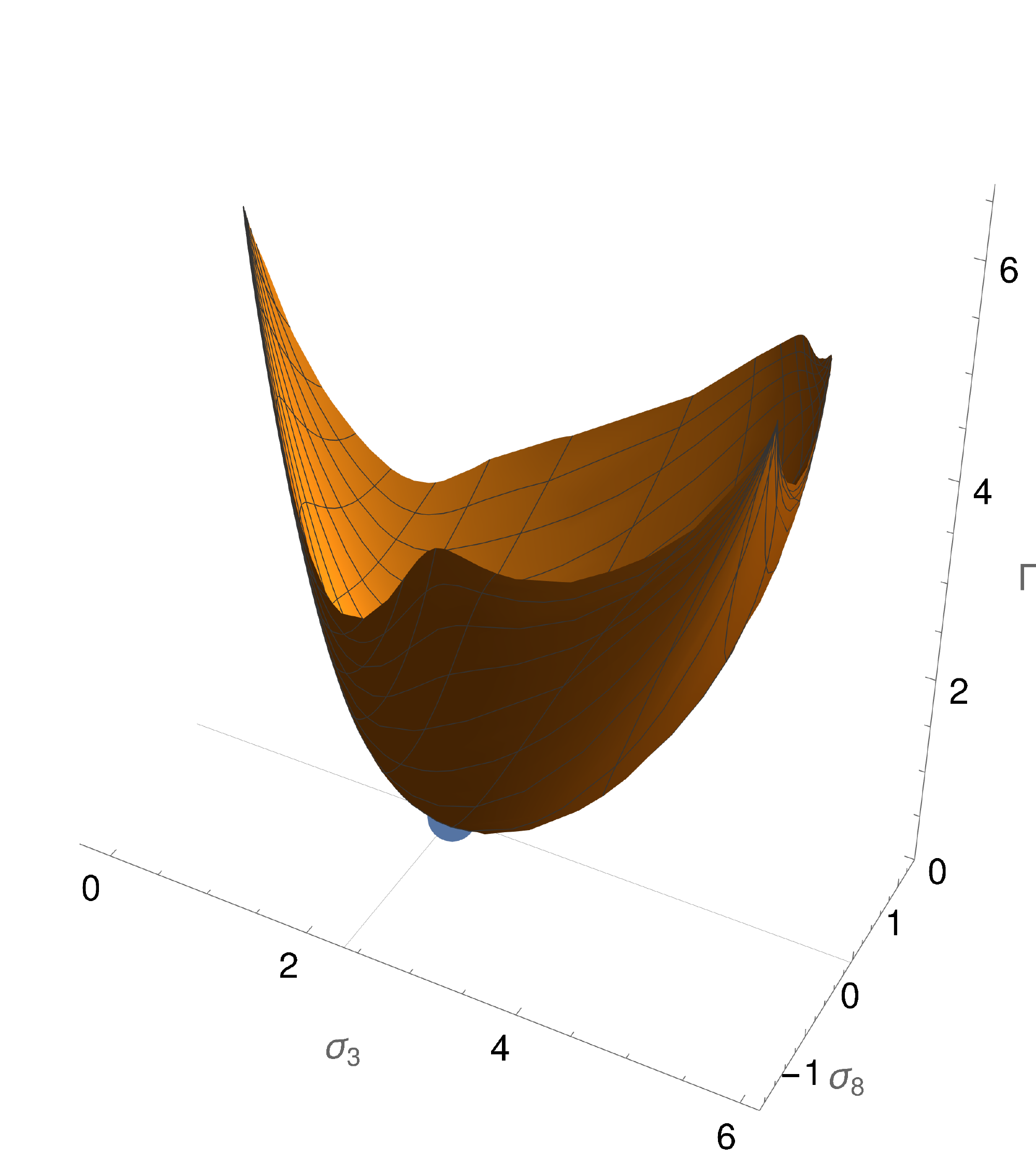}
    \caption{$V=20^4$}
  \end{subfigure}
  \begin{subfigure}[b]{0.19\textwidth}
    \includegraphics[width=\linewidth]{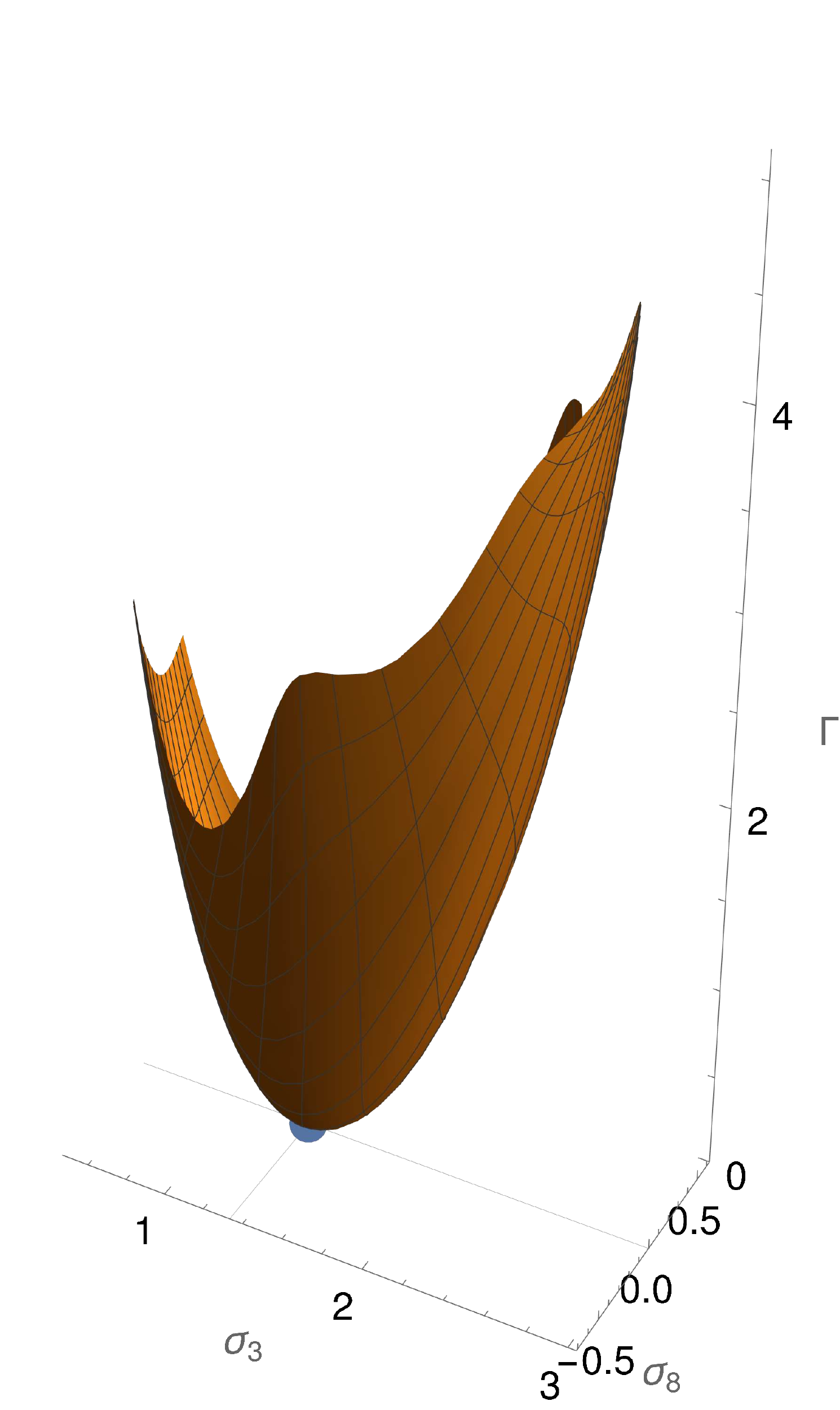}
    \caption{$V=24^4$}
  \end{subfigure}
  \caption{The quantum effective potential $\Gamma$ deep in the broken region close to $\gamma\approx 0$. The equilibration problems for increasing volumes result in steeper potentials with the minimum moving towards smaller classical field values $(\sigma_3,\sigma_8)$.}
  \label{fig:equilibration_qep}
\end{figure*}

There is, however, a very marked feature in some panels of figure \ref{fig:equilibration_vev}, e.g.\ at a volume of $20^4$ and $\beta=9.5$. After some time of gradual change there is a relatively instantaneous jump, after which the results stabilize. This feature can be traced back to Monte Carlo trajectories, which equilibrate on very different time scales. In fact, we observe either of two possibilities. One, which is the majority of Monte Carlo trajectories, which equilibrate slowly, and few suitable updates are found, and thus requiring more time to update. Hence, their Monte Carlo trajectories are short. These show a drift, and generically a $\U[1]\times\U[1]$ behaviour. The other type equilibrates very quickly, and show a marked and stable $\SU[2]\times\U[1]$ pattern. Since the updates are cheaper, there are more configurations available on these trajectories, and they therefore dominate the late Monte-Carlo time behaviour. However, since they are rare, they do not dominate early Monte-Carlo times. The jump reflects the crossover between both cases. This behaviour is illustrated in figure \ref{fig:app:trajectories}. As those with a $\U[1]\times\U[1]$ pattern show a slow drift towards the ones with $\SU[2]\times\U[1]$ pattern, it appears likely that the latter are the actually equilibrated case, and the others will eventually behave the same. As an additional check, simulations on a $4^4$ and $6^4$ lattice of the $\beta=10.0$ set have been carried out using two different algorithms, i.e.\ a multi-hit-metropolis and a new heatbath algorithm \cite{newhb:2023} that is currently under development. Running these simulations for exceedingly long Monte-Carlo trajectories showed that equilibration for these sets can be reached at very small volumes. In particular the obtained order parameters like the plaquette and the scalar length, agreed within error bars for these two methods, but have been slightly larger than the ones obtained from the unequilibrated simulations. In addition, after gauge-fixing the so-obtained configurations have primarily been very close to the $\SU[2]\times\U[1]$ pattern. Thus, at the moment we interpret this feature as supporting to be close to the $\SU[2]\times\U[1]$ pattern, but not more, with exceedingly long trajectories necessary to achieve this pattern.

For the calculation of the classical fields and effective potential in \cref{fig:qep} the equilibration errors cannot be compensated simply by increasing the error bars or by leaving out unequilibrated data, as this would leave us with no data at all in some cases. Therefore, we decided to show in \cref{fig:qep} data obtained on a smaller volume $V=16^4$ where the data is sufficiently enough equilibrated. For the \ZZ[2]-unbroken phase there is no significant difference in the potentials throughout different volumes except for the exact location of the minimum. Since the minimum of the quantum effective potential is always located at the classical field values $(\sigma_3,\sigma_8) = (\expval{\Sigma_3},\expval{\Sigma_8})$ this position is directly proportional to the value of $w$, which goes to zero for $V\to\infty$. For the \ZZ[2]-broken phase where we observe critical slowing down, we do know however what the result of the bad equilibration is, namely that the configurations are not yet fully equilibrated towards (almost) $\SU[2]\times\U[1]$-like configurations. This in turn means that the $\theta_0$ distributions look more Gaussian/flat centred around $0$ instead of doubly-peaked close to $\pm\pi/6$, as can be seen in \cref{fig:theta}, as well as the vev itself being smaller than expected, see \cref{fig:equilibration_vev}. This results in the effect that the effective potential becomes more localized (i.e. not flat in any direction), steeper and also moves to smaller values of the classical fields $\sigma_3$ and $\sigma_8$. This, can be seen in \cref{fig:equilibration_qep}.

\begin{figure}[t!]
  \centering
  \includegraphics[width=0.45\textwidth]{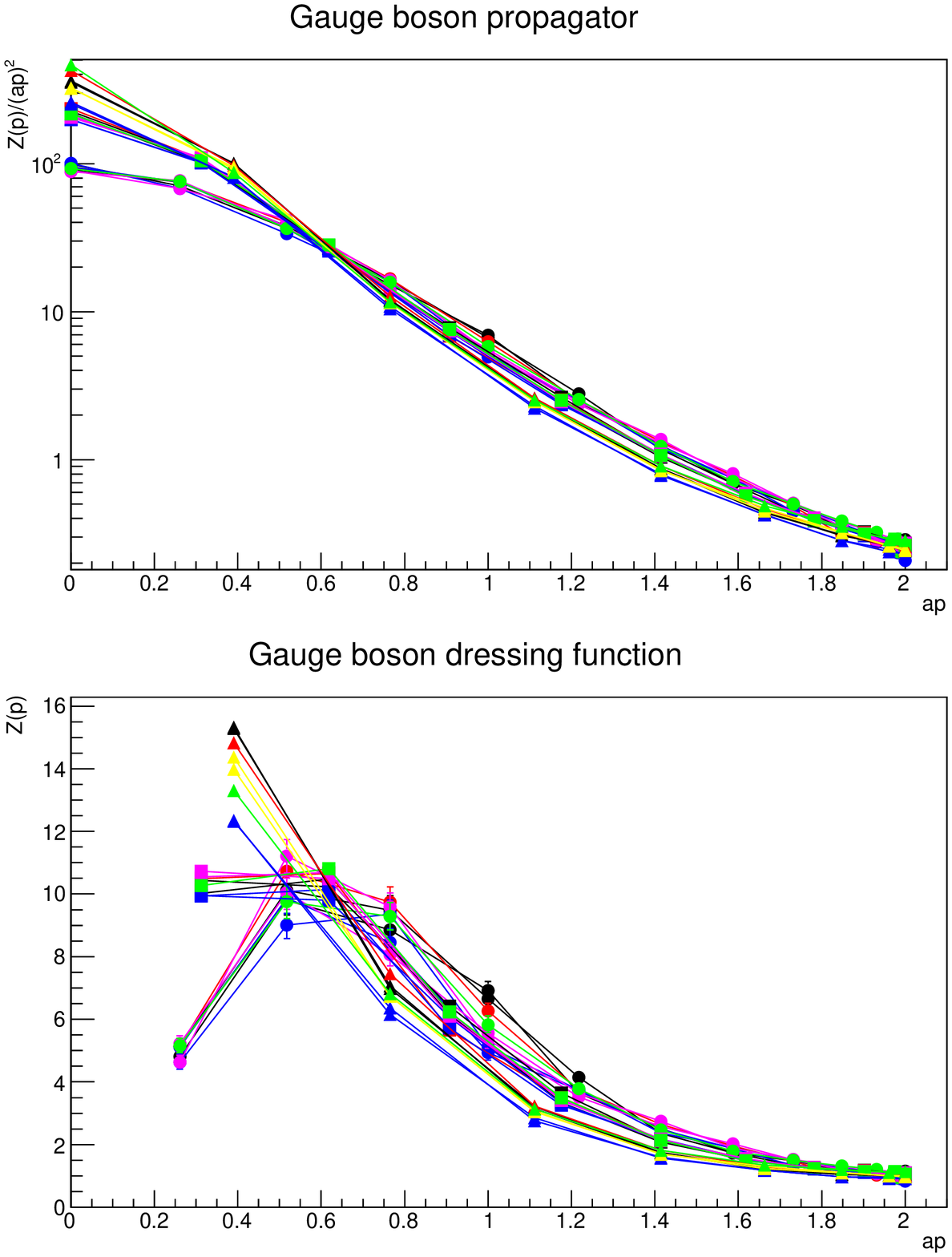}
  \includegraphics[width=0.45\textwidth]{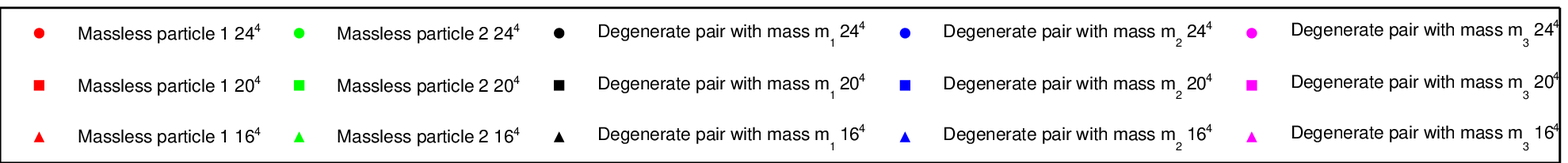}
  \caption{The gauge boson propagator (top panel) and the gauge boson dressing function (bottom panel) at $\beta=10$, $\kappa=1$, and $\gamma=0.05$ for the different lattice volumes 16$^4$, $20^4$, and $24^4$.}
  \label{fig:gp-app}
\end{figure}

At the same time, we also observe issues with the gauge-dependent correlation functions, most pronounced at $\beta=10$, as is shown in figure \ref{fig:gp-app}. On the smallest displayed volume, $16^4$, where the critical-slowing down process seems to be under control, it appears that there is indeed a splitting, but it is still of $\U[1]\times\U[1]$ type. Also, except for zero momentum, the massless gauge bosons are still behaving similarly to the massive ones, and the splitting between degenerate pairs seem not to be consistent with the tree-level values (\ref{eqn:boson_masses}). This pattern is already on the $20^4$ no longer present. On the other hand, the increase of volume drastically and qualitatively changes the propagator, an effect known very well from strongly-interacting theories \cite{Maas:2011se}. Thus, at the present time it is not possible to decide, whether physical effects, finite volume effects, or critical slowing down effects are the source for the loss of degeneracy. However, the results on $16^4$ volume suggests strongly that this is not alone a lattice artefact.

\subsection{Ghost propagator}\label{a:ghost}

\begin{figure*}[t!]
  \centering
  \begin{tabular}{cccc}
    \includegraphics[width=0.42\linewidth]{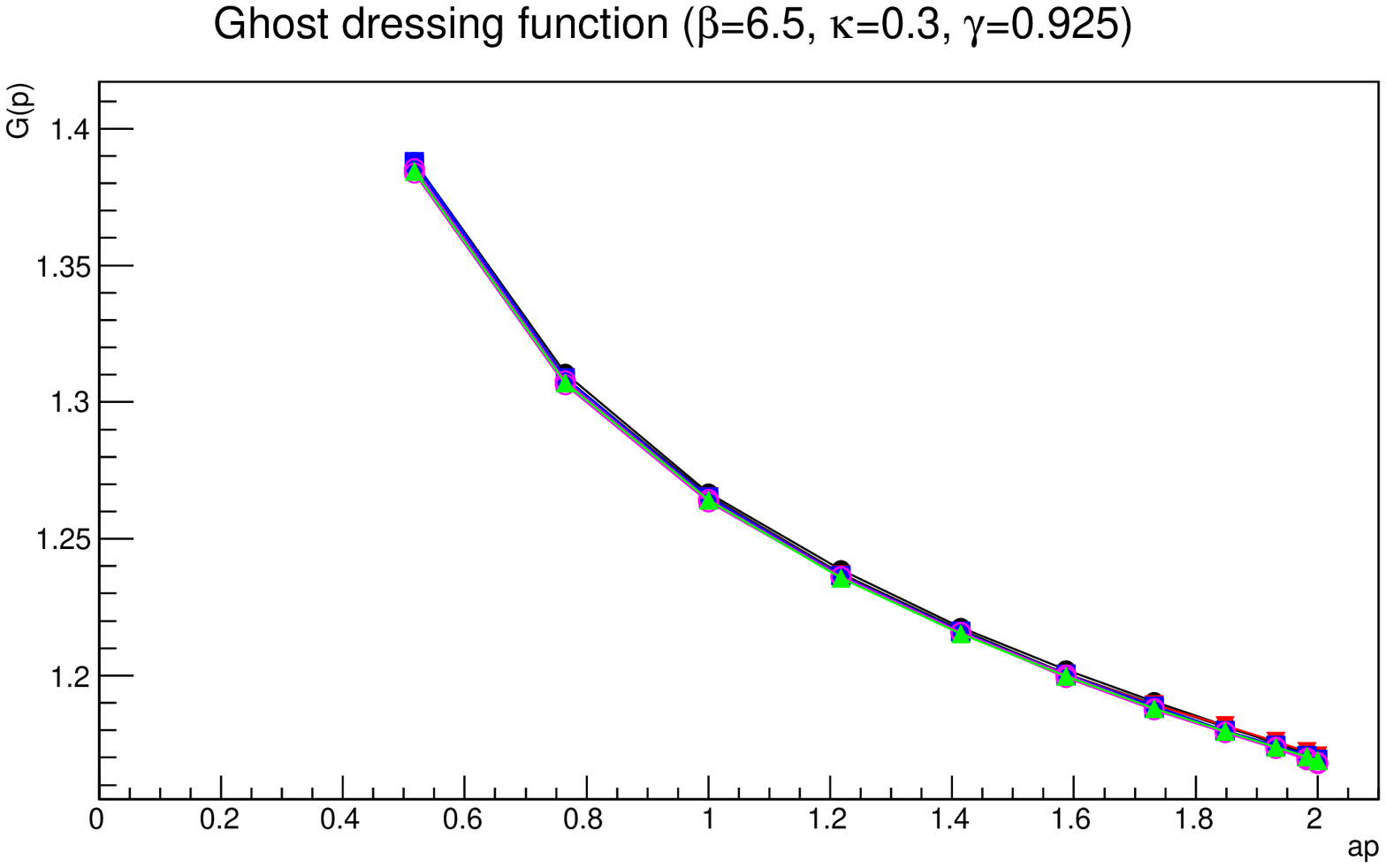}  &
    \includegraphics[width=0.42\linewidth]{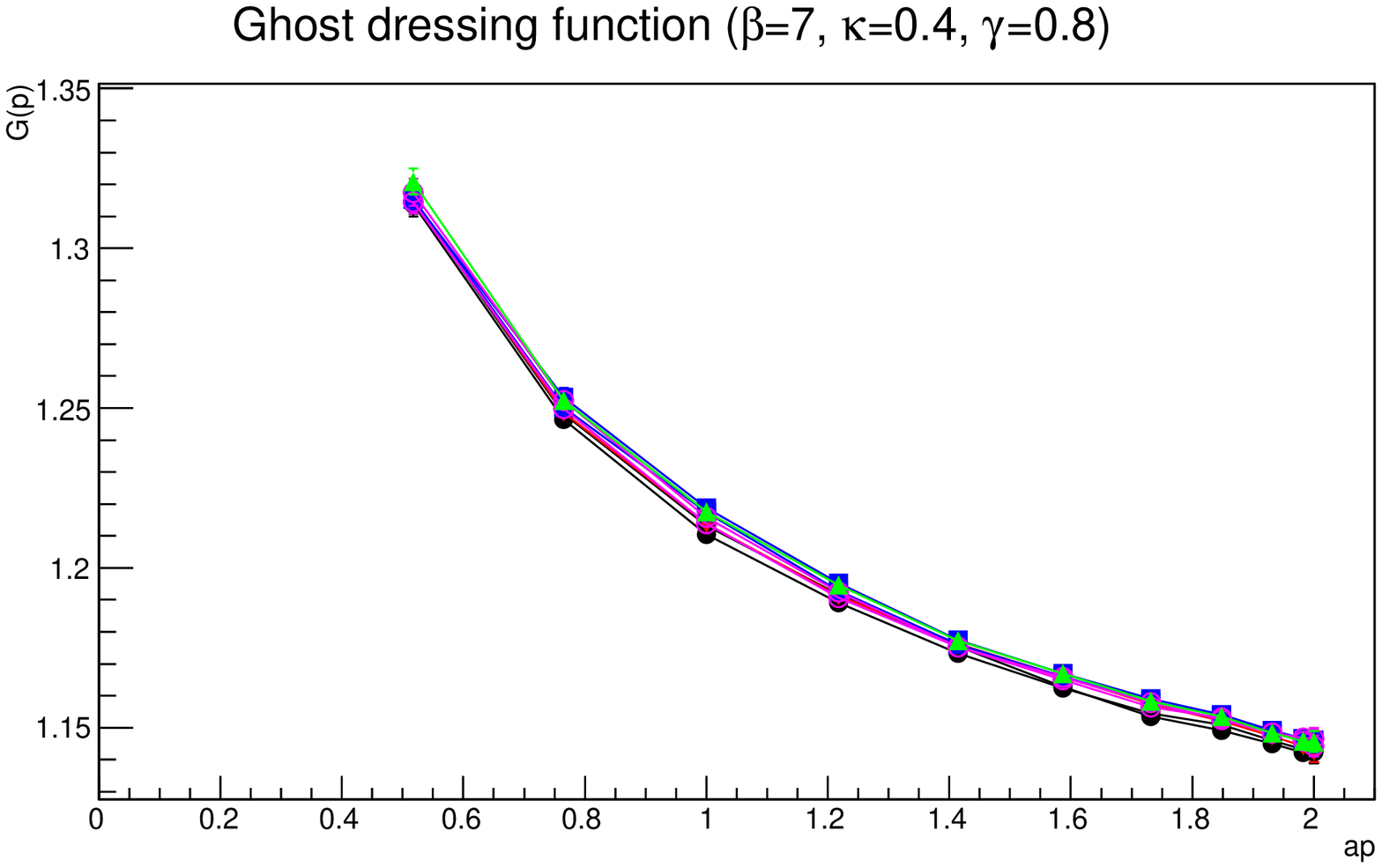}    & \\
    \includegraphics[width=0.42\linewidth]{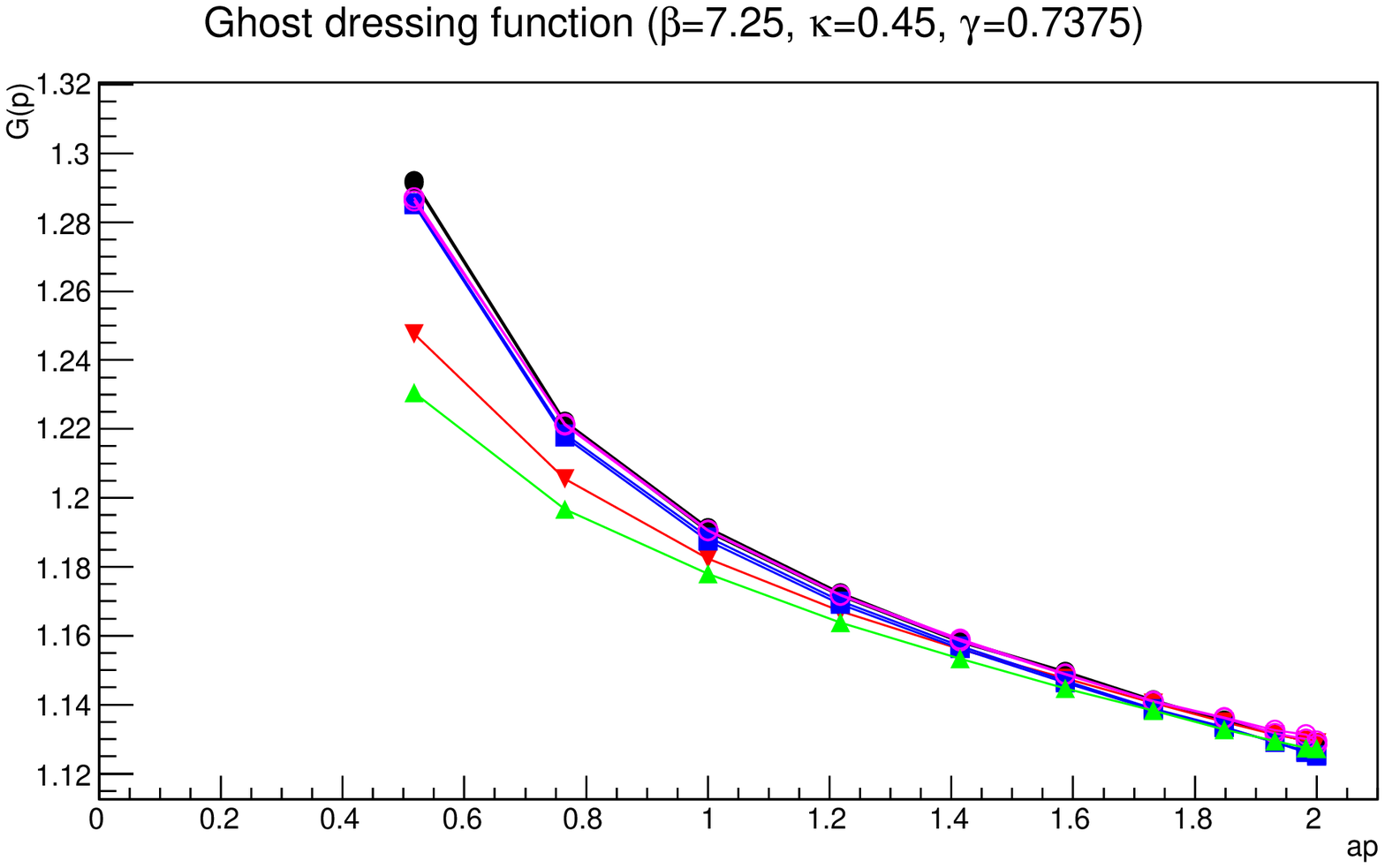} &
    \includegraphics[width=0.42\linewidth]{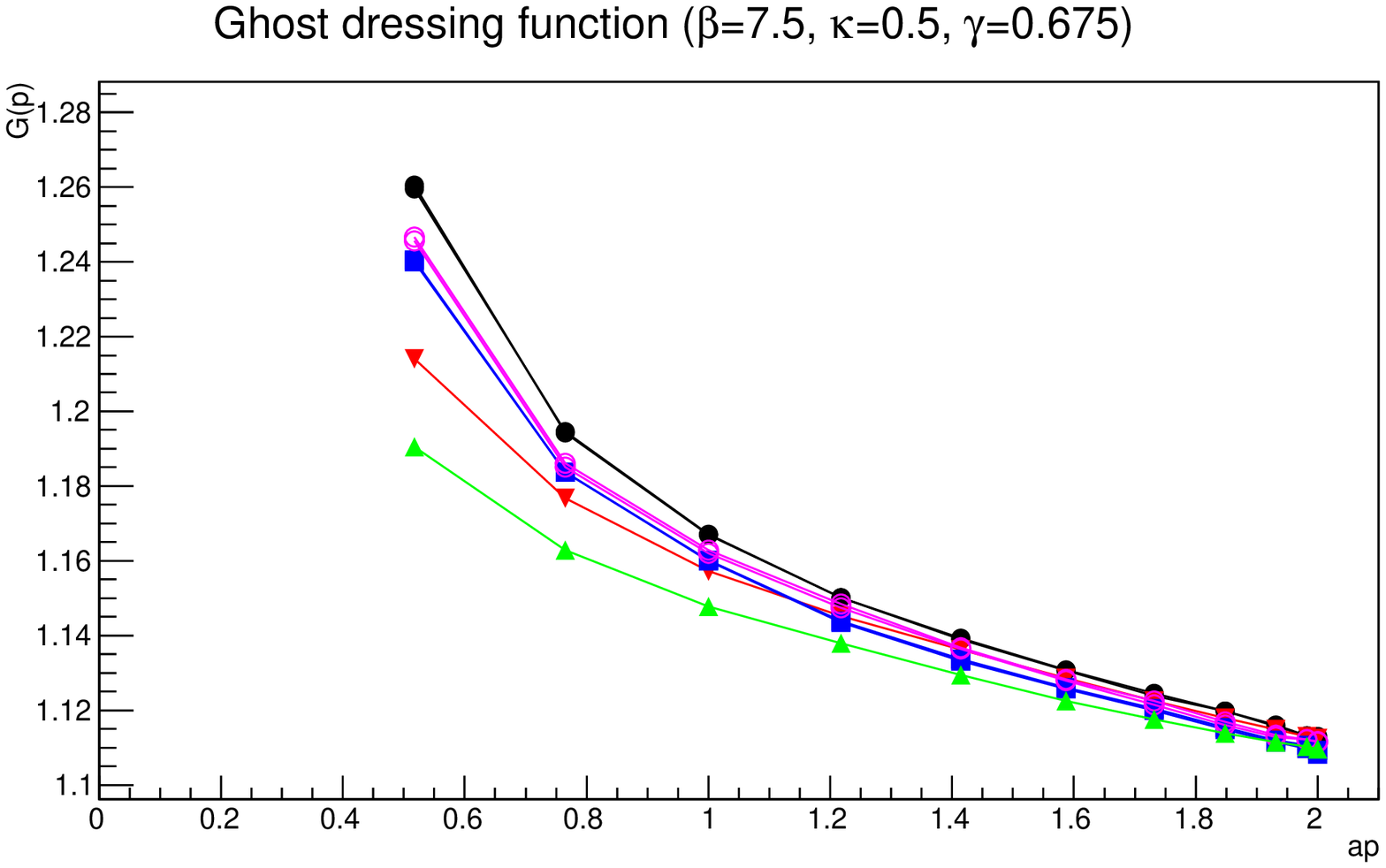}    \\
    \includegraphics[width=0.42\textwidth]{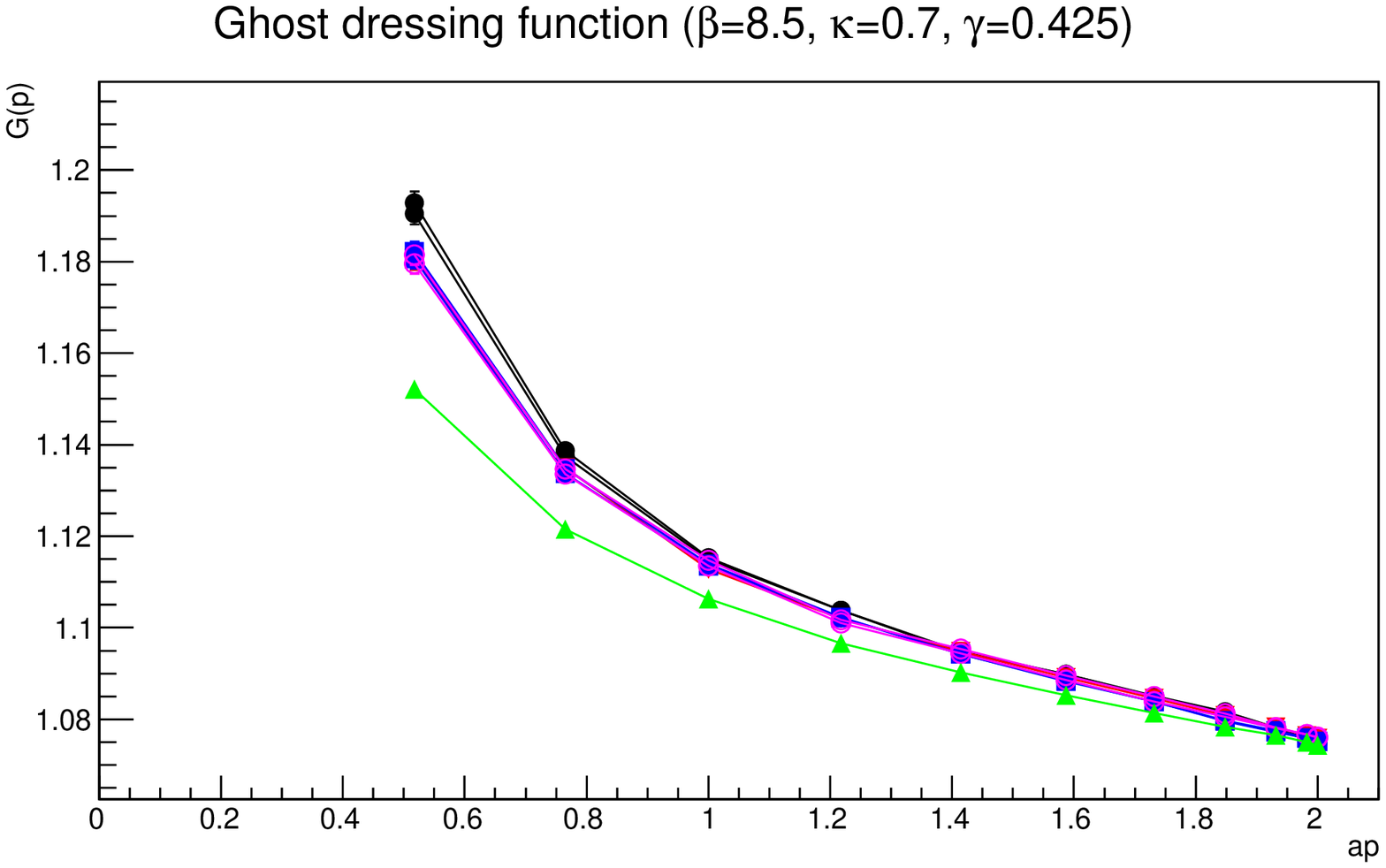}  &
    \includegraphics[width=0.42\textwidth]{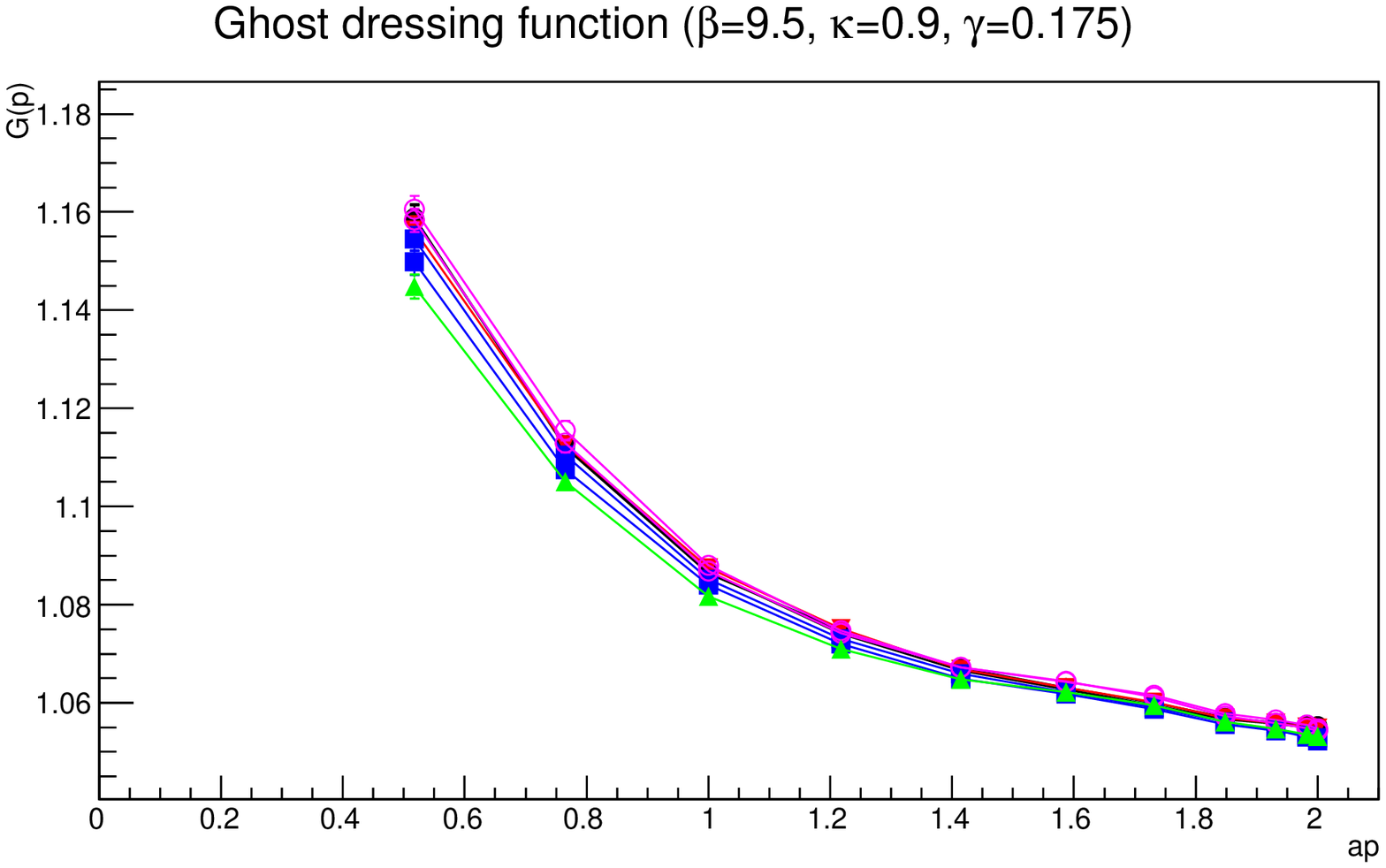}  & \\
    \includegraphics[width=0.42\textwidth]{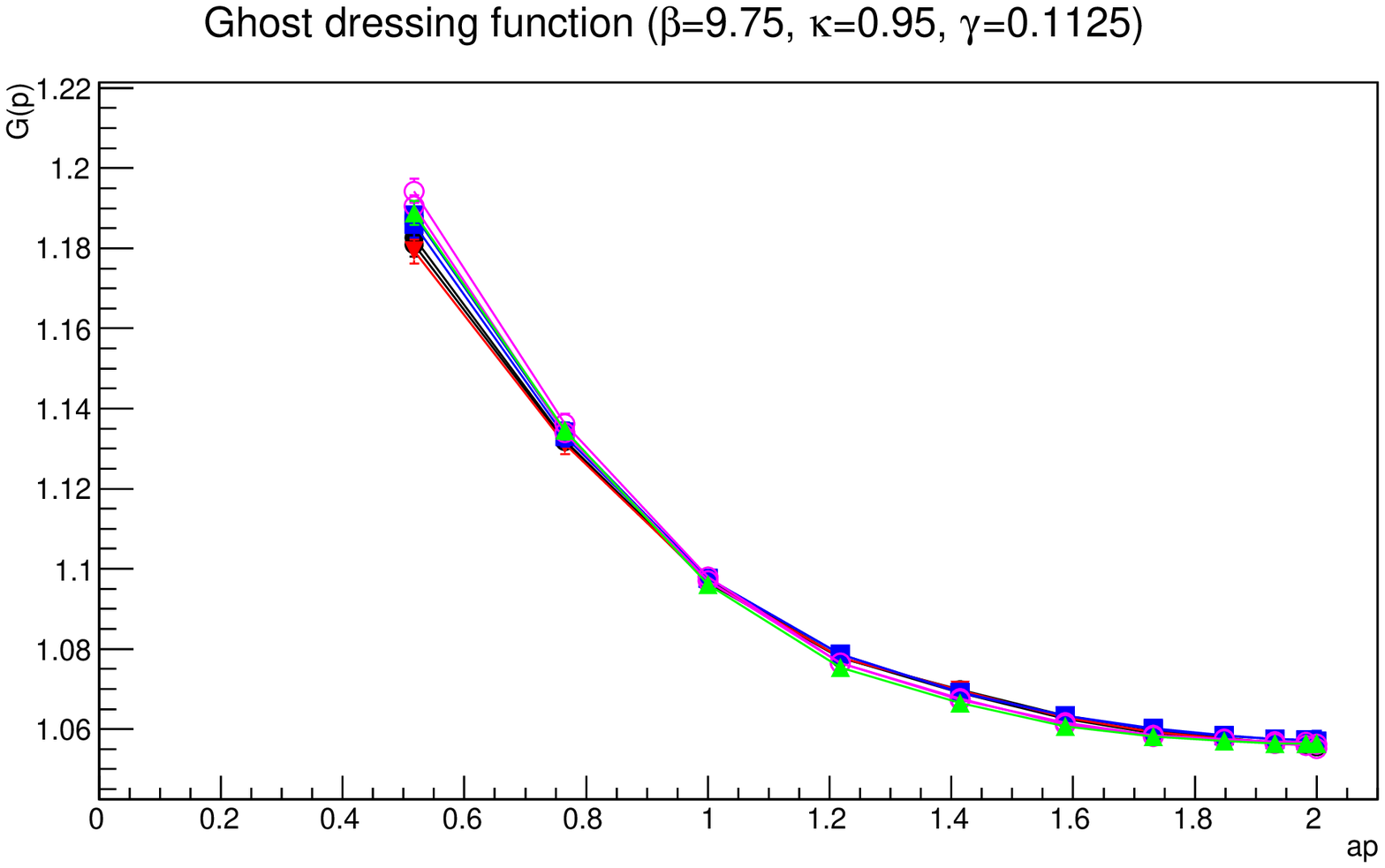} &
    \includegraphics[width=0.42\textwidth]{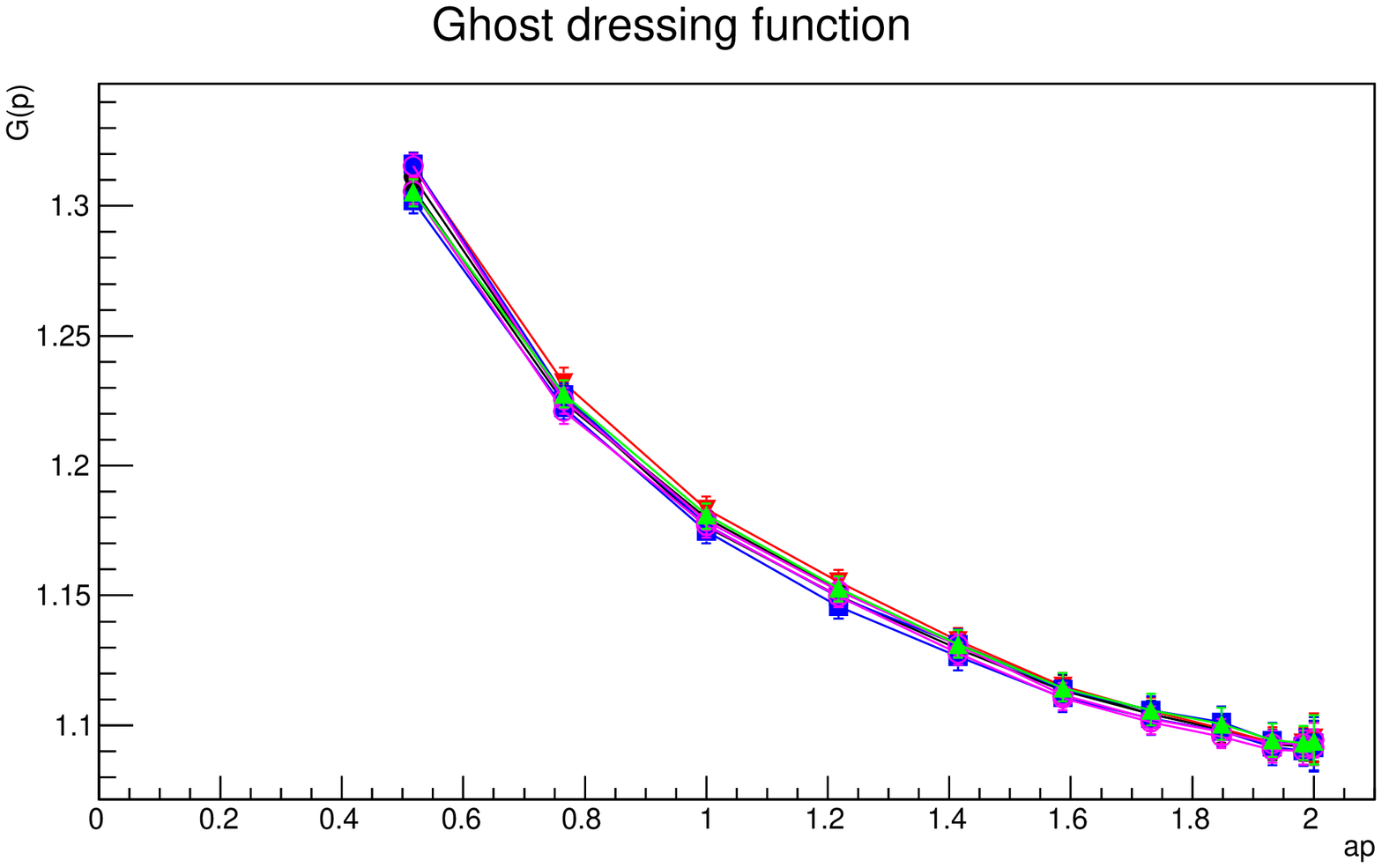}
  \end{tabular}\\
  \includegraphics[width=\textwidth]{{leg-gb}.eps}
  \caption{The ghost dressing function for the eight different charges moving from deep inside the unbroken phase ($\beta=6.5$, top right) to close to both sides of the \ZZ[2] phase transition ($\beta=7$ and $\beta=7.25$) through the broken phase ($\beta=7.5$, $\beta=8.5$, $\beta=9.5$, $\beta=9.75$) to the smallest numerically accessible value of $\gamma$ ($\beta=10$) in the lower-right panel. All results from the $24^4$ lattice, the momenta are along the $x$-axis, which minimizes finite-volume effects, and the first non-zero momentum point is suppressed due to remaining finite-volume effects.}
  \label{fig:ghp}
\end{figure*}

For completeness, in figure \ref{fig:ghp} also the ghost dressing function is shown, which together with the gauge boson dressing function of figure \ref{fig:gp} yields the running coupling (\ref{rcoupling}) in \cref{fig:alpha}. As has already been observed previously \cite{Maas:2014pba,Maas:2018xxu}, the ghost is much less impacted by the BEH effect. Of course, this is likely due to the fact that in Landau--'t Hooft gauge the ghosts are all massless at tree-level \cite{Bohm:2001yx}. Thus, only at loop level the degeneracy can be broken. Other than that, the ghosts show the same trends as a function of their parameters as the gauge bosons.

\bibliographystyle{apsrev4-2-mod.bst}
\bibliography{refs.bib}

\end{document}